\documentclass[useamsfonts]{pasj01}
\usepackage{graphicx}
\usepackage{natbib}
\Received{$\langle$reception date$\rangle$}
\Accepted{$\langle$acception date$\rangle$}
\Published{$\langle$publication date$\rangle$}

\begin{document}

\title{First-generation Science Cases for Ground-based Terahertz
Telescopes}

\author{Hiroyuki~\textsc{Hirashita}\altaffilmark{1}}
\author{Patrick~M.~\textsc{Koch}\altaffilmark{1}}
\author{Satoki~\textsc{Matsushita}\altaffilmark{1}}
\author{Shigehisa~\textsc{Takakuwa}\altaffilmark{1}}
\author{Masanori~\textsc{Nakamura}\altaffilmark{1}}
\author{Keiichi~\textsc{Asada}\altaffilmark{1}}
\author{Hauyu~Baobab~\textsc{Liu}\altaffilmark{1}}
\author{Yuji~\textsc{Urata}\altaffilmark{1,2}}
\author{Ming-Jye~\textsc{Wang}\altaffilmark{1}}
\author{Wei-Hao~\textsc{Wang}\altaffilmark{1}}
\author{Satoko~\textsc{Takahashi}\altaffilmark{3,4}}
\author{Ya-Wen~\textsc{Tang}\altaffilmark{1}}
\author{Hsian-Hong~\textsc{Chang}\altaffilmark{1}}
\author{Kuiyun~\textsc{Huang}\altaffilmark{5}}
\author{Oscar~\textsc{Morata}\altaffilmark{1}}
\author{Masaaki~\textsc{Otsuka}\altaffilmark{1}}
\author{Kai-Yang~\textsc{Lin}\altaffilmark{1}}
\author{An-Li~\textsc{Tsai}\altaffilmark{2}}
\author{Yen-Ting~\textsc{Lin}\altaffilmark{1}}
\author{Sundar~\textsc{Srinivasan}\altaffilmark{1}}
\author{Pierre~\textsc{Martin-Cocher}\altaffilmark{1}}
\author{Hung-Yi~\textsc{Pu}\altaffilmark{1}}
\author{Francisca~\textsc{Kemper}\altaffilmark{1}}
\author{Nimesh~\textsc{Patel}\altaffilmark{6}}
\author{Paul~\textsc{Grimes}\altaffilmark{6}}
\author{Yau-De~\textsc{Huang}\altaffilmark{1}}
\author{Chih-Chiang~\textsc{Han}\altaffilmark{1}}
\author{Yen-Ru~\textsc{Huang}\altaffilmark{1}}
\author{Hiroaki~\textsc{Nishioka}\altaffilmark{1}}
\author{Lupin~Chun-Che~\textsc{Lin}\altaffilmark{1}}
\author{Qizhou~\textsc{Zhang}\altaffilmark{6}}
\author{Eric~\textsc{Keto}\altaffilmark{6}}
\author{Roberto~\textsc{Burgos}\altaffilmark{6}}
\author{Ming-Tang~\textsc{Chen}\altaffilmark{1}}
\author{Makoto~\textsc{Inoue}\altaffilmark{1}}
\author{Paul~T.~P.~\textsc{Ho}\altaffilmark{1,7}}

\altaffiltext{1}{Institute of Astronomy and Astrophysics, Academia Sinica,
P.O. Box 23-141, Taipei 10617, Taiwan}
\altaffiltext{2}{Institute of Astronomy, National Central University,
Chung-Li 32054, Taiwan}
\altaffiltext{3}{Joint ALMA Observatory, Alonso de Cordova 3108, Vitacura, Santiago, Chile}
\altaffiltext{4}{National Astronomical Observatory of Japan, 2-21-1 Osawa, Mitaka, Tokyo 181-8588, Japan}
\altaffiltext{5}{Department of Mathematics and Science, National Taiwan Normal University, Lin-kou District, New
Taipei City 24449, Taiwan}
\altaffiltext{6}{Harvard-Smithsonian Center for Astrophysics, Cambridge, MA 02138, USA}
\altaffiltext{7}{East Asian Observatory, 660 N. Aohoku Place, University Park,
Hilo, Hawaii 96720, USA}
\email{hirashita@asiaa.sinica.edu.tw}


\maketitle

\begin{abstract}
Ground-based observations at terahertz (THz) frequencies are
a newly explorable area of astronomy for the next ten years.
We discuss science cases for a first-generation 10-m class
THz telescope, focusing on the Greenland Telescope as
an example of such a facility. We propose science cases and
provide quantitative estimates for each case.
The largest advantage of ground-based THz telescopes
is their higher angular resolution ($\sim 4$ arcsec for
a 10-m dish), as compared to space or airborne THz telescopes.
Thus, high-resolution mapping
is an important scientific argument. In particular,
we can isolate zones  of interest for Galactic and
extragalactic star-forming regions.
The THz windows are suitable for
observations of high-excitation CO lines and
[N \textsc{ii}] 205 $\mu$m lines, which are scientifically
relevant tracers of star formation and stellar feedback.
Those lines are the brightest lines in the THz windows,
so that they are suitable for the initiation of ground-based
THz observations. THz polarization of star-forming regions
can also be explored since it traces the dust population
contributing to the THz spectral peak.
For survey-type observations, we focus on
``sub-THz'' extragalactic surveys, whose uniqueness
is to detect galaxies at redshifts $z\sim 1$--2, where
the dust emission per comoving volume is the
largest in the history of the Universe.
Finally we explore possibilities of flexible
time scheduling, which enables us to monitor
active galactic nuclei, and to target
gamma-ray burst afterglows. For these objects,
THz and submillimeter wavelength ranges have not yet been
explored.
\end{abstract}

\section{Introduction}

The access to the terahertz (THz) frequency range or
far-infrared (FIR) wavelength range from the ground is
mostly limited by the absorption of water vapor in the Earth's
atmosphere. Therefore, the THz region is one of the remaining unexplored wavelength
ranges from the ground. Space, balloon-borne, and airborne observations
have so far been used to explore THz astronomy.
There are only limited sites on Earth where the
THz windows are accessible, and ground-based THz astronomy is, indeed,
possible. Greenland (Section \ref{subsec:glt}),
high-altitude ($>$5,000 m) Chilean sites
\citep{matsushita99,paine00,matsushita03,peterson03}, and
Antarctica \citep{yang10a,tremblin11}
are examples of suitable
places for ground-based THz astronomy.

Even in those locations suitable for THz observations,
the time of excellent weather is still limited. Therefore,
observations need
to be planned well in order to maximize the
scientific output within the limited amount of observing
time. Moreover, ``first-generation'' observations are
especially important because they determine the direction of
subsequent THz science. There have already been
some pioneering efforts of ground-based THz observations
such as with the Receiver Lab Telescope (RLT; \citealt{marrone04}),
the Atacama Pathfinder Experiment (APEX; \citealt{wiedner06}), and
the Atacama Submillimeter Telescope Experiment (ASTE; \citealt{shiino13}).
However, the attempts by these
existing facilities have been difficult and sparse due to
challenging weather conditions. A dedicated telescope
at an excellent site -- as proposed for the GLT -- is, thus,
paramount for successful THz observations.
In practice, such first-generation
observations are associated with the development of THz
detectors. Searching for targets that are relatively easy
to observe but scientifically pioneering, is of fundamental
importance to maximize the scientific value of the
instrumental development.
The first aim of this paper is, thus, to search for
scientifically important and suitable targets for
first-generation THz science cases.

The largest advantage of ground-based THz telescopes
compared with space telescopes (e.g., \textit{Herschel};
\citealt{pilbratt10}), airborne telescopes (e.g.,
the Stratospheric Observatory for Infrared Astronomy (SOFIA);
\citealt{young12}), and balloon-borne telescopes
is that it is possible to operate large dishes with a high
diffraction-limited resolving capability. To clarify this advantage,
it is convenient to have a specific telescope in mind
in discussing science cases. In this paper, we focus on
the Greenland Telescope (GLT). The GLT project is planning to
deploy an
Atacama Large Millimeter/submillimeter (ALMA)-prototype 12-m antenna 
to the Summit Station in Greenland
(3,200 m altitude) for the purpose of using it as part of
submillimter (submm) very long baseline interferometry (VLBI)
telescopes \citep[][see also Section \ref{sec:glt}]{inoue14a}.
Data of atmospheric transmission at the Summit Station have been accumulated over the 
past four years through our continuous monitoring campaign
\citep[][see also Section \ref{subsec:atmosphere}]{martin14}.
Our analysis indicates that site conditions are suitable for 
a first-generation ground-based THz facility. 
Therefore, we mainly target the GLT in this paper, noting
that the THz science cases will likely be similar or even common for all 
the first-generation THz telescopes except for their sky coverages.


This paper also provides a first basis for ground-based
THz observations for later generations of telescopes with larger
dishes, such as the initially proposed Cerro Chajnantor Atacama
Telescope
(CCAT)\footnote{http://www.ccatobservatory.org/} or any successor project.
With a diameter twice as large as the GLT, these telescopes will
have even better sensitivity and resolution to further push the scientific 
results achieved by the GLT.


One of the major scientific advantages of the THz
regime is that thermal dust continuum emission will be
measured around its peak in the spectral energy
distribution (SED). Moreover, the angular resolution
(4$''$ at 1.5 THz for the diffraction-limited primary beam)
enables us to spatially resolve the individual
star-formation sites within nearby molecular clouds,
typically located within a distance of $\sim$300 pc
(Section \ref{subsubsec:THz_SF}).
This high resolution is also an advantage for
extragalactic observations where star-formation activities
within nearby galaxies will be resolved.
(Section \ref{subsec:extragal}). We emphasize that a resolution of
$\sim 4''$ is comparable to what is achieved with submm
interferometers such as the Submillimeter Array
(SMA; \citealt{ho04}),\footnote{The SMA is a joint project between the Smithsonian Astrophysical
Observatory and the Academia Sinica Institute
of Astronomy and Astrophysics and is funded by
the Smithsonian Institution and the Academia Sinica.}
although ALMA can have a substantially better spatial resolution.
 {No THz facility, including non-ground-based observatories, 
has ever routinely had such a resolution for imaging.}
For dust continuum, combining submm
interferometric data
with new 1.5 THz data will significantly improve
dust temperature estimates.

A wealth of interesting but unexplored emission lines
are also in the THz windows
(Section \ref{subsec:line}). Highly exited rotational molecular lines (e.g., CO, HCN) will probe ``extremely
hot'' (300--500 K) molecular regions, which have been missed in observations at
longer wavelengths.  Line profiles will reveal gas motions in 
these regions. Additional lines accessible in the THz windows
will be groups of atomic fine-structure lines tracing diffuse transitional regions in the interstellar
medium (ISM) from
ionized or atomic gas to molecular gas,
and pure rotational lines (e.g., CH) tracing chemically
basic light molecules. 

In summary, the THz frequencies are suitable for tracing some
key chemical species in gas and solid materials often associated
with star-formation activities. Therefore, the grand aim of
THz science is to trace
the processes in the ISM that lead to
star formation (see Section \ref{sec:key} for more detailed
discussion).

Since the weather conditions necessary for THz observations
are only realized for a small fraction of the winter time
(typically $\sim 10$\%; Section \ref{subsec:glt}),
it is worth considering ``sub-THz'' observations as well
(Section \ref{sec:subTHz}).
Some survey observations in the 850 GHz and 650 GHz windows can be
unique for THz telescopes, because these relatively high-frequency
submm bands still remain difficult to be fully explored at those sites where
the current submm telescopes are located.

We will also discuss the possibility of flexible scheduling of
observing time, because the GLT will be capable of doing this.
One of the largest advantages of
such flexible time allocations is that we will be able
to execute time-consuming surveys. We will also
allocate monitoring and target-of-opportunity (ToO) observations.
In the field of very-high-energy (VHE) phenomena, THz continuum
observations can help to constrain the mechanism and region of
origin of the VHE in active galactic nuclei (AGNs) and
gamma-ray bursts (GRBs) (Section \ref{sec:monitoring}).  In particular,
current multi-frequency monitoring campaigns from optical to X- and
gamma-rays lack observations at THz or sub-THz frequencies for a complete SED
to constrain the underlying physics of the origin of VHE.
THz continuum observations will provide clean
measurements of the source intensity, without being affected by
scintillation and extinction.

In this paper, we aim at exploring the scientific importance
of ground-based THz observations, providing
some quantitative estimates.
For clarification, we focus on the GLT as one of the first-generation
ground-based facilities, but the scientific discussions are
general enough to be applied to any 10-m class THz
telescope. This paper is organized as follows: we start by
describing the features of the GLT as an example of future
first-generation
THz telescopes in Section \ref{sec:glt}. We discuss the key THz science cases
in Section \ref{sec:key}, and some sub-THz cases in
Section \ref{sec:subTHz}. We also describe some science
cases that are making maximal use of a flexible time
allocation in Section \ref{sec:monitoring}. Based on the
science cases discussed in this paper, we also
summarize possible developments of
future instruments in Section \ref{sec:development}. Finally,
we give a summary in Section \ref{sec:summary}.
We use $(h,\,\Omega_\mathrm{m},\,\Omega_\Lambda)
=(0.71,\,0.27,\,0.73)$
for the cosmological parameters.

\section{Project Overview and General Requirements}\label{sec:glt}

As emphasized above, the scope of this paper is not limited
to a specific future project, but rather 
evaluating quantitatively some key science cases suitable
for the first-generation ground-based THz telescopes.
Here, we introduce the GLT as a typical telescope for this
purpose, only to make clear what kind of capabilities would
be expected in the near future for THz observations. We
refer the interested reader to \citet{inoue14a} and 
\citet{grimes14} for the
details of specifications and possible instruments
for the GLT.

 {We also emphasize that, by considering the ``realistic''
case of the GLT, we will be able to extract the important aspects
and requirements in
performing ground-based observations. Indeed, as we
demonstrate below, the following
advantages of the GLT are of fundamental importance in
realizing ground-based THz observations:
(i) Atmospheric transmission (in the winter time,
the atmospheric condition is statistically better than
the Mauna Kea site \citep{martin14}, and comparable to the ALMA site);
(ii) stability of the atmospheric condition
(in the winter time, there is no daylight and only minor
daily temperature variations, so that the weather
condition suitable for THz observations can last $>$ a day);
(iii) flexible scheduling
(except for the time occupied by the VLBI observations,
the GLT can be used as a dedicated telescope for THz
observations). In what follows, we will show that the
GLT really has these advantages that are
fundamentally important to the success of ground-based
THz observations.}

\subsection{Brief introduction to the Greenland Telescope (GLT)}
\label{subsec:glt}

The GLT is an ALMA -- North-American 12-m
Vertex prototype antenna \citep{mangum06} that is being
retrofitted  to arctic conditions, with the goal of 
deploying it to the Greenland Summit Station\footnote{http://www.summitcamp.org/} 
at an altitude
of 3,200 m (located at latitude 72\fdg57N
and longitude 38\fdg46W). 
Primary operating conditions for the GLT -- defined as 
extending over 90\% of the weather conditions -- cover an ambient
temperature down to $-50^{\circ}$C, wind speeds of up to 11 m s$^{-1}$ and a vertical 
temperature gradient (due to an inversion layer over the Greenlandic plateau)
of $+7$ K over the antenna height from bottom to top. Furthermore, an ambient temperature 
change rate of up to 2 K hr$^{-1}$ is taken into account.
Absolute (non-repeatable) pointing errors of 2\arcsec with a final goal of 1\farcs4 are targeted
for primary conditions, together with offset pointing and tracking errors of 0\farcs6
and a final 0\farcs4. An antenna surface accuracy of 10~$\mu$m is targeted. For THz
observations around 1.5~THz this yields an antenna surface efficiency of about 65\%. 
This efficiency will drop to 40\% and 20\% for 15~$\mu$m and 20~$\mu$m surface accuracies.
Secondary operating conditions -- additionally covering the 90--95\% range of weather
occurrence -- extend the ambient temperature further down to $-55^{\circ}$C and the 
wind speeds up to 13 m s$^{-1}$. A degraded antenna performance is accepted under these
conditions.

The exceptionally dry atmospheric conditions at this site
are suitable for submm and even for FIR (THz) observations
(Section \ref{subsec:atmosphere}).
The principal purpose of the GLT is to provide a 
submm-VLBI station that can be correlated with other telescopes such
as the SMA, the James Clerk Maxwell Telescope (JCMT),
and ALMA in order to achieve extremely long
intercontinental baselines \citep{inoue14a}. This will greatly enhance the angular resolution at
submm wavelengths.
The key scientific goal of this interferometry is to measure the shadow caused
by the strong gravitational
field around the central supermassive black hole
in M87. This project will provide a unique opportunity to study the strong
general relativistic
effects immediately surrounding the black hole's event horizon.
Related scientific topics include measuring the black hole spin in M87, constraining
the nature of its accretion flow and determining the launching mechanism
of its relativistic jet. Additional science cases are the very high energy
emission in AGNs and the exploration of dark energy through high-precision measurements
of locations and velocities of maser spots in galaxies.

The GLT will likely be used for VLBI observations only during
a short period within each year (roughly 1--2 months),
because many telescopes must be made available simultaneously.
Therefore, we expect that a significant fraction of
observing time can be used for single-dish observations. Below, we
describe the capability of the GLT as a single-dish telescope.

\subsection{Atmospheric condition}\label{subsec:atmosphere}

\begin{figure}
\begin{center}
\includegraphics[width=0.43\textwidth]{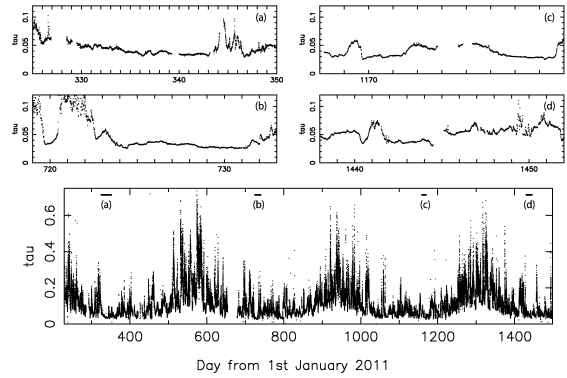}
\end{center}
\caption{\bf
Atmospheric opacity at 225 GHz from August 17th 2011 to February 7th 2015.
Selected low opacity epochs of several days are indicated with the labels (a), (b),
(c), (d) in the bottom plot. The zoomed-in plots are shown in the top
four panels. 
The relation between the 225 GHz opacity $\tau$ and the PWV (in units of mm)
was calculated with {\it am} \citep{paine2012},
and expressed as
$\tau = 0.049\mathrm{PWV} + 0.018$ (valid in October--May for $\tau < 0.075$)
or
$\tau = 0.047\mathrm{PWV} + 0.015$ (valid in June--September for $\tau < 0.2$).
}
\label{fig:transmission}
\end{figure}

\begin{figure}
\begin{center}
\includegraphics[width=0.45\textwidth]{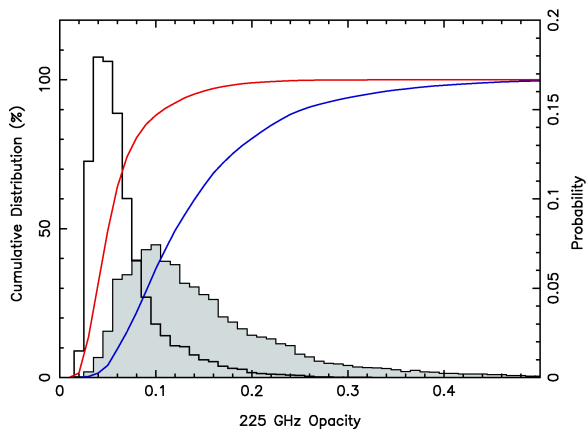}
\end{center}
\caption{\bf
Histograms of the 225 GHz opacity and cumulative distribution functions
for the data shown in Fig.~\ref{fig:transmission}.
The blue line and shaded histogram show the statistics in the summer
months (May--October) while the red line and white histogram present
the statistics in the winter months (November--April).
}
\label{fig:histogram}
\end{figure}

In 2011, we deployed a radiometer to Greenland to monitor
the sky and weather conditions at the Summit Station 
\citep{martin14}. The average temperature is low at around $-$30 to $-40^{\circ}$C, down to $-70^{\circ}$C in winter. 
Because of such low ambient temperatures, the water content in the air is exceedingly low. 
 {Every 10 minutes, a tipping measurement is performed to take
opacity data.
The measured 225 GHz atmospheric opacity as a function of time is
shown in Figure \ref{fig:transmission}.The corresponding histogram together with its cumulative distribution function (CDF) is depicted in
Figure \ref{fig:histogram}.}
We confirmed that the opacities lower quartile in these months can get as low as 0.047, with occasional opacities as low as 0.030 in the winter regime (November to April; see also the zoomed plots in Figure \ref{fig:transmission}).
Based on the monitoring opacity ($\tau$) data at 225 GHz, we estimated the precipitable water vapor (PWV), and then converted that into an atmospheric opacity at 1.5 THz 
using the ``{\it am}'' program and climatological data from Summit Camp for temperature, relative humidity and ozone profiles \citep[]{paine2012}.
 {For convenience, we give the conversion
formula between $\tau$ and PWV in Figure \ref{fig:transmission}.}
The observational rate with THz opacity ($\tau_\mathrm{THz}$) condition less than 2.0 and 3.0 were 1.8\% and 11.4\% in the winter regime (November to April).  
These fractions are similar to those expected for the ALMA site (2.0\% for $\tau_\mathrm{THz}<2.0$ and 16.8\% for $\tau_\mathrm{THz}<3.0$) in the winter season \citep[]{matsushita2011}. 
It is noteworthy that THz opacity conditions giving $\tau_\mathrm{THz}<3.0$ can last from a few hours to half a day in many cases.   {In some cases, they last up to 3--7 days (see Figure \ref{fig:transmission}a, b), which is only possible in the polar regions where there is no daylight in the winter time.
If we focus on the months when such long durations of THz weather
are achieved, the statistics of atmorpheric opacity at the
Greenland Summit site are as good as at the Antarctic site.
}
We also estimated the opacity condition less than 0.6 and 1.0 at sub-THz bands.
Those were 8.2\% and 32.9\% at 675 GHz and 3.9\% and 24.2\% at 875 GHz, respectively. 

 {\citet{tremblin12} have presented the atmospheric conditions
at various sites suitable for submm astronomy using satellite data,
including Greenland. However, their data
do not have a time resolution of less than $\sim$ a day, while
time variations within a day are important for THz
observations. Moreover, at the polar sites, the statistics
focusing on the winter season is meaningful because
the night continues over the whole winter period.
We, therefore, think that our new opacity statistics is more
reliable for the THz observing conditions.
}

 {Although Dome A in Antarctica would be better than
the Greenland Summit Station, the GLT can play a unique
role because it is still the best site in the northern hemisphere.
In other words, the GLT and telescopes in Antarctica are complementary
in their sky coverages. We emphasize that the science cases discussed
in this paper are applicable to any THz telescope with a
different fraction of time suitable for THz observations, as long
as we adopt the same criterion for the THz weather ($\tau_\mathrm{THz}<2.0$
or 3.0).
}

\subsection{Typical performance of the GLT}\label{subsec:capability}


The expected observational performance of the GLT at
the frequencies of atmospheric windows is summarized
in Table~\ref{tab:capability}, which is revised from \citet{grimes14},
for the frequency range of interest in this paper (350--1,500 GHz).
For this calculation, we assume the surface accuracy to be 15 $\mu$m.
Indeed, the telescope was demonstrated to have a surface accuracy
of 16 $\mu$m when it was at the evaluation phase as the ALMA prototype
antenna \citep{mangum06}. 
Using the best 10\% and median opacity conditions for the frequencies lower than 1 THz and the best 1.5\% and the best 10\% opacity conditions for the frequency higher than 1 THz at the winter season and
the expected receiver performance (i.e., receiver temperature $T_\mathrm{Rec}$),
we estimated the system temperature ($T_\mathrm{sys}$) for each frequency band,
and corresponding continuum noise equivalent flux density for 1 second
integration (NEFD) and spectral NEFD for 1 km s$^{-1}$ resolution.

\begin{table*}
\tbl{Expected performance of the GLT.}{%
\begin{tabular}{lcccccc}
\hline
Frequency & Resolution & Transmission & $T_\mathrm{Rec}$ & $T_\mathrm{sys}$ & Continuum NEFD & Spectral NEFD \\
(GHz)     & (arcsec)   & (\%)         & (K)              & (K)              & (mJy s$^{1/2}$)    & [Jy s$^{1/2}$ (km s$^{-1}$)$^{1/2}$]\\
\hline
 230 & 27   & 0.96 (0.91) &  40 &     89 (110)    &     33 (41)     &   3.39 (4.18) \\
 345 & 18   & 0.89 (0.82) &  75 &    177 (192)    &     68 (73)     &   5.61 (6.06) \\
 675 &  9.2 & 0.53 (0.28) & 110 &    663 (1,510)  &    285 (648)    &  17.3 (39.4) \\
 875 &  7.1 & 0.48 (0.24) & 160 &  1,060 (2,450)  &    511 (1,190)  &  27.1 (63.0) \\
1020 &  6.1 & 0.16 (0.06) & 525 &  9,110 (26,500) &  5,120 (14,900) & 245 (712) \\
1350 &  4.6 & 0.16 (0.06) & 650 & 13,500 (36,100) &  9,600 (25,600) & 413 (1,100) \\
1500 &  4.1 & 0.15 (0.06) & 750 & 20,300 (55,100) & 17,900 (48,700) & 718 (1,950) \\
\hline
\end{tabular}}\label{tab:capability}
\begin{tabnote}
Note: This table is an updated version of \citet{grimes14}; in this paper, we used the up-to-date opacity statistics mentioned in the previous subsection.
Opacity, $T_\mathrm{sys}$, and NEFDs are given for two cases:
the best 10\% and median (in parenthesis) weather for the frequency lower than 1 THz, and the best 1.5\% and the best 10\% (in parenthesis) for the frequency higher than 1 THz. Both cases are based on the opacity statistics of the winter season (from November to April).
\end{tabnote}
\end{table*}



In practice, we typically consider to observe the sky area with
a zenith angle $<60^\circ$. With the latitude of the site,
$72^\circ$, therefore, we limit the declination to
$\delta >12^\circ$.

\subsection{Advantages of ground-based telescopes for THz astronomy}

The advantage of ground-based observations is that we can use a
large telescope. As shown in Section \ref{subsec:capability},
the angular resolution achieved is about 4\arcsec at 1.5 THz
for a 10-m class ground-based telescope, while
the diffraction limit of a 3.5-m-class space telescope,
such as \textit{Herschel}, is $\sim 12$\arcsec
around 1.5 THz. Thus, the obvious new parameter space
to be explored by ground-based THz facilities is the
angular resolution with an improvement of about an 
order of magnitude in beam area.
 {In particular, no THz facility
has ever routinely had such a resolution for imaging.}
Additional advantages are:
(i) we can directly control and maintain the telescope;
and (ii) we can run long-term projects which is difficult
for non-ground-based facilities with limited lifetimes
or limited observational durations.

As far as sensitivity is concerned, ground-based facilities
cannot win over airborne or space telescopes even at
the Summit Station in Greenland because
of the atmospheric opacity. For example, a $\sim$30 times
smaller
continuum NEFD is achieved by the currently available
airborne facility, SOFIA. 
Therefore, the first-generation ground-based
THz telescopes need to focus on relatively bright objects
where high-angular-resolution observations have the potential to do
pioneering work.

\subsection{Operations at sub-THz}

Since the weather conditions suitable for THz observations
are not guaranteed for all the winter time (typically $\sim 10$\%;
Section \ref{subsec:atmosphere}), operations at somewhat lower
frequencies should also be considered. In order to
be different from
existing submm telescopes, we are taking advantage of the 
good atmospheric conditions to target relatively high frequencies,
referred to as sub-THz in this paper. In particular,
atmospheric windows around
850 GHz (350 $\mu$m) and 650 GHz (450 $\mu$m) are viable
targets. Operations at submm/millimeter (mm)-wavelength sites like Mauna Kea
have been rare in these frequency bands. The Greenland site will
provide a significant fraction of time (Section \ref{subsec:atmosphere})
for these windows. Therefore, the second purpose
of this paper is to identify compelling science cases in the
shortest submm or sub-THz ranges.

A special advantage of the GLT is that it allows for flexible
time scheduling. Time-consuming survey-type observations
at sub-THz will
be unique, because such observations are extremely difficult
for practically all other telescopes in the northern hemisphere.
This flexibility increases the chance
to carry out monitoring and
target-of-opportunity (ToO) observations. The
scientific importance of
this possibility is also explored in this paper.

\section{THz Observations}\label{sec:key}

\subsection{Important lines at THz}\label{subsec:line}

In the THz regime there are a number of interesting but
unexplored spectral lines, which should be useful for
studies of the ISM and star formation.
In line observations, sky background subtraction is much easier
than in continuum observations, and thus observations of intense THz lines
should be the starting point of our THz astronomical experiment.
Table \ref{tab:lines} summarizes representative THz lines whose
rest-frame wavelengths are accessible in the THz atmospheric windows. These lines
can be classified into three categories; (i) very high-$J$ ($J$ is the
rotational excitation state) molecular lines
(CO, HCO$^{+}$, and HCN); (ii) atomic fine-structure lines ([N \textsc{ii}]); and (iii) pure rotational lines of
light molecules
(H$_{2}$D$^{+}$, HD$_{2}$$^{+}$, and CH).

\begin{table*}
\tbl{Representative Tera-Hertz Lines.}{%
\begin{tabular}{lccc}
\hline
 Species & Frequency (THz) & Transition & Excitation energy (K) \\
\hline
  CO &1.037--1.497 & (9--8)--(13--12)  & 248.87486--503.134028 \\ 
  HCO$^{+}$ &1.070--1.337 & (12--11)--(15--14) & 333.77154--513.41458\\
  HCN &1.0630--1.593 & (12--11)--(18--17) & 331.68253--726.88341 \\
  H$_{2}$D$^{+}$ &1.370 &1$_{0,1}$--0$_{0,0}$ &65.75626\\
  N\,\textsc{ii} &1.461 &$^{3}$P$_{1}$--$^{3}$P$_{0}$ & --- \\
  CH &1.471 &N=2, J=3/2-3/2, F=2$^{+}$--2$^{-}$ &96.31131\\
  HD$_2^+$ &1.477 &1$_{1,1}$--0$_{0,0}$ &70.86548 \\
\hline
\end{tabular}}\label{tab:lines}
\end{table*}

\subsubsection{Tracers of star-forming places}\label{subsubsec:THz_SF}

\citet{kawamura02} observed several positions in the Orion Molecular
Cloud (OMC-1) by the ground-based 10-m Heinrich Hertz Telescope on Mount
Graham, Arizona. They targeted the CO $J=9$--8 rotational line at
1.037 THz. They detected some regions with such a high excitation
along the ridge of OMC-1. It has been clarified that there are some
warm regions with kinetic temperatures $\gtrsim$130 K and hydrogen
number density $\gtrsim 10^6$ cm$^{-3}$ in the molecular cloud.
They also emphasized the importance of high angular resolutions
achieved by ground-based telescopes in specifying the locations of
the emission.

With the GLT, we aim at observing higher-excitation CO lines at
a higher frequency, 1.5 THz,
where CO $J=13$--12 (1.4969 THz) line is present.
With the very high-$J$ molecular lines,
``extremely hot'' ($\gtrsim$300 K) molecular regions
in the vicinity of the forming protostars can be traced.
Moreover, CO is the most strongly emitting species in such a
region, enabling us to trace lower-mass or more distant objects
than other species.
Profile shapes
of those lines are probes of gas motions in such regions.
\citet{wiedner06} built a 1.5 THz heterodyne receiver,
CO N$^+$ Deuterium Observations Reciever (CONDOR),
installed it in the APEX, and performed ground-based THz line observations.
With a total on-source time of 5.8 min, they detected CO ($J=13$--12) line
toward Orion FIR 4, a cluster-forming region in the Orion Giant Molecular Cloud.
The main beam brightness temperature and the line width of the THz CO line
are $\sim 210$ K and $\sim 5.4$ km s$^{-1}$, respectively. As compared to
the CO $J=9$--8 and 7--6 lines, there is no line wing component with
$\gtrsim 10$ km s$^{-1}$
in the THz CO line. Multi-transitional analysis of the ``quiescent'' component of
the CO lines show that the gas temperature and density traced by the THz line
are $380\pm 70$ K and $(1.6\pm 0.7)\times 10^{5}$ cm$^{-3}$, respectively.
These results show that the THz CO line traces the very hot molecular gas in
the vicinity of the protostars, without any contamination
from the extended outflow component. The bulk of the outflows are apparently
at lower temperatures. With the GLT, we aim at a systematic survey of
similar objects hosting highly excited CO lines.

Observations of submm and THz HCN lines toward Sgr B2(M) by
\textit{Herschel}/the Heterodyne Instrument for the Far-Infrared (HIFI)
show that the submm HCN $J=6$--5 and 7--6 lines exhibit ``blue-skewed''
profiles suggestive of infall, while the HCN $J=8$--7 and
the THz HCN $J=12$--11 (1.06 THz) lines
exhibit ``red-skewed'' profiles suggestive of expansion \citep{rolffs10}.
The submm HCN lines with the blue-skewed profiles
originate from gas with temperature $\sim 100$ K
while the THz line is emitted from gas
with higher excitation $\sim 330$ K. These results
suggest
that the lower-$J$ HCN lines
trace the outer infalling motion toward the massive stars, whereas
the THz line traces the inner
expansion driven by the radiation pressure and the stellar wind from
the central massive stars.
All of the above results imply that high-$J$ THz molecular lines can be unique tracers
to probe the very inner parts of protostars and to study the gas motions there.

To quantify the physical conditions traced by THz high-$J$ molecular lines,
we performed statistical equilibrium calculations of the CO lines based on
the large velocity gradient (LVG) model \citep{goldreich74,scoville74}.
For the calculations, the rotational energy levels, line frequencies,
and the Einstein $A$ coefficients of CO are adopted from
Leiden Atomic and Molecular Database (LAMDA; \citealt{schoier05}),
and the collisional transition rates with ortho-H$_{2}$ from
\citet{yang10b}. The rotational energy levels up to $J=41$ (4,513 K)
are included in the calculations.
The formulae of the line profile function and photon escape probability
of a static, spherically symmetric and homogeneous cloud are adopted
\citep{osterbrock06}.
Our calculations are analogous to those of RADEX
\citep{vandertak07},
and we confirmed that our codes provide the same
results as those by RADEX.

\begin{figure}
\begin{center}
\includegraphics[width=0.45\textwidth]{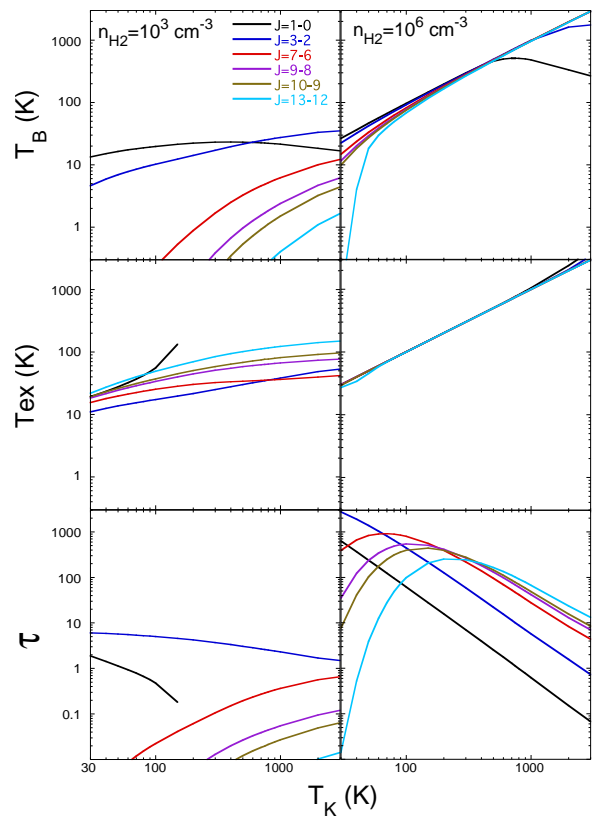}
\end{center}
\caption{Results of the statistical equilibrium calculations of the CO lines.
The brightness temperatures (\textit{upper panels}), excitation
temperatures (\textit{middle panels}), and the optical depths of the CO lines
(\textit{lower panels})
are plotted as a function of the gas kinetic temperature.
In the left and right panels
the number density of H$_2$ is set to be 10$^{3}$ cm$^{-3}$ and
10$^{6}$ cm$^{-3}$, respectively. The cloud size and velocity width
are assumed to be 0.1 pc and 1.0 km s$^{-1}$, respectively,
and the CO abundance relative to hydrogen is fixed as
9.5 $\times$ 10$^{-5}$ \citep{crapsi04}.
\label{fig:lvg}}
\end{figure}

Figure \ref{fig:lvg} shows the calculated line brightness temperatures
($T_\mathrm{B}$), excitation temperatures ($T_\mathrm{ex}$), and the
optical depths ($\tau$) of the various CO lines as a function of
the gas kinetic temperature ($T_{K}$)
at an H$_2$ ($n_{\rm H_2}$) number density of 10$^{3}$ cm$^{-3}$
and 10$^{6}$ cm$^{-3}$.
In the case of the diffuse molecular gas
($n_{\rm H_2}$ = 10$^{3}$ cm$^{-3}$),
the brightness temperatures of the THz CO lines
($J=9$--8, 10--9, and 13--12) are too low
to be detected even at the very high gas temperatures,
while those of lower-$J$ CO lines ($J=1$--0; 3--2) are high.
This is because at the low gas density the higher $J$ levels
are not populated and the THz CO lines are too optically thin, as shown in
the lower-left panel. In contrast, at $n_{\rm H_2}$ = 10$^{6}$ cm$^{-3}$,
typical of dense cores (dense-gas condensations in molecular clouds),
all CO transitions are thermalized
and the excitation temperatures follow the gas kinetic
temperature (middle-right panel). Consequently the optical
depths, so the brightness temperatures, become high even for the THz lines. These results
show that the THz CO lines are an excellent tracer of dense
($\gtrsim$10$^{6}$ cm$^{-3}$) and hot ($\gtrsim$100 K) molecular gas.

According to radiation-hydrodynamic models of protostar formation,
such dense and high-temperature molecular 
regions should be present at the latest stage of protostellar collapse
within $\sim 500$ AU from the forming central star \citep{masunaga00}.
For typical distances of protostellar sources ($d\sim 300$ pc), the
diameter ($\sim$1,000 AU) corresponds to 3.3 arcsec, which roughly matches
the angular resolution of the GLT at 1.5 THz ($\sim$ 4 arcsec;
Section \ref{subsec:capability}).
With the GLT, thus, very hot molecular gas in the vicinity of
the forming protostars can be observed.
The line profile shape can then be used to trace the gas motions in such regions.


%
%
%

\begin{table*}
\tbl{Luminous Protostellar Sources for the THz Line Experiment}{%
\begin{tabular}{lccccc}
\hline
  Name & $L_\mathrm{bol}$ & $D$ & $\alpha$(J2000) & $\delta$(J2000) & Ref. \\
            &  ($L_{\odot}$) & (pc) & (h m s) & (\arcdeg~\arcmin~\arcsec) & \\
\hline
    L1448-mm     &4.4 &250  &03 25 38 87  &30 44 05.4  &1,2 \\
NGC1333 IRAS 2A      &19.0   &250  &03 28 55.58  &31 14 37.1  &1,2 \\
SVS 13          &32.5  &250  &03 29 03.73 &31 16 03.80 &2,3 \\
NGC1333 IRAS 4A     &4.2  &250  &03 29 10.50  &31 13 31.0  &1,2 \\
L1551 IRS 5  &22  &140  &04 31 34.14  &18 08 05.1  &4,5 \\
L1551 NE      &4.2 &140  &04 31 44.47  &18 08 32.2  &5,6 \\
L1157             &5.8  &325  &20 39 06.28  &68 02 15.8  &1,7  \\ 
\hline
  \end{tabular}}\label{tab:line_sources}
  \begin{tabnote}
References: 1) \citet{jorgensen07}; 2) \citet{enoch09};
3) \citet{chen09}; 4) \citet{takakuwa04}; 5) \citet{froebrich05};
6) \citet{takakuwa12}; 7) \citet{shirley00}.
\end{tabnote}
\end{table*}

In Table \ref{tab:line_sources}, we summarize possible target
protostellar objects for the first-generation THz line experiments.
We selected nearby (with distance $D\lesssim 300$ pc) and luminous
($\gtrsim 4~\mathrm{L}_{\odot}$) protostellar sources with
ample previous (sub)mm molecular-line studies.

Now we estimate the expected THz intensity of these sources.
Considering that there is a linear correlation between
the bolometric luminosities and the intensities of
the submm CS (7--6) line toward protostellar sources \citep{takakuwa11},
we simply assume that there is also a linear correlation between the
intensity of the THz CO line and the source bolometric luminosity
($L_\mathrm{bol}$). The CONDOR result above \citep{wiedner06} shows
that the THz CO intensity is $\sim$210 K toward Orion FIR 4
($L_\mathrm{bol}$= 50 $L_{\odot}$). The lowest bolometric luminosity
among our target is 4.2 $L_{\odot}$
(L1551 NE), and the expected THz CO intensity is
210 K $\times$ 4.2 $L_{\odot}$/50 $L_{\odot}$
= 17.6 K ($\sim$400 Jy within primary beam).
Using the typical sensitivity discussed in
Section \ref{subsec:capability} (see also Table \ref{tab:capability}),
we should be able to achieve a 5 $\sigma$ detection of this source
with a velocity resolution of 0.2~km~s$^{-1}$ and an on-source
integration time of 34 minutes.
Therefore, the GLT
can collect a systematic sample of hot regions in the vicinity of
protostars.

\subsubsection{The diffuse ISM: molecules}

Some atomic or molecular lines in the diffuse medium can be targeted
at THz frequencies. Fine-structure lines of fundamental atoms and ions,
which trace diffuse ($\sim$30--100 cm$^{-3}$),
transitional regions from ionized or atomic gas
to molecular gas in the ISM, and thus photo-dissociation
regions (PDRs), H \textsc{ii} regions, and surfaces of molecular clouds.
Among them
[N\,\textsc{ii}] (1.46 THz) 
can be observed in the THz atmospheric windows.
The THz line profiles provide kinematical information, which is a clue
to the cloud formation mechanism,
or the disruption mechanism of clouds by the newly formed stars.

There are several pure rotational lines of
``chemically basic'', light molecules in the THz region.
Previous mm molecular-line studies have shown that
toward dense cores without known protostellar sources
there are abundant carbon-chain
molecules, while toward dense cores with known protostellar sources
those carbon-chain molecules are deficient \citep{suzuki92}.
This result suggests that abundances of carbon-chain molecules can
be used to trace evolutionary stages of dense cores toward
star formation. CH is considered to be the starting point of
carbon-chain chemistry \citep{leung84}, and observation of
the CH line at 1.47 THz in molecular clouds is useful to
understand carbon-chain chemistry and the evolution of dense starless cores
into star-forming cores. CH is also considered to be a key molecule
to form organic molecules in protoplanetary disks \citep{najita11}.
Observations of the CH line toward disks around young stellar objects
should be important to understand the formation mechanism of complex organic
molecules, which can play a role in the origin of life.
H$_{2}$D$^{+}$ and HD$_{2}^{+}$ molecules control deuterated
chemistry in molecular clouds \citep{caselli08}. In contrast
to carbon-chain molecules,
deuterated molecules are considered to be abundant in later
evolutionary stages of dense cores, just prior to the onset of
the initial collapse \citep{hirota01,hirota03}.
Hence, observation of CH and those deuterated species
is important to trace evolutionary sequence of dense cores until
the initial collapse leading to protostar formation.

\begin{figure}
\begin{center}
\includegraphics[width=0.45\textwidth]{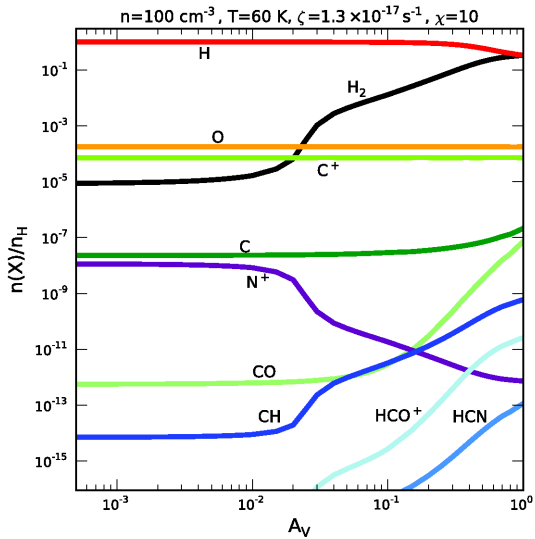}
\end{center}
\caption{Fractional abundance of several interesting species in the
THz band with respect to
the total abundance of hydrogen atoms as a function of the depth into
the cloud, indicated by $A_V$, in a plane-parallel PDR model with
illumination from one side (to the left in the plot). 
The physical conditions of the PDR model are: gas density, $n=100$ cm$^{-3}$,
temperature, $T=60$ K, cosmic-ray ionization rate,
$\zeta=1.3\times10^{-17}$ s$^{-1}$, and intensity of
the FUV radiation field $\chi=10$ in units of the
\citet{draine78} standard radiation field.
\label{fig:PDR_Oscar}}
\end{figure}

In Figure \ref{fig:PDR_Oscar}, we present the fractional abundance of
several species calculated in a plane-parallel PDR model with illumination
from one side, in order to show in which region the species of interest
(Table \ref{tab:lines}) reside.
For the details of treatment of chemical reactions, etc.,
see \citet{morata08}. We adopted gas density, $n=100$ cm$^{-3}$,
gas temperature, $T=60$ K, cosmic-ray ionization rate, $\zeta=1.3\times10^{-17}$
s$^{-1}$, and intensity of the FUV radiation field $\chi=10$ in units of the
\citet{draine78} standard radiation field.
Since we are interested in the diffuse medium, we
focus on the regions with $A_V\lesssim 1$, where $A_V$ is the extinction
at $V$ band. Indeed, H$_2$ dominates the hydrogen
species at $A_V>1$ as shown in Figure \ref{fig:PDR_Oscar}
so that molecular clouds are favorably formed at
$A_V>1$.
We observe that both [N\,\textsc{ii}] and CH are fair THz
tracers of the diffuse ISM (note that CO requires a dense
state to be excited enough to emit THz lines;
Figure \ref{fig:lvg}). Moreover, the excitation
energy for the 1.5 THz CH line is $\sim 100$ K, matching the
temperature of the cold neutral medium (CNM) \citep{mckee77},
outside of molecular clouds.
Some other species in Table \ref{tab:lines} also exist:
both HCO$^+$ and HCN favor shielded regions and are more
suitable for the molecular gas tracer.

\citet{godard08} examined the
case of turbulent dissipation regions (TDRs) in the CNM instead of PDRs.
In TDRs, the above species (CH and HCO$^+$) are also key species.
The abundances of these species relative to hydrogen is
$\sim 10^{-8}$--$10^{-9}$ for CH and $10^{-10}$--$10^{-9}$ for
HCO$^+$, slightly larger than the above prediction for the PDR.

High-excitation CO lines are tracers of dense regions located
near to the ionizing source. Indeed, \citet{perez12} observed
M17 SW, a giant molecular cloud
illuminated by a cluster of OB stars, with
the SOFIA/German Receiver for Astronomy at Terahertz (GREAT)
and detected the CO $J=13$--12 emission line, which indicates
a dense
region near to the ionizing source in the nebula. We could probably use
THz CO lines to exclude the contamination of dense regions and
choose really diffuse regions;
that is, we avoid the lines of sights where THz CO lines are
detected to choose purely diffuse regions.

In reality observations of other species such as CH and HCO$^+$
in emission require much higher sensitivity than
those species with strong lines such as CO and [N\,\textsc{ii}].
If we assume a similar level of intensity to other weak
THz lines detected by Herschel ($\sim 0.3$ K),
$\sim 100$ hr integration on source would be necessary
for a 3 $\sigma$ detection.
Therefore, it is recommended that
we first observe lines easier to detect such as
[N\,\textsc{ii}] 205 $\mu$m (Section \ref{subsec:Gal_NII}) in the
first generation ground-based THz detectors.
Alternatively, we observe bright background continuum sources and try to detect
the THz lines of minor species in absorption
\citep{gerin12}.

\subsubsection{The diffuse ISM: [N\,{\footnotesize II}] 205 $\mu$m}
\label{subsec:Gal_NII}

Since the ionization potential of nytrogen (14.5 eV) slightly exceeds
1 Ryd, [N \textsc{ii}] lines trace the ionized medium.
In star-forming regions or star-forming galaxies, H \textsc{ii}
regions are prevalent because young massive stars emit
a large amount of ionizing photons.
Indeed, hydrogen
recombination lines emitted from H \textsc{ii} regions are
often used as indicators of star formation activity
\citep[e.g.,][]{kennicutt98}.

Recently, \citet{inoue14} have investigated possibilities
of using FIR (THz) nebular lines as a tracer of
star-forming galaxies.
Indeed, FIR fine-structure lines could be a useful
star formation indicator of high-redshift galaxies,
since ALMA is capable of detecting redshifted FIR
fine-structure lines \citep[e.g.,][]{nagao12}.
Because of the small Einstein
coefficients, FIR fine-structure lines are usually
optically thin. They expressed the line
luminosity, $L_\mathrm{line}$, as a function of
star formation rate (SFR):
\begin{eqnarray}
L_\mathrm{line}=C_\mathrm{line}(Z,\, U,\, n_\mathrm{H})\,
\mathrm{SFR},\label{eq:NII_SFR}
\end{eqnarray}
where $C_\mathrm{line}$ is the proportionality constant
that depends on the metallicity $Z$, ionization parameter
$U$, and hydrogen number density $n_\mathrm{H}$. They
calculated $C_\mathrm{line}$ using the photoionization
code \textsc{cloudy} \citep{ferland13} and
calibrated the values of $U$ and $n_\mathrm{H}$ by
comparing with [O\,\textsc{iii}] 88 $\mu$m line
luminosities of nearby galaxies. As a consequence they obtained the
values of $C_\mathrm{line}$ as a function of metallicity.

Among the lines, [N\,\textsc{ii}] 205~$\micron$ at redshift 0 can be
observed from the ground in the 1.5 THz atmospheric window.
Thus, we are able to examine the applicability of
[N \textsc{ii}] 205~$\mu$m line strength to the SFR estimate
by observing Galactic H\,\textsc{ii} regions.
The critical density of this transition is only 44 cm$^{-3}$
at 8000 K \citep{oberst06}, so that the collisional excitation
is efficient enough.
However, the small critical density also means that the [N\,\textsc{ii}] 205 $\mu$m
line is more weighted for the diffuse regions; thus, we keep in mind
that, when we compare the [N \textsc{ii}] 205 $\mu$m
line with other star-formation indicators in a resolved star-forming
regions, the correlation may not be good in dense star-forming regions
\citep{wu15}. Indeed, \citet{wu15} showed, using spatially resolved
[N \textsc{ii}] 205 $\mu$m
data of M83 taken by \textit{Herschel}, that the [N \textsc{ii}] intensity
has a much shallower dependence on the SFR surface density than
expected from a linear relation. They interpret this shallow dependence
as the [N \textsc{ii}] emission being more diffuse than other star
formation indicators.

On an entire-galaxy scale, at least, the [N \textsc{ii}] 205 $\mu$m
emission is potentially a good star-formation indicator
\citep{zhao13}. \citet{zhao13} examined the relation between
the [N \textsc{ii}] 205 $\mu$m line luminosity observed in nearby
starburst galaxies and their FIR luminosity, which is
known to be a good indicator of SFR \citep[e.g.,][]{kennicutt98,inoue00}.
They excluded galaxies with significant contributions from AGNs and
included normal star-forming galaxies, for which
the [N \textsc{ii}] 122 $\mu$m luminosities are observed
by \textit{ISO} \citep{brauher08}, after converting
the [N \textsc{ii}] 122 $\mu$m luminosity to the
[N \textsc{ii}] 205 $\mu$m luminosity using the theoretical
ratio. As a consequence, they found that there is a tight
relation between the [N \textsc{ii}]
205 $\mu$m luminosity and the FIR luminosity:
\begin{eqnarray}
\log\mathrm{SFR}~(M_\odot~\mathrm{yr}^{-1}) & = &
(-5.31\pm 0.32)\nonumber\\
& & +(0.95\pm 0.05)\log L_\mathrm{N\,II}~(L_\odot ),\label{eq:zhao13}
\end{eqnarray}
which is consistent with a linear relation with
$C_\mathrm{NII}=10^{5.31\pm 0.32}~L_\odot/(M_\odot~\mathrm{yr}^{-1})$
(equation \ref{eq:NII_SFR}; when we consider the [N \textsc{ii}]
205 $\mu$m line, we denote $C_\mathrm{line}$ as $C_\mathrm{N\,II}$).
\citet{wu15} also show the integrated [N \textsc{ii}] 205 $\mu$m
luminosity and SFR of M83 matches this relation.
\citet{inoue14} derived
$C_\mathrm{N\,II}=10^{6.23}$ and $10^{5.48}~L_\odot /(M_\odot~\mathrm{yr}^{-1})$
for 1 $Z_\odot$ and 0.4 $Z_\odot$, respectively.
Since equation (\ref{eq:zhao13}) is derived for bright galaxies,
the metallicity is probably nearly solar. If the metallicity is solar,
\citet{inoue14}
tends to overproduce $C_\mathrm{N\,II}$. To examine this possible
discrepancy,
further tests are necessary with a fixed-metallicity
sample. We propose
the following tests using Galactic H \textsc{ii} regions,
whose metallicities can be assumed to be solar.

A representative indicator of SFR (or H \textsc{ii}
regions) is H$\alpha$ emission \citep[e.g.,][]{kennicutt98},
which is linked to the SFR under the solar metallicity
and the same initial mass function (IMF) as adopted
by \citet{inoue14} by the following equation \citep*{hirashita03}:
\begin{eqnarray}
\mathrm{SFR}=C_\mathrm{H\alpha}L_\mathrm{H\alpha}^\mathrm{c},
\label{eq:SFR_Ha}
\end{eqnarray}
where $C_\mathrm{H\alpha}=7.9\times 10^{-42}
(\mathrm{M}_\odot~\mathrm{yr}^{-1})/(\mathrm{erg~s}^{-1})$,
and $L_\mathrm{H\alpha}^\mathrm{c}$ is the H$\alpha$ luminosity
corrected for extinction.

Inserting equation (\ref{eq:SFR_Ha}) into
equation (\ref{eq:NII_SFR}), we obtain
\begin{eqnarray}
L_\mathrm{N\,II}=C_\mathrm{H\alpha ,N\,II}
L_\mathrm{H\alpha}^\mathrm{c},\label{eq:NII_Ha}
\end{eqnarray}
where $C_\mathrm{H\alpha ,N\,II}\equiv C_\mathrm{N\,II}\,C_\mathrm{H\alpha}.$
(As we noted above, we denote $C_\mathrm{line}$ for the [N \textsc{ii}]
205 $\mu$m emission as $C_\mathrm{N\,II}$.)
For the [N \textsc{ii}] 205 $\mu$m line, \citet{inoue14} obtained
$C_\mathrm{N\,II}(Z,\, U,\, n_\mathrm{H})=10^{39.82}$
(erg s$^{-1}$)/(M$_\odot$ yr$^{-1}$) for the solar
metallicity, by adopting $\log U=-3$ and $\log n=1$.
In this case, $C_\mathrm{H\alpha ,N\,II}=0.0521$.

Equation (\ref{eq:NII_Ha}) can be checked or calibrated with a sample of
Galactic H \textsc{ii} regions whose H$\alpha$
luminosities are already measured. Note that the
proportionality constant does not change even if
we use the surface brightness, i.e.,
$I_\mathrm{N\,II}=C_\mathrm{H\alpha , N\,II}
I_\mathrm{H\alpha}^\mathrm{c}$.

The major disadvantage of H$\alpha$ line is that H$\alpha$ photons
are subject to dust extinction. A way of resolving this problem is
that the extinguished portion of the H$\alpha$ luminosity is corrected
using the FIR dust emission \citep{kennicutt07}.
Alternatively, radio free--free emission, which is
free from dust extinction, can also
be used as a tracer of H \textsc{ii} region,
although contamination with synchrotron or dust radiation
might be a problem depending on the frequencies.
\citet{mezger74} relate the
radio free--free flux $S_\nu$ to the number of ionizing
photons radiated per time $N_\mathrm{Lyc}$
(referred to as ionizing photon
luminosity) as
\begin{eqnarray}
N_\mathrm{Lyc} & = &
4.8\times 10^{48}a(\nu,\, T_e)^{-1}\left(
\frac{\nu}{\mathrm{GHz}}\right)^{0.1}\left(
\frac{T_e}{\mathrm{K}}\right)^{-0.45}\nonumber\\
& & \times\left(
\frac{S_\nu}{\mathrm{Jy}}\right)\left(
\frac{D}{\mathrm{kpc}}\right)^2~\mathrm{s}^{-1},
\end{eqnarray}
where $a(\nu,\, T)$ is a slowly varying function, which
is assumed to be 1 for the electron temperatures of
interest ($T_e\sim 10^4$ K), and $D$ is the distance to
the object. Using the relation
$(N_\mathrm{Lyc}/\mathrm{s}^{-1})=7.3\times 10^{11}
[L_\mathrm{H\alpha}/(\mathrm{erg~s^{-1}})]$
\citep{deharveng01}, the above equation is equivalent with
\begin{eqnarray}
L_\mathrm{H\alpha}^\mathrm{c} & = & 6.5\times 10^{36}
a(\nu,\, T_e)^{-1}\left(
\frac{\nu}{\mathrm{GHz}}\right)^{0.1}\left(
\frac{T_e}{\mathrm{K}}\right)^{-0.45}\nonumber\\
& & \times\left(
\frac{S_\nu}{\mathrm{Jy}}\right)\left(
\frac{D}{\mathrm{kpc}}\right)^2~\mathrm{erg~s^{-1}}.
\end{eqnarray}
Further, using equation (\ref{eq:NII_Ha}),
we obtain
\begin{eqnarray}
L_\mathrm{N\, II} & = & 6.5\times 10^{36}\,
C_\mathrm{H\alpha ,N\,II}\,
a(\nu,\, T_e)^{-1}\left(
\frac{\nu}{\mathrm{GHz}}\right)^{0.1}\nonumber\\
& & \times\left(
\frac{T_e}{\mathrm{K}}\right)^{-0.45}\left(
\frac{S_\nu}{\mathrm{Jy}}\right)\left(
\frac{D}{\mathrm{kpc}}\right)^2~\mathrm{erg~s^{-1}}.
\end{eqnarray}
The expected flux of [N \textsc{ii}] 205 $\mu$m,
$f_\mathrm{N\, II}$ is then
\begin{eqnarray}
f_\mathrm{N\, II} & = & \frac{L_\mathrm{N\,II}}{4\pi D^2\Delta\nu}
\nonumber\\
& = & 5.4\times 10^{7}\, C_\mathrm{H\alpha ,NII}\,
a(\nu,\, T_e)^{-1}\left(
\frac{\nu}{\mathrm{GHz}}\right)^{0.1}\nonumber\\
& & \times\left(
\frac{T_e}{\mathrm{K}}\right)^{-0.45}\left(
\frac{S_\nu}{\mathrm{Jy}}\right)\left(
\frac{\Delta\nu}{10^8~\mathrm{Hz}}\right)^{-1}~\mathrm{Jy},
\end{eqnarray}
where $\Delta\nu$ is the assumed frequency width of
the line ($10^8$ Hz corresponds to a velocity width of
20 km s$^{-1}$ at 1.5 THz).
Here we assume that the instrumental frequency resolution
is smaller than the physical broadening of the line.
By adopting $a(\nu ,\, T_e)=1$, and taking the normalization
at $\nu =5$ GHz (wavelength 6 cm,
as we use later) and $T_e=10^4$ K, the above equation is reduced
to
\begin{eqnarray}
f_\mathrm{N\, II} & = & 1.0\times 10^6\,C_\mathrm{H\alpha ,NII}\,
\left(\frac{\nu}{5~\mathrm{GHz}}\right)^{0.1}\left(
\frac{T_e}{10^4~\mathrm{K}}\right)^{-0.45}\nonumber\\
& & \times\left(
\frac{S_\nu}{\mathrm{Jy}}\right)\left(
\frac{\Delta\nu}{10^8~\mathrm{Hz}}\right)^{-1}~\mathrm{Jy}.
\label{eq:fNII}
\end{eqnarray}

\begin{table*}
  \tbl{Galactic H \textsc{ii} regions to be observed by
  the GLT \citep{mathis00}}{%
  \begin{tabular}{lccccccc}
  \hline
  Name & $(l,\, b)\,^\mathrm{a}$ & $\alpha$(J2000) & $\delta$(J2000) &
  Size$^\mathrm{b}$ & $S_\nu$(6\,cm) &
  $f_\mathrm{N\, II}\,^\mathrm{c}$
  & $f_\mathrm{N\, II,\,beam}\,^\mathrm{d}$ \\
    &   & (h m s) & (\arcdeg~\arcmin~\arcsec) & & (Jy) & (MJy) &
    (Jy/beam) \\
    \hline
  W 3 & 133.7+1.2 & 02 25 30 & +62 05 19 & $1\farcm7\times 1\farcm5$ & 80
  & 0.38 & 420\\
  W 51 & 49.5$-$0.4 & 19 23 48 & +14 30 46 & complex & 400
  & -- & -- \\
  DR 21 & 81.7+0.5 & 20 39 15 & +42 19 11 & $0\farcm3\times 0\farcm4$ & 19
  & 0.090 & 2400 \\
  NGC\,7538 & 111.5+0.8 & 23 13 21 & +61 28 32 & $2\farcm3\times 1\farcm9$ & 26
  & 0.12 & 90 \\
  \hline
  \end{tabular}}\label{tab:Gal_HII}
\begin{tabnote}
$^\mathrm{a}$Galactic coordinate.\\
$^\mathrm{b}$Size in the directions of $\alpha$ and $\delta$.\\
$^\mathrm{c}$[N \textsc{ii}] 205 $\mu$m flux estimated by
$f_\mathrm{N\, II}=4.7\times 10^3(S_\nu /\mathrm{Jy})$ Jy,
which is derived with $C_\mathrm{H\alpha ,N\,II}=4.0\times 10^{-3}$
and $T_e=7,000$ K (these values are appropriate for Carina I and
Carina II) in equation (\ref{eq:fNII}).\\
$^\mathrm{d}$[N \textsc{ii}] 205 $\mu$m flux per beam, where
the beam size is assumed to be $4''\times 4''$.
\end{tabnote}
\end{table*}

It is possible to calibrate $C_\mathrm{H\alpha, NII}$ with
Galactic H \textsc{ii} regions. Ground-based observations
of [N \textsc{ii}] 205 $\mu$m were already performed for
the Carina Nebula by \citet{oberst11} with
the South Pole Imaging Fabry-Perot Interferometer (SPIFI) at
the Antarctic Submillimeter Telescope and Remote Observatory
(AST/RO). The beam size was 54$''$ FWHM.
For Carina I and Carina II, they obtained
[N \textsc{ii}] 205 $\mu$m brightness of
$1.87\times 10^{-7}$ and $1.62\times 10^{-7}$ W m$^{-2}$ sr$^{-1}$,
respectively. Multiplying the intensity with the consistent area
with radio observations by \citet{huchtmeier75},
that is,
$4\farcm 5\times 6\farcm 3$ and $5\farcm 2\times 4\farcm 8$
for Carina I and
Carina II, respectively, we obtain
$F_\mathrm{NII}=3.5\times 10^{-10}$
and $2.7\times 10^{-10}$ erg s$^{-1}$ cm$^{-2}$,
respectively, where $F_\mathrm{N\,II}$ is the integrated flux
over all the frequency range
(i.e., $F_\mathrm{N\,II}=f_\mathrm{N\,II}\Delta\nu$).
\citet{huchtmeier75} obtained
70.2 and 53.7 Jy at 8.87 GHz for Carina I and
Carina II, respectively. By adopting
$\nu =8.87$ GHz and $T_e=7000$ K \citep{huchtmeier75} in
equation (\ref{eq:fNII}), we obtain
$C_\mathrm{H\alpha ,NII}=4.0\times 10^{-3}$ for
both Carina I and Carina II. This is an order of magnitude smaller
than the value above obtained based on \citet{inoue14} (0.0521),
probably because of the
different values of $U$ and $n$. Indeed, the
[N \textsc{ii}] 205 $\mu$m intensity is sensitive
to those quantities; for example, a larger
value of $n$ or a larger value of $U$
appropriate for the Carina Nebula
significantly reduces the [N \textsc{ii}] 205 $\mu$m
intensity \citep[e.g.,][]{nagao12}.
It is interesting to point out that we obtained the same
value of $C_\mathrm{H\alpha ,NII}$ for both
Carina I and Carina II; indeed,
\citet{oberst11} show that they have similar
gas density and radiation field intensity from
line ratio analysis (see their figure 17).

A larger sample taken by the GLT would help to derive
more general conclusions for $C_\mathrm{H\alpha ,NII}$
for the solar metallicity environments.
In Table \ref{tab:Gal_HII}, we list representative Galactic H \textsc{ii}
regions whose declination ($\delta$) is larger than 12$^\circ$.
The sample is taken from \citet{mathis00}. Observations of these objects
provide a local calibration of [N\,\textsc{ii}] 205 $\mu$m luminosity
as an indicator of star formation rate.
Using the sensitivity discussed in Section \ref{subsec:capability}, we estimate the
on-source integration time necessary to detect
the faintest H \textsc{ii} region in Table \ref{tab:Gal_HII} (NGC 7538) with 5$\sigma$ as
13 min
with a spectroscopic resolution of 2 km s$^{-1}$.
Because of the extended nature of H \textsc{ii}
regions, a multi-pixel detector is desirable.

Metallicity dependence of [N \textsc{ii}] emission is addressed by
observing nearby galaxies \citep{cormier15}. In fact,
the high angular resolution of the GLT is useful for
resolving or separating extragalactic H \textsc{ii}
regions (Section \ref{subsec:NII_extragal}).
With these efforts for [N \textsc{ii}] 205 $\mu$m observations
of nearby objects, we will be able to provide a firm local calibration
of FIR fine-structure lines as a star-formation indicator. As mentioned
above, fine-structure lines are already used as a star formation
indicators accessible from ALMA bands for high-redshift galaxies.

\subsection{Dust continuum}\label{subsec:cont}

\subsubsection{Importance of THz bands}\label{subsec:hiro_Tdust}

To derive the dust temperature, we need two or more bands
in the FIR. In the wavelength range where dust emission
dominates, angular resolution comparable to
GLT THz observations can only be achieved by submm
interferometers or \textit{Herschel} PACS at
its shortest wavelength $\sim 70~\mu$m 
($\sim 6''$; \citealt{foyle12}).
The emission at such a short wavelength as
70 $\mu$m
is contaminated by very small grains which are
not in radiative equilibrium with the interstellar
radiation field \citep{draine85}. Thus, the
combination of THz observations by GLT with
interferometric submm observations is the
best way to make dust temperature maps in galaxies.

Here we examine how useful adding 1.5 THz data is
to determine the dust temperature. To quantify the goodness
of dust temperature estimate, we carry out
a simulation by the following steps
(see Appendix for details): First, we give a certain
dust temperature ($T_\mathrm{real}$) and produce
observed flux at given wavelengths by adding noise,
which mimics the observational error. Next,
we fit the data and determine the dust temperature
$T_\mathrm{obs}$, which is compared with $T_\mathrm{real}$.
The results are summarized as follows:
\begin{enumerate}
\item Addition of THz data to submm data improves the dust
temperature estimate, especially for the range of
dust temperatures ($\lesssim$30 K) typical of nearby galaxies 
(Figure \ref{fig:Tdust_err}).
In particular, with only submm bands, 450 $\mu$m and
850 $\mu$m, the error of dust temperature is larger than
the dust temperature itself at $>20$ K
even if the flux uncertainty is less than 30\%.
(Figure \ref{fig:Tdust_450_850}).
A shorter wavelength (i.e., $\sim 200~\mu$m; 1.5 THz), nearer to the peak of dust SED, is better
than a longer wavelength (i.e., $\sim 300~\mu$m; 1 THz).
\item If we only use THz bands, 200 and 300 $\mu$m,
the temperature estimate is not as good as in the case of
having submm data points as well (compare
Figures \ref{fig:Tdust_err}a and d). This is
because these two bands are so close that the
temperature estimate is not robust against the
errors in the flux measurements. If we only use the THz windows for the
dust temperature estimates, the measurement error
should be smaller than 10\% (Figure \ref{fig:Tdust_200_300}).
\item If submm measurements are precise such that the error is
within 3\%, the dust temperature is
determined quite well without THz data (20\% error in
the dust temperature at $<$40 K; Figure \ref{fig:Tdust_450_850}).
However, addition of a THz data point, if the THz
measurement error is within 10\%, actually
improves a temperature significantly, especially
at $T_\mathrm{dust}\gtrsim 40$ K
(Figure \ref{fig:Tdust_450_850}), and the error of the
dust temperature is suppressed to $\sim$10\% even at 50 K.
Such a temperature is of significant importance
to actively star-forming regions/galaxies
\citep[e.g.,][]{hirashita08}.
\end{enumerate}

\subsubsection{Continuum observations of star-forming regions: filamentary molecular cluods and their structures}
\label{subsubsec:cont_SF}

\begin{figure*}
\begin{center}
\includegraphics[width=0.8\textwidth]{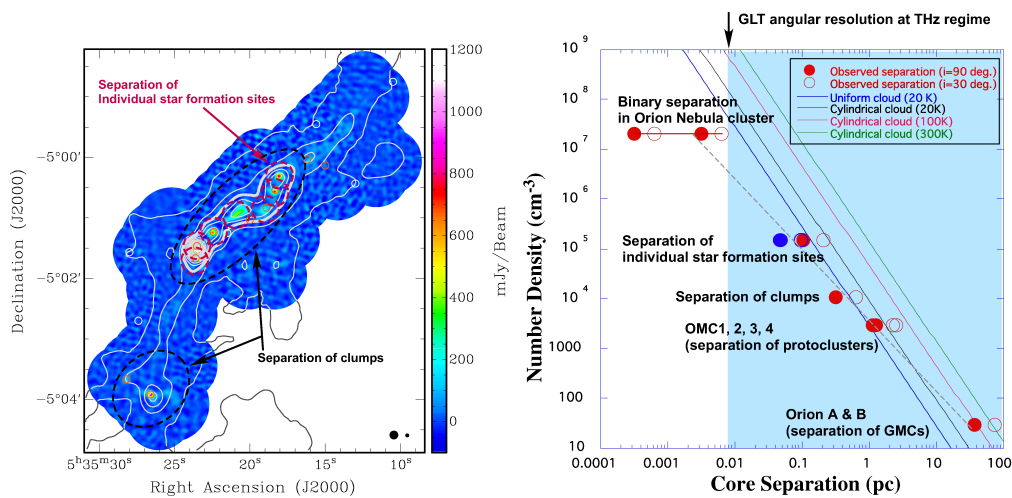}
\end{center}
\caption{Hierarchical fragmentation suggested in the Orion Giant Molecular
Cloud A.  
Left panel: The 850 $\mu$m continuum image obtained with the SMA (color)
overlaid with the continuum at same wavelength obtained with the JCMT
(white contours; \citealt{johnstone99}). Black open circles present
clumps identified by JCMT observations,
in which the reported clumps line up 
along the filaments with a spacing of $\sim$0.3 pc \citep{takahashi13}. 
Pink open circles present identified individual star forming sites with the SMA
\citep{takahashi13}.
The average separation of between detected sources are $\sim$0.05 pc.
Black filled circles in the bottom right corner show the
JCMT and SMA beam sizes of 14$\arcsec$ and $4\farcs5$, respectively.
At the distance of the cloud (414 pc; \citealt{menten07}), $1'$ corresponds to
0.12 pc.
Right panel: Observed separations of clouds/clumps/cores as a function of the mean number density of the parental cloud. 
The filled circles show the separation with $i=90^{\circ}$
(filament is in the plane of the sky), 
and the open circles show the separation with $i=30^{\circ}$ (filament closer to being perpendicular to the plane of the sky). 
The doted line shows the best fit of the observed results (power law index of $-1.4$). 
The solid lines with various colors show expected maximum instability size derived from Jeans length calculation with various temperatures indicated in the small panel.
The size scale achieved by the GLT angular resolution at THz ($\sim 4\farcs 0$ or 0.008 pc at the distance of the Orion cloud, 414 pc) is indicated by an arrow in the upper part of the horizontal axis. The size scale which can be mapped with the GLT is highlighted in blue.
\label{fig:omc3}}
\end{figure*}

How the ISM structures change associated with star formation
processes should provide a key to the dominant physical
processes of star formation. It has been known that
parsec-scale filamentary structures are common for molecular clouds
\citep[e.g.,][]{schneider79,myers09}.
Recent \textit{Herschel} observations have revealed the ubiquity of such filamentary structures both in low- and high-mass star-forming regions with a high dynamic range in mass \citep{andre10,konyves10,arzoumanian11,hill12}.
These filamentary molecular clouds often show hierarchical structures and fragment into $\sim$0.1 pc scale clumps, which likely corresponds to coherent-velocity filaments and isothermal cores
\citep{pineda10,hacar13}.
They will presumably fragment into star-forming cores. Therefore, multi-scale fragmentation processes are likely a key process that determines the prestellar and protostellar core mass function, which may directly lead to the stellar initial mass function. Here we examine a possibility of tracing the ISM structures associated with the star formation processes by the continuum emission.

Previous studies suggest that observations of low-mass-star-forming
regions are consistent with thermal Jeans fragmentation \citep{alves07},
while observations of massive-star-forming regions are better described
by turbulent fragmentation \citep{pillai11}. Recent interferometric
observations have started to spatially resolve the individual
star-forming sites and suggested multi-scale periodical structures
within filaments toward several different regions
\citep{zhang09,wang11,liu12,takahashi13,wang14}.

As an example, \citet{takahashi13} detected quasi-periodical separations
between sources with a spacing of $\sim$0.05 pc in the Orion Molecular Cloud-3 region (OMC-3; Figure \ref{fig:omc3}).
This spatial distribution is part of a larger hierarchical structure, which also includes fragmentation scales of GMCs ($\simeq$37 pc), representative star forming regions of OMC-1, -2, -3, and -4 ($\simeq$1.3 pc), and clumps ($\simeq$0.3 pc).
This suggests that hierarchical fragmentation operates within the OMC regions.
The measured separations of GMCs and representative star-forming regions are larger than those expected from the thermal fragmentation length, suggesting turbulent dominant fragmentation.
The measured separations of clumps are comparable to the thermal fragmentation length, and those
of spatially resolved individual cores are smaller than those expected from the thermal fragmentation length, which could be explained by local collapse within the clumps (gravitational focusing of the edges of the elongated structure) or helical magnetic fields.
These results imply that the multi-scale turbulent dissipation and consequent fragmentation processes may play an essential role in determining the initial physical condition of cores and the census of filamentary molecular clouds.
Furthermore, SMA 1.3 mm continuum observations toward OMC-1n, which is located $\simeq$2.2pc south from OMC-3 and connected to the most massive part of the cloud OMC-1, show a double peaked distribution of the source separations, corresponding to quasi-equidistant lengths of $\sim$0.06 pc for the clumps and 0.012 pc for the individual star-forming cores \citep{teixeira15}.
Comparison between OMC-3 and OMC-1n clearly demonstrates that the source separation varies along the OMC filaments. This suggests that the initial condition of cores and their census could vary within the parsec-scale filamentary molecular clouds.

Sampling individual star-forming sites, which typically have a size of $\leq$0.1 pc (Figure \ref{fig:omc3}) and are mainly observed with the (sub)mm interferometers, is limited by targeted observations.
On the other hand, imaging of parsec-scale clouds are mainly performed by ground-based single-dish (sub)mm telescopes or space infrared telescopes. These single dish facilities are not able to spatially resolve individual star-forming sites except the nearest low-mass star-forming regions ($\sim$150 pc).
Thus, it has been difficult to study multi-scale structures in filamentary molecular clouds systematically with a uniform data set.

The GLT, with its typical resolution of 4\arcsec
(Section \ref{subsec:capability}), will be able to
resolve individual star formation sites up to distances of $\sim$500 pc.
THz observations will have an advantage of observing continuum emission originating from prestellar cores and T Tauri sources at the wavelengths close to their SED peaks.
Multi-wavelength continuum detectors anticipated to be onboard the GLT enable us to directly measure dust temperature and dust emissivity within filamentary molecular clouds such as demonstrated by the recent SCUBA-2 observations
\citep{hatchell13,rumble15,salji15}.
As discussed in Section \ref{subsec:hiro_Tdust}, adding THz observations to existing data sets will help us to estimate those quantities more precisely.
Estimated dust temperatures will be directly used for estimating the thermal fragmentation lengths, which will be compared with the separations of spatially resolved sources. This will tell us what kind of fragmentation process is dominant on each size scale and in different parts of a filament. Moreover, the correct dust temperature serves to obtain a good estimate of the total dust mass (or the total gas mass with an assumption of dust-to-gas ratio), which is crucial to study the census of filamentary molecular clouds.

The SMA observations by \citet{takahashi13} had a similar
beam size to the GLT. The peak brightnesses that they detected were
in the range of $\sim 100$--4000 mJy beam$^{-1}$. If we assume a power-law
spectrum expected for the Rayleigh-Jeans side of the SED
with an index of $\alpha =2.5$, we obtain
$(1500/350)^{2.5}$ times 100 mJy beam$^{-1}$ for the faintest
case, i.e., 3.8 Jy beam$^{-1}$. Aiming at a root mean square (rms) of 0.4 Jy beam$^{-1}$
($\sim 10\sigma$ detection for the peak), a source requires
an on-source integration time of 34 min (Section \ref{subsec:capability}).
Currently the specs for THz bolometers to be equipped on the GLT are not clear yet. However, if we assume the similar spec as ArTeMiS (Architectures de bolometres pour des Telescopes a grand champ de vue dans le domaine sub-Millimetrique au Sol)\footnote{http://www.apex-telescope.org/instruments/pi/artemis/} equipped on APEX has at 350 $\mu$m \citep{reveret14}, the field of view (FOV) of bolometer will be expected as $\sim 5\farcm 0\times2\farcm 5$. Assuming the same FOV, mapping of a GMC which has a similar angular size to the northern part of the Orion A molecular cloud of $\sim 35'\times 75'$ \citep{salji15} retuires 210 mosaic images. As estimated above, the on-source time on each FOV is 34 min; thus, naive estimation of the total on-source observing time would be 34 min $\times$ 210 = 119 hr.

\subsubsection{Importance for polarization studies}\label{subsubsec:pol}

Polarization continuum observations in the THz frequency range are
an unexplored domain.
Dust continuum emission peaks at THz frequencies.
Lower frequency observations in the submm regime by the JCMT
at 850~$\micron$ \citep[e.g.,][]{matthews09},
the Caltech Submillimeter Observatory (CSO) at 350 $\micron$
\citep[e.g.,][]{dotson10},
the SMA at 870 $\micron$ \citep[e.g.,][]{rao09,tang10,zhang14} and CARMA at 1.3 mm \citep[e.g.,][]{hull14}
reveal typically dust polarized emission at a level 
of a few percent up to about 10\% of Stokes $I$.
The dust continuum polarized emission is thought to result from dust particles being aligned
with their shorter axes parallel to the magnetic field \citep[e.g.,][]{hildebrand00,lazarian07}. 
The magnetic field is being recognized as a key component in star formation theories 
\citep[e.g.,][]{mckee09}.
Polarization observations
toward star-forming regions provide, therefore, a unique tool to map the sky-projected 
magnetic field morphology.

\begin{figure}
\begin{center}
\includegraphics[width=0.42\textwidth]{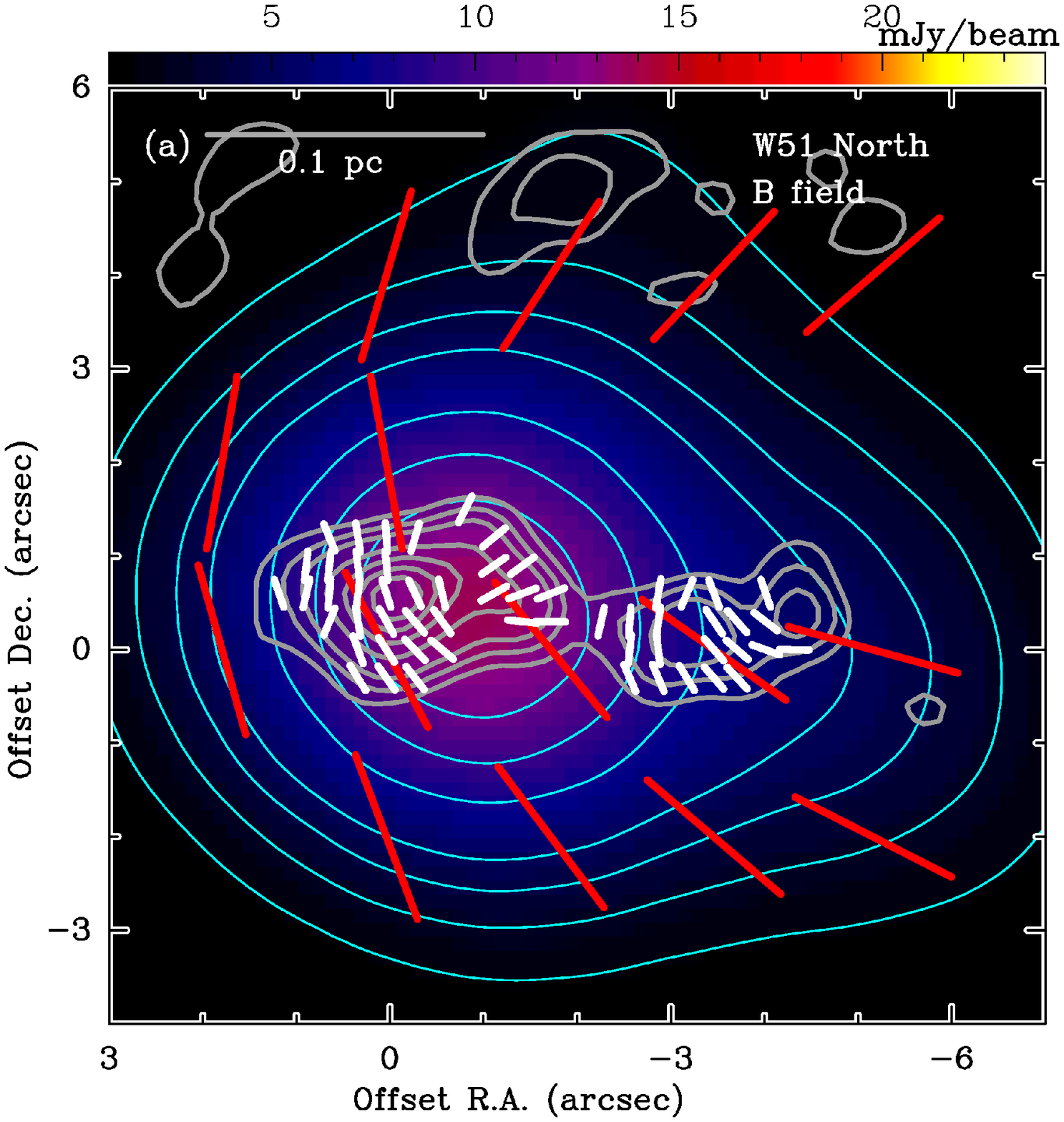}
\includegraphics[width=0.45\textwidth]{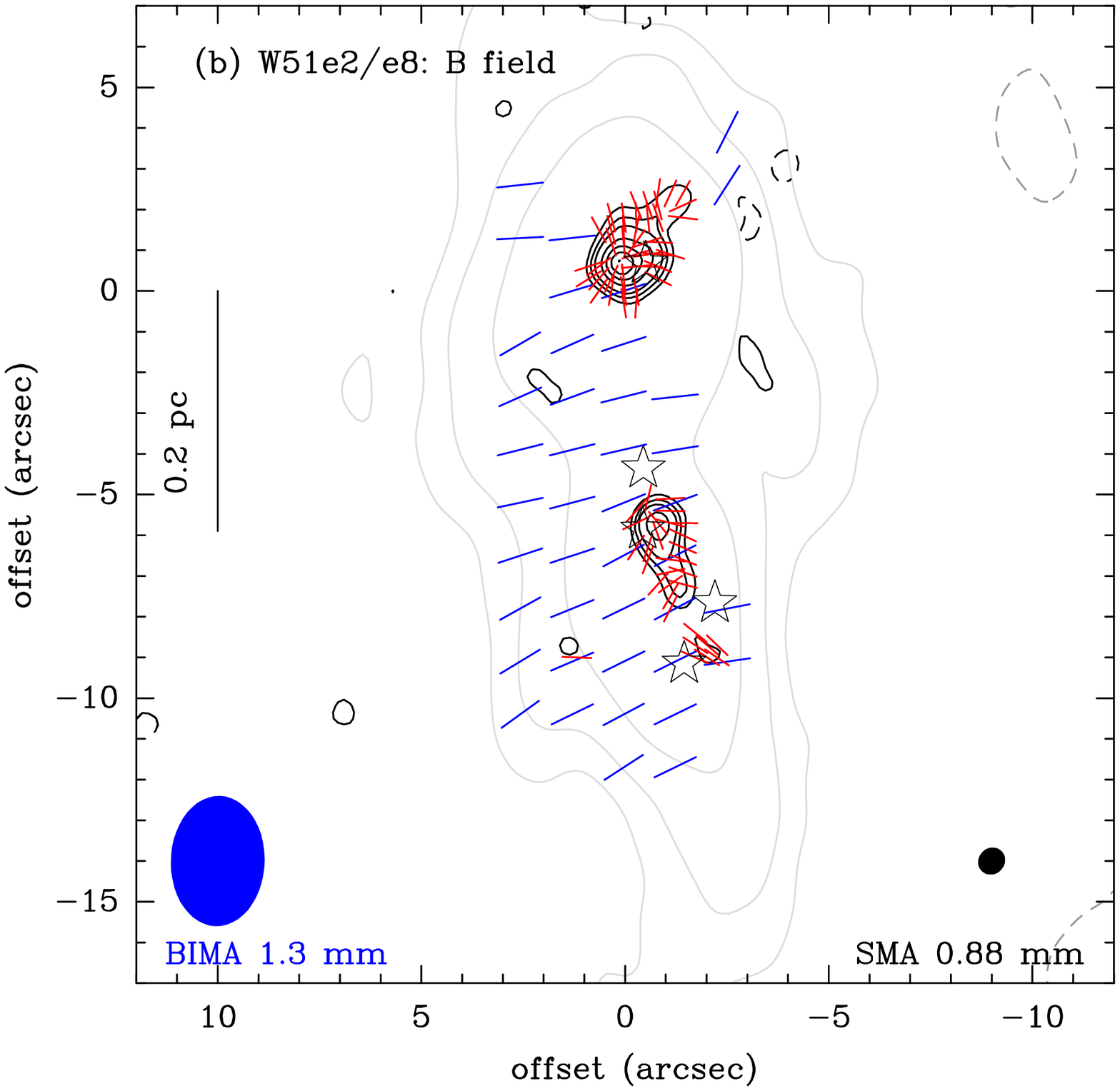}
\end{center}
\caption{(a): SMA dust polarization observation at 870~$\micron$ (345 GHz)
toward the star forming region W51 North \citep{tang13}. The color coding
and gray contours are the Stokes $I$ dust continuum emission from subcompact
(synthesized beam resolution $\sim 4\arcsec$) and extended ($\sim 0\farcs7$)
array configurations, respectively. The corresponding magnetic field orientations are shown with red and white segments, respectively. The 4$\arcsec$ resolution observation seems to trace the envelope of the collapsing higher-resolution cores, with the magnetic field channeling material from both the north and south leading to the formation of cores along an east-west axis. (b): Reproduced based on \citet{tang09b}. The filamentary structure observed by the Berkeley-Illinois-Maryland Association (BIMA) array with $\sim 3\arcsec$ resolution from \citet{lai01} around W51 e2/e8 with dust continuum in gray contours. The magnetic field is shown with blue segments mostly perpendicular to the filament’s longer axis. The SMA higher resolution observations ($\sim 0\farcs7$ resolution) resolve the two cores e2 and e8 (dust continuum intensity in black contours and magnetic field in red segments) that probably appear at locations defined by the larger scale field and filament. The synthesized beam resolutions are shown in the lower-left and lower-right corners. The stars mark ultracompact H \textsc{ii} regions.
\label{fig_sma_pol}}
\end{figure}

\begin{figure}
\begin{center}
\includegraphics[width=0.47\textwidth]{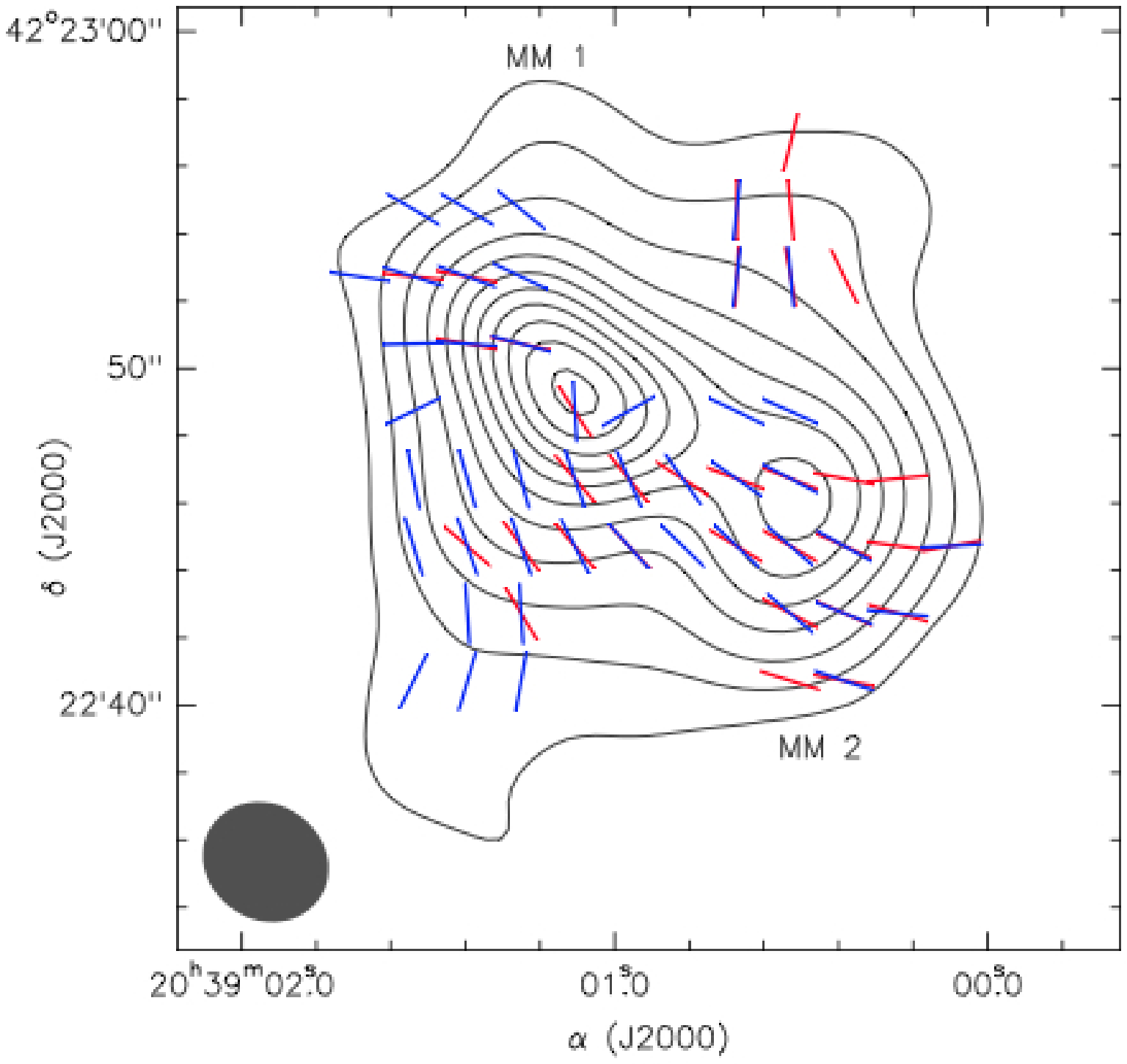}
\includegraphics[width=0.45\textwidth]{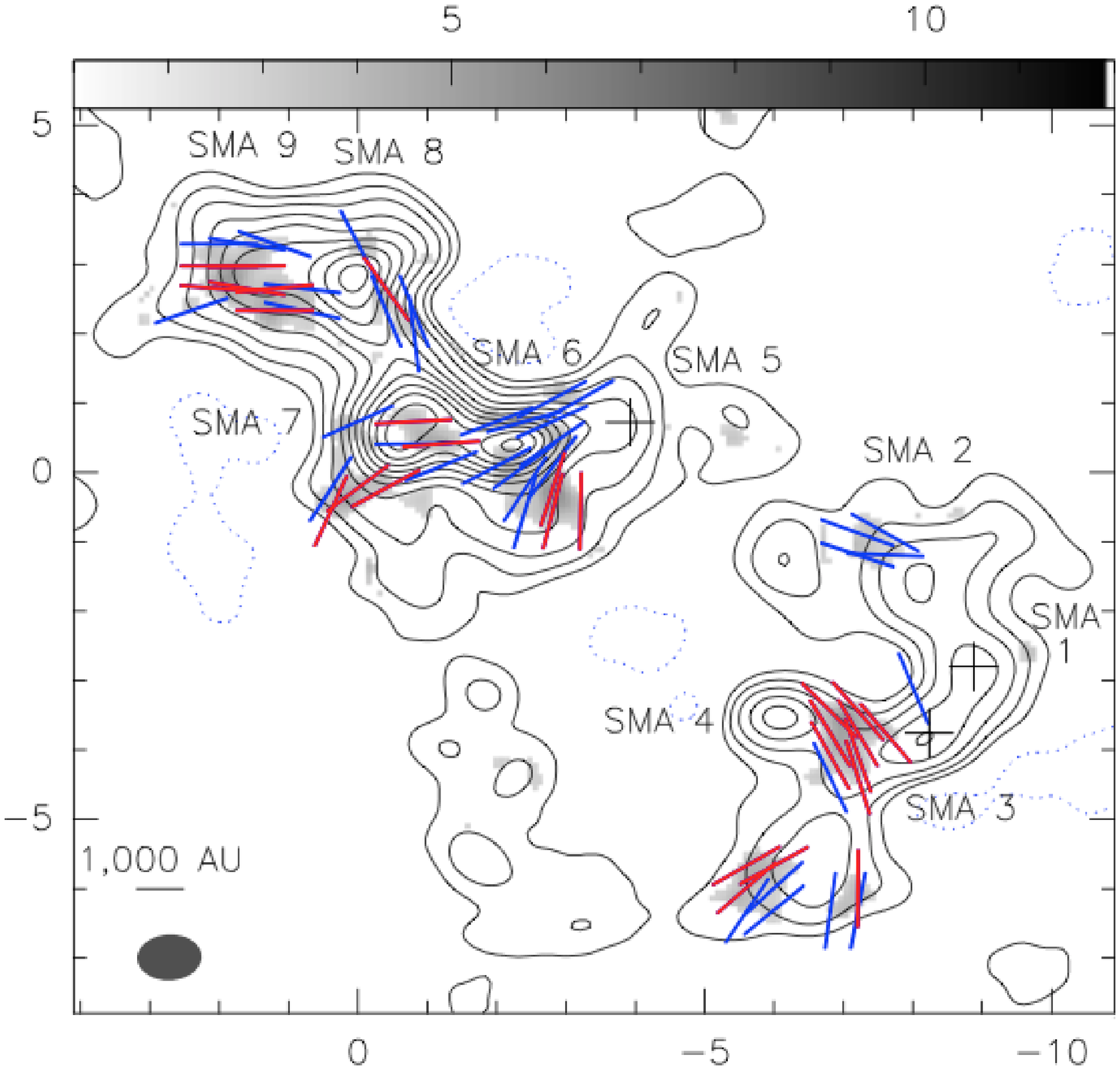}
\end{center}
\caption{Upper panel: DR21(OH) dust continuum contours observed with the SMA
and convolved with a $\sim 3\arcsec$ resolution to match earlier BIMA observations from \citet{lai03}. Magnetic field segments from the SMA and BIMA are in blue and red, respectively. 
Lower panel: The SMA higher resolution observations ($\sim 0.7\arcsec$ resolution) zooming in onto MM 1 and MM 2 from the upper panel. Highly fragmented structures become apparent. Contours display the dust continuum and magnetic field segments in red and blue are detections at more than $3\sigma$ and between 2.5--3$\sigma$.
Figures are reproduced from \citet{girart13}.
\label{fig_pol_more}}
\end{figure}

Unlike the, so far, rather isolated Zeeman effect observations,
dust polarization observations can yield coherent
magnetic field structures over an extended area. Recent observations 
with the SMA \citep[e.g.,][]{girart06,girart09,tang09b,tang13,qiu13}
are revealing hourglass-like magnetic field structures in collapsing
cores. These high-resolution observations ($\sim 0.7\arcsec$)
in combination with new analysis tools \citep{koch12a,koch13,koch14} are now leading to new insights into
the role of the magnetic field in the star-formation process. At the same time, these observations
also show that it is crucial to not only resolve the field structures in the collapsing cores, but 
that it is equally important to trace the magnetic field at intermediate scales ($\sim 2$--$5\arcsec$)
in order to fully understand the role of the field in star formation across all scales (Figure \ref{fig_sma_pol}, upper panel).
These intermediate scales -- in contrast to the large scales
($\sim 10$--$20\arcsec$) traced by the JCMT and CSO --
often reveal the structure of the core-surrounding envelope. Consequently, they provide information
on how the material is transported close to the cores before a collapse is initiated. Figure \ref{fig_sma_pol} (upper panel)
illustrates this for W51 North where the magnetic field in the envelope appears to channel
material that leads to the formation of denser cores along a central axis. 
The filamentary envelope around W51 e2/e8 (Figure \ref{fig_sma_pol}, lower panel) is another example where a uniform field morphology (from a resolution of about $3\arcsec$; \citealt{lai01}) perpendicular to the filament's major axis is likely guiding and defining the locations of the denser collapsing cores e2 and e8 \citep{tang09b}. The high-mass star-forming region DR21(OH) displays a fragmentary structure when observed with the SMA with a resolution of about $0.7\arcsec$ \citep{girart13}. Zooming out shows again more coherent and regular field patterns in its sourrounding environment (Figure \ref{fig_pol_more}). 
Thus, THz observations with a 12-m single-dish antenna, providing resolutions around $\sim 4\arcsec$, will ideally sample these envelope scales. Moreover, these scales will be probed without the zero-spacing missing flux that is present in an interferometer like e.g., the SMA.

Yet another relatively unexplored topic in dust polarization studies is
the role of different dust grain populations as they presumably exist
in different shapes, sizes, and compositions. Typically, it is assumed that dust particles are aligned with their shorter axes parallel to the 
magnetic field, so that detected polarization orientations rotated by 90$^{\circ}$
reveal the magnetic field morphology \citep{cudlip82, hildebrand84, hildebrand88, lazarian07, andersson12}. Although this is the common practice, this only reflects
a crude picture of a ``universal'' dust grain presumably coupling to the magnetic field while
a lot of detailed dust physics is left out. Dust polarized emission is generally
thought to be a complex function of many parameters like dust size and shape, 
dust populations with different temperatures, dust paramagnetic properties, external magnetic field
etc. \citep[e.g.,][]{lazarian07_single}. Current dust polarization observations in the 
submm regime are typically limited to one observing frequency at a telescope, which 
necessarily limits information, sampling likely only a certain population
of dust particles. Combining dust polarization observations from several frequencies
can shed light on the composition of dust populations.

\begin{figure}
\begin{center}
\includegraphics[width=0.46\textwidth]{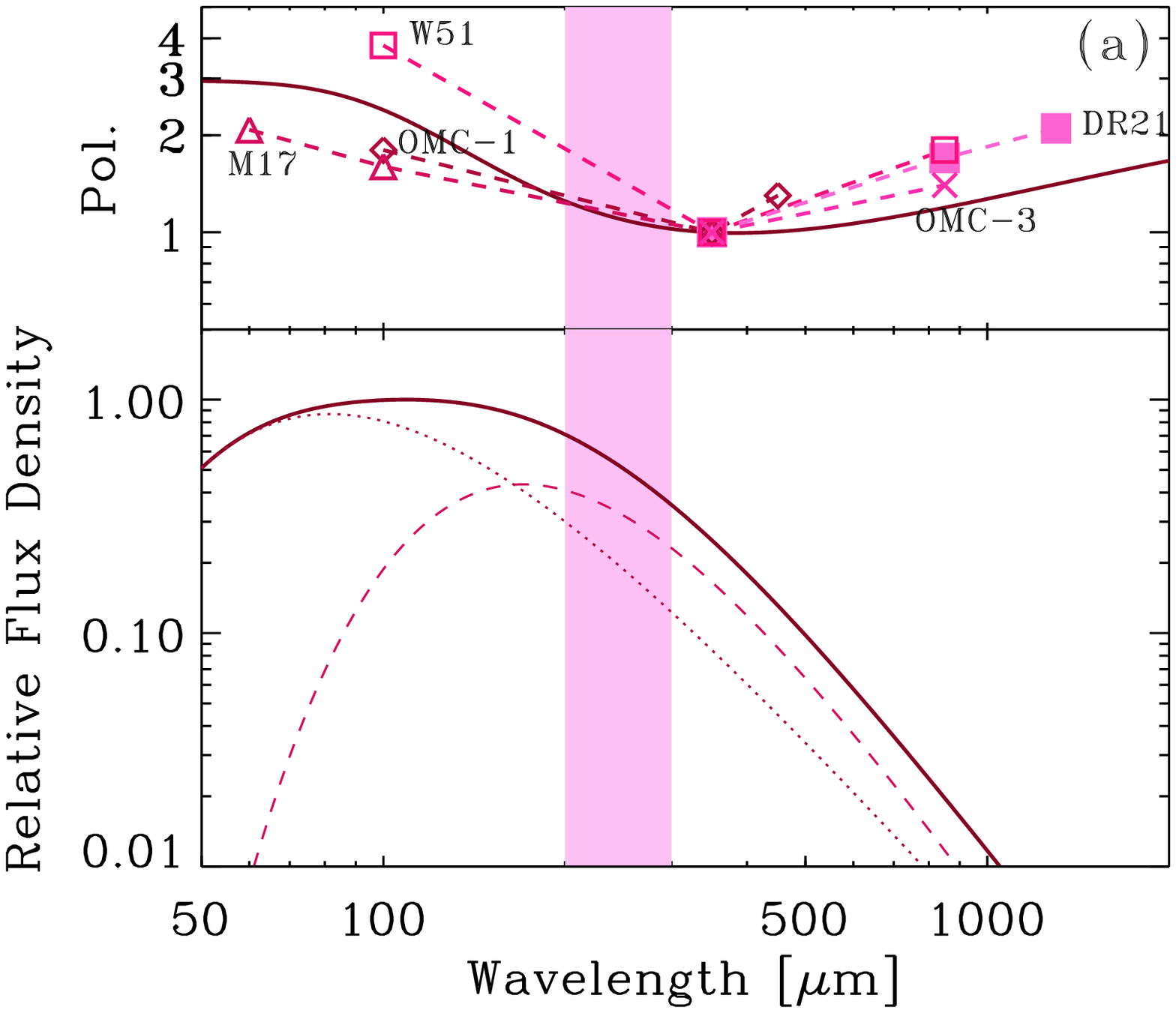}
\includegraphics[width=0.46\textwidth]{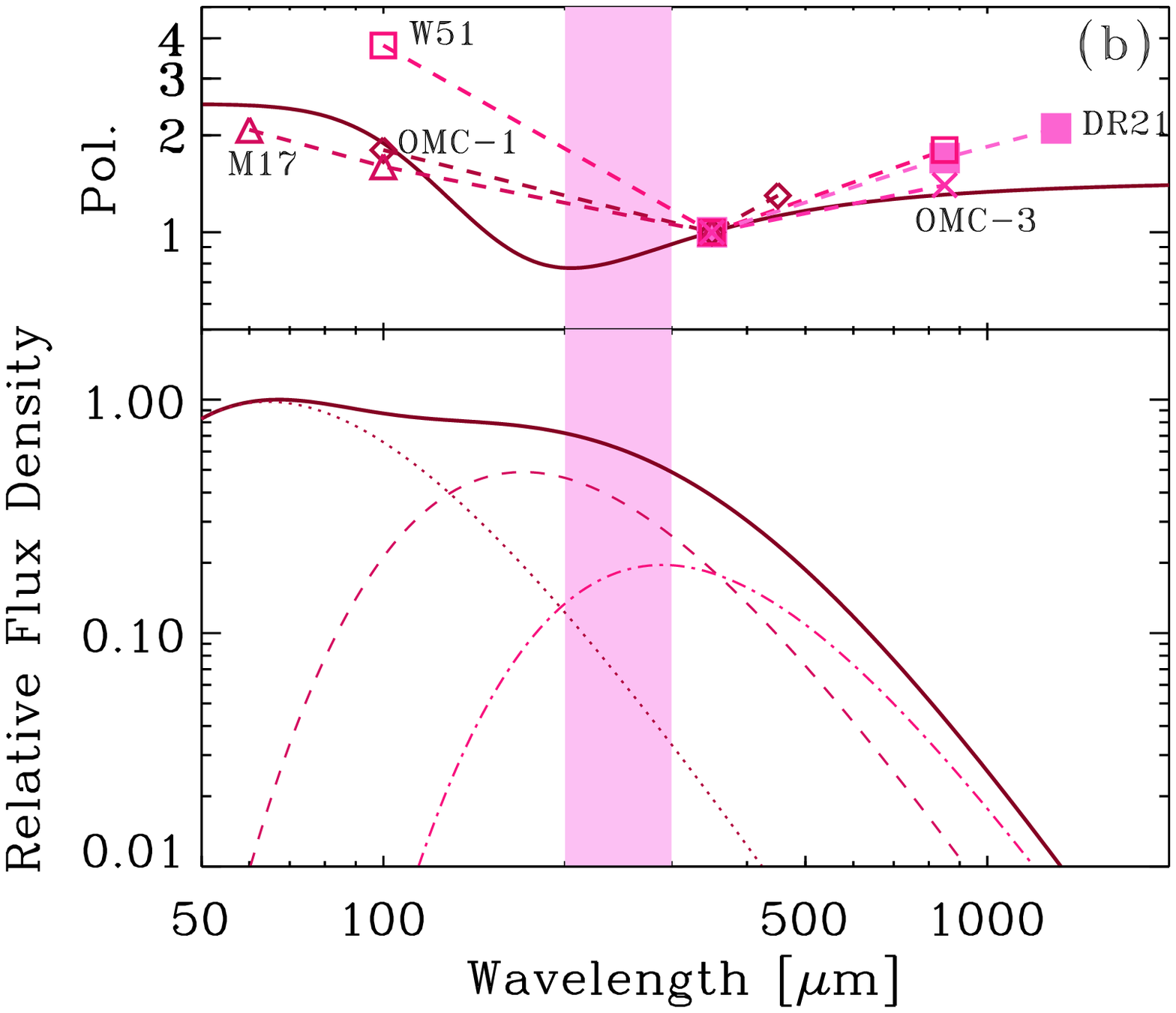}
\end{center}
\caption{\label{spectrum}
Upper windows in each panel: Polarization spectra calculated by
the models in the text, compared with measured polarization spectra
for molecular cloud envelopes, normalized at 350 $\micron$.
The observational data are taken from \citet{vaillancourt08}
and references therein.
Data at 850 $\micron$ are from the JCTM. The wavelength range between 100 and 350$\micron$ is unexplored.
Lower windows in each panel: Corresponding SEDs. The solid
line is the total (the peak is normalized to 1) and the components are shown by
dotted, dashed, and dot-dashed lines.
The shaded regions in the upper and lower panels show
the THz coverage expected for the GLT.
Panel (a):
A 2-temperature component SED with $T_{1,2}=45$ and 17~K,
$\beta_{1,2}=1$ and 2, and $p_{1,2}=4$ and 0\%.
Panel (b): A 3-temperature component SED with $T_{1,2,3}$= 45, 17 and 10 K, $\beta_{1,2,3}=2$, 2, and 2, and $p_{1,2,3}=4$, 0, and 3 (only the relative polarization strangths among the components are important).
Differences and changes in slopes are apparent at $\sim$100--300 $\micron$. 
}
\end{figure}

Recent studies by \citet{vaillancourt12}
comparing polarization percentages and polarization position angles between JCMT 850 $\mu$m and 
CSO 350 $\mu$m sources provide hints for mixtures of different grain properties and polarization efficiencies. In the densest part of molecular cloud cores,
the polarization spectrum rises monotonically with wavelength
\citep{vaillancourt11}, which is consistent with the effect of optical
depth.
In cloud envelopes, where the FIR--submm emission is optically
thin, the polization spectrum falls with
wavelength up to $\sim 350~\mu$m, while it rises at longer wavelengths
(Figure \ref{spectrum}). This non-monotonic behavior requires
multiple components whose polarization degrees are correlated
with the dust temperature and/or spectral index
\citep{hildebrand99,vaillancourt11}.

For more quantification of the importance of THz observations,
we adopt an empirical multi-component SED--polarization model
in \citet{hildebrand99} and \citet{vaillancourt02}.
If the emitting dust consists of components with
different emission and polarization properties, the
resulting polarization spectrum, $P_\mathrm{tot}(\lambda )$, is estimated
as a flux-weighted average of each component:
\begin{eqnarray}
P_{\nu ,\mathrm{tot}}=\sum_i p_iX_i(\lambda ),
\end{eqnarray}
where $i$ is the index of the components, $p_i$ is
the polarization degree of component $i$, and
$X_i(\lambda )$ is the ratio of the flux at
wavelength $\lambda$ given by
$X_i(\lambda )\equiv F_{\nu ,i}(\lambda )/F_{\nu,\mathrm{tot}}(\lambda )$
with $F_{\nu ,i}(\lambda )$ being the flux density
emitted by component
$i$ and $F_{\nu ,\mathrm{tot}}(\lambda )\equiv\sum_iF_{\nu ,i}(\lambda )$.
The flux, $F_{\nu, i}(\lambda )$, is assumed to be described by a modified
black body with an emissivity index of $\beta_i$:
\begin{eqnarray}
F_{\nu ,i}(\lambda )=W_i\nu^{\beta_i}B_\nu (T_i),\label{eq:F_comp}
\end{eqnarray}
where $B_\nu (T_i)$ is the Planck function at
frequency $\nu$ and temperature $T_i$ and $W_i$ describes the
weight for the component $i$.\footnote{Using the mass absorption
coefficient of dust $\kappa_\nu =\kappa_0(\nu /\nu_0)^\beta$
($\kappa_0$ is the value at $\nu =\nu_0$), the dust mass
$M_\mathrm{d}$, and the distance to the object $D$, the
flux is expressed as $F_\mathrm{\nu}=\kappa_\nu M_\mathrm{d}B_\nu (T)/D^2$
(the subscript $i$ is omitted for brevity). Comparing this with
equation (\ref{eq:F_comp}), the
weight factor $W$ represents $\kappa_0M_\mathrm{d}/(\nu_0^\beta D^2)$;
that is, the weight $W$ is proportional to the
mass absorption coefficient times the dust mass for the
component of interest.}

For the purpose of showing representative examples, we first adopt
the same parameters as in \citet{vaillancourt11}. We examine two models:
one is a two-component model in which $\beta_i$
is different between the two components
($\beta_1=1$ and $\beta_2=2$), while
the other is a three-component model
in which
$\beta_i$ is the same ($\beta_i=2$) for all the three components
($i=1$, 2, and 3).
Each component has a different dust temperature,
45, 17, and 10 K for $i=1$, 2, and 3, respectively,
and a different polarization degree, $p_i=4$, 0, and
3 for $i=1$, 2, and 3, respectively
(the last component does not exist in the two-component
model). In this analysis we are only interested in the
relative strength of polarization, and the polarization is
always normalized to the value at 350 $\micron$
(this is why we give $p_i$ without units).
In Figure \ref{spectrum}, we show the results for
the polarization spectrum ($P_{\nu ,\mathrm{tot}}(\lambda )$)
and the flux ($F_{\nu ,i}(\lambda )$ and
$F_{\nu ,\mathrm{tot}}$) for these typical choices of values
given in \citet{vaillancourt11}. The flux is normalized so that the
maximum of $F_{\nu ,\mathrm{tot}}$ is 1.

As shown in Figure \ref{spectrum}, we confirm the previous results
that the non-monotonic behavior of the polarized spectrum is
reproduced both by the 2- and 3-component models.
The observational studies are currently limited to a few wavelengths
with no observations in the THz atmospheric window (1--1.5 THz) regime
(Figure \ref{spectrum}). The falling spectrum between 60 and 350 $\micron$
is indicative of a dust model where warmer grains
are better aligned than colder grains \citep{hildebrand99,vaillancourt02}. The change in slope
with the rising spectrum at $\geq 350~\micron$ requires a second dust component with a colder
temperature or a lower spectral index than the warmer component. 
Generally, a varying polarization spectrum will require at least
2 dust components at different temperatures.
Every change from a falling to a rising (or vice versa) spectrum and every (noticeable) change in 
the slope will point toward an additional component. Measurements in the THz regime around 
200 $\mu$m will fill in the gap between 100 and 350 $\mu$m which can
distinguish between a 2-temperature and a 3-temperature component model
(Figure \ref{spectrum}).
Additionally, in combination with ALMA, which is expected to 
eventually cover polarization observations 
up to about 900 GHz, we will
be able to study dust grain polarization properties in the full frequency range
around THz.

\begin{figure}
\begin{center}
\includegraphics[width=0.46\textwidth]{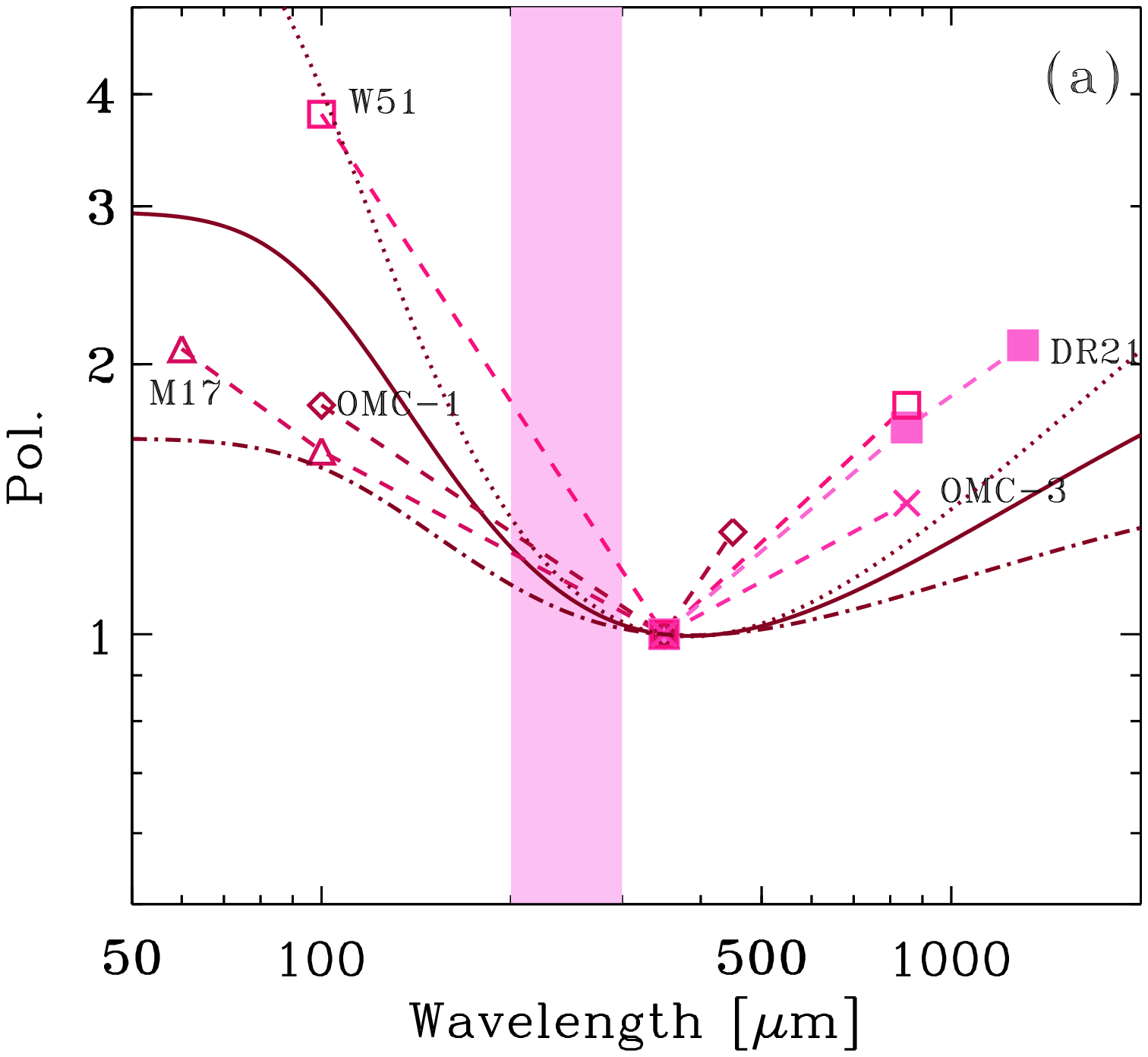}
\includegraphics[width=0.46\textwidth]{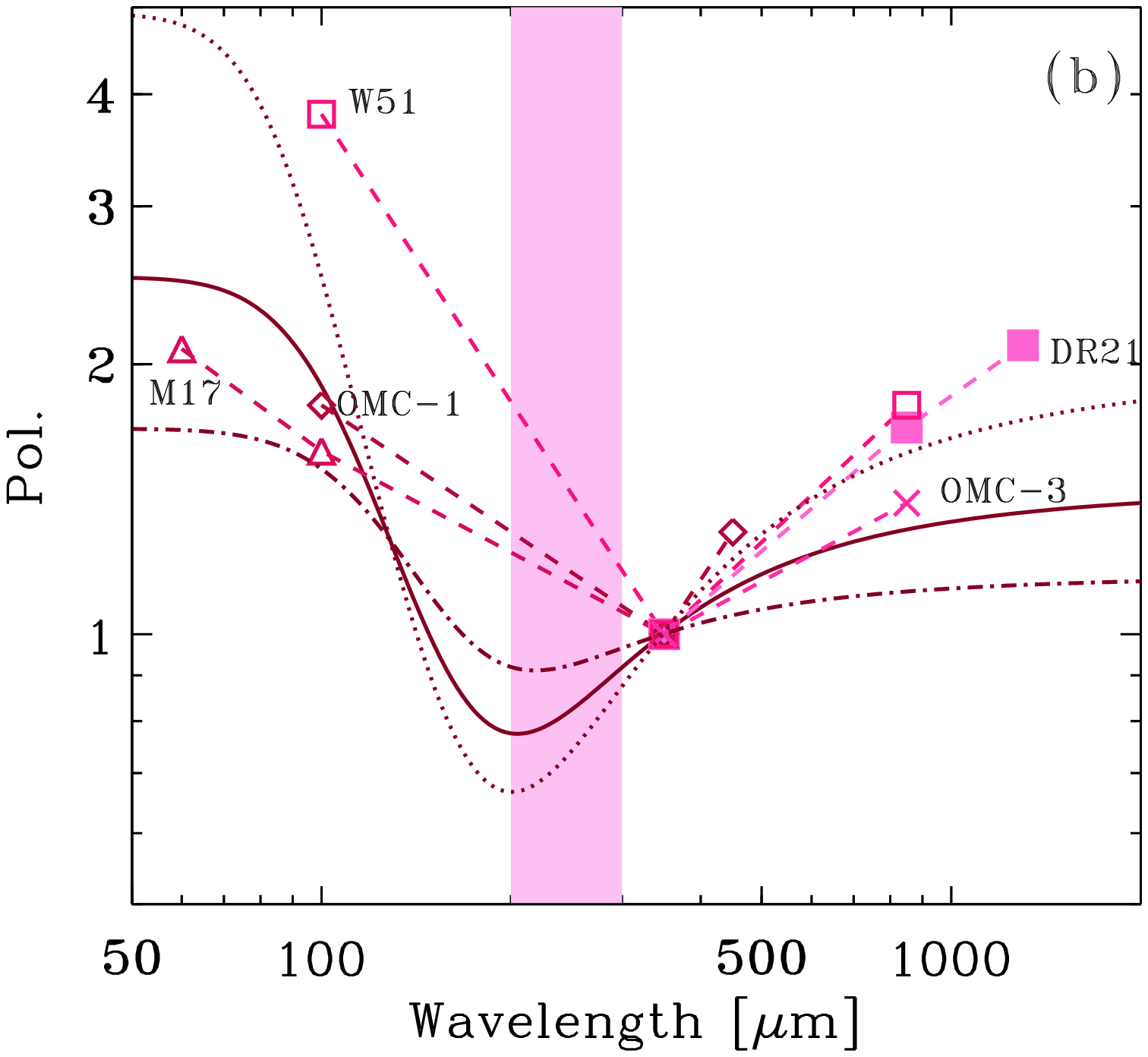}
\end{center}
\caption{\label{fig:pol_spectrum}
Same as the upper windows in each panel of Figure \ref{spectrum}, but
for various fractions of the 17-K component (component $i=2$),
which contributes to the THz band the most. The solid,
dotted, and dot-dashed lines show the same fraction as in
Figure \ref{spectrum}, an increased fraction of component
$i=2$ by a factor of 3, and a decreased fraction of component $i=2$
by a factor of 3, respectively.
}
\end{figure}

To emphasize more the importance of the GLT THz wavelengths,
we present the polarization spectra for various fractions of
component $i=2$, which contributes to the flux at THz
wavelengths the most, in Figure \ref{fig:pol_spectrum}.
We observe a clear distinction in the polarization spectra
between the 2- and 3-component models.
As mentioned above, in the 2-component model,
the difference tends to appear at
wavelengths out of the range where the contribution of
component $i=2$ (with $T=17$ K; i.e., the component peaking
at the THz frequencies) to the flux is dominated.
In contrast, the 3-component model predicts that the
THz polarization is sensitive to the different contribution
from component $i=2$. Therefore, adding a data point at
THz wavelengths to both the SED and the polarization spctrum
provides a strong test for the multi-component dust models.
Within the GLT THz band (i.e., the shaded region in
Figure \ref{fig:pol_spectrum}), the shortest wavelength
(i.e., 200 $\mu$m; or frequency 1.5 THz) has the largest
advantage in distinguishing the difference given the 350 $\mu$m data,
simply because it is the farthest from 350 $\mu$m.

Careful modeling of radiative transfer will be an additional asset.
The different frequencies possibly sampling different grain populations are likely also presenting
different optical depths.
By consistently treating the optical depth effects with
the temperature variation, a complete frequency coverage
including the THz domain will eventually also 
allows us to do dust polarization tomography with an ultimate goal of understanding the 3-dimensional
magnetic field structure.

The polarization capability of the THz detectors discussed in
Section \ref{sec:development} is not well established.
For the purpose of roughly estimating time and efficiency for
a typical continuum polarization observation
in the THz regime, we can use a comparison with the SMA that has been conducting polarization observations
over the past decade. The GLT, operating at $\lambda\sim 200~\mu$m, will win over
the SMA ($\lambda\sim 870~\mu$m) by a factor $\sim (870/200)^{2.5}\sim 30$,
assuming dust emission to scale in frequency with 
a power-law index $\alpha=2.5$. The GLT will loose in collecting area by a factor 2 as well as in system
temperature by a factor of $\sim$100 (Table 1) compared with the SMA.
All together, the GLT is likely to be a little slower in integration time
than the SMA by a factor $\sim 3$. This further builds on anticipating a 
multi-beam receiver, covering a field-of-view similar to the SMA ($\sim 30\arcsec$). Currently, 
the SMA is reaching a $\sim 1$~mJy beam$^{-1}$ sensitivity, allowing $\sim 7\sigma$ detection on a typical
Galactic source in one night. We, therefore, project similar numbers for the GLT, i.e., 1--2 source
detections per night. In order to maximize scientific impact and to be able to
perform statistical studies, we are aiming for a dedicated program with a sample of $\sim$50 star-forming regions. 
Including overhead (e.g., technical issues, bad weather) we expect $\sim 3$ months to carry out 
such a program.

SOFIA/High-resolution Airborne Wideband Camera
(HAWC)\footnote{https://www.sofia.usra.edu/Science/instruments/\\
instruments\_hawc.html}, which covers 50--240 $\mu$m, also has a capability of
THz polarimetry. Since the atmospheric transparency is much better for
SOFIA than for the GLT, HAWC has better sensitivity than the GLT. However, the
beam size of SOFIA (19$\arcsec$) may be too large to trace the
interesting structures shown in Figures \ref{fig_sma_pol} and \ref{fig_pol_more}.

In summary, the uniqueness of THz dust continuum polarization observations
with a 10-m class antenna such as the GLT lies in what follows:
We can trace the polarization of the dust component that have the
largest contribution to the spectral peak. Compared with airborne and space
THz observatories, the ground-based THz telescopes provide higher angular
resolutions, which correspond to intermediate scales of a few arcsec
in Galactic star-forming regions.
THz dust polarization observations will probe different dust grain populations
and/or different  optical depths from existing submm observations.
Thus, adding the THz polarization data point provides
additional constraints to separate different dust components. Furthermore,
in combination
with the SMA and ALMA covering a frequency range up to about 900 GHz, various optical depths or various dust temperature layers will 
be probed, which will allow us to do dust polarization tomography well into the THz regime.

\subsection{Extragalactic THz cases}
\label{subsec:extragal}

Dust emission from a galaxy is often used as an indicator of
SFR since young massive stars dominate the total stellar light in
star-forming galaxies and ultraviolet radiation originating from
massive stars is most efficiently absorbed by dust
compared with longer-wavelength radiation \citep[e.g.,][]{buat96}.
The overall luminosity and the peak frequency of the dust thermal
emission spectrum, which characterize the amount and the temperature of the
interstellar dust, can be obtained by fitting the SED at
$\lambda\sim 20$--1000 $\mu$m.
For example, tremendous attention has been given to the ground based telescope surveys
\citep[][and references therein]{blain03,blain04} of the submm selected
$z\sim 1$--3 galaxies, which revealed the intriguing starburst objects with exceptionally luminous ($\sim 10^{11-14}$ L$_\odot$) FIR emission from relatively warm (30--60 K) dust.

Despite the more abundant ancillary data to trace the young stellar clusters
with high angular resolution, the FIR dust imaging of starburst regions
in galaxies remains difficult because the ground-based observations demand
conditions with extremely low water vapor and the observations by
space telescopes only marginally resolved a limited number of
the nearest galaxies. Analysis of the \textit{AKARI} satellite
\citep{murakami07} images of M81 with an angular resolution of
$\sim 40\arcsec$ by \citet{sun11} demonstrated that regions with
high dust temperatures trace star-forming regions associated with H\,\textsc{ii} regions but that the dust temperatures and
FIR colors obtained do not reflect the correct (or physical) dust emission
properties in the regions because the high-dust-temperature regions are
not resolved.
Compared with the \textit{Herschel} SPIRE mapping (with
an angular resolution of $18\arcsec$ at 250 $\micron$;
\citealt{pohlen10,foyle12}), a higher angular resolution
$\sim 4\arcsec$ should be the advantage of the GLT.

The high angular resolution of the GLT will enable us to
resolve nearby (with distance $d<30$ Mpc) galaxies with
spatial resolutions comparable to
or better than our above \textit{AKARI} observation of M81
($d\sim 3.6$ Mpc; \citealt{freedman94}). This is especially important
to the lowest metallicity end in our sample, which are generally compact
(see Section \ref{subsubsec:BCD}). The comparable spatial resolution with
our previous M81 analysis is the minimal requirement to avoid the unreliable dust temperature estimate because of a mixture of various dust-temperature components
\citep{sun11}, and the GLT (or ground-based THz telescopes) is the only facility that can meet this requirement.
The resolved nearby templates can be used to gauge the uncertainties
in fitting the SEDs of distant unresolved galaxies.

We also emphasize that the spatial resolution achieved by the GLT is
indeed suitable for resolving the individual star-forming
complexes. Observing the well-known nearby starburst galaxy NGC 4038/4039
(the antennae galaxies) with the SMA, \citet{wei12} presented
132 marginally resolved giant molecular star-forming complexes
(see also \citealt{espada12} for ALMA observations). The reported cloud
statistics in the antennae galaxies appears bimodal, and can be divided
into two populations according to
the molecular gas mass, $M_\mathrm{mol}$:
(i) $\log (M_\mathrm{mol}/M_{\odot})\ge 6.5$, with
a mean cloud radius of 180 pc (i.e., a diameter of 360 pc), and
(ii) $\log (M_\mathrm{mol}/M_{\odot})<6.5$, with
a mean cloud radius of 57 pc (i.e., a diameter of 114 pc).
The mean molecular gas mass of the sample (i) is
$\sim$100 times higher than the molecular gas mass of the other
sample. The higher mass clouds are likely the candidates to
form super star clusters. The angular resolution (4$\arcsec$) of
the GLT corresponds to a $<$390 pc
spatial scale for $d<20$ Mpc galaxies, which is suitable for
diagnosing the heating by stars in these clouds. 

In order to maximize the scientific output within the limited time
of suitable weather for THz observations,
we make more focused discussions below.

\subsubsection{Possible targets I: high surface brightness galaxies with
a wide metallicity range}\label{subsubsec:BCD}

Although survey observations of an ``unbiased'' nearby galaxy sample
is the ultimate goal to obtain a general understanding of
the dust distribution in galaxies, it is more realistic to
start with high surface brightness galaxies, which are relatively easy to detect.
In particular, we focus on galaxies which host compact starbursts.
Our nearby galaxy sample described later
covers a wide range in metallicity (which may be related to the dust abundance), so that we can also test how the dust mass estimate in the entire galaxy depends on the metallicity.
The metal poor starbursts in the nearby Universe could be used as a proxy
of high-$z$ primeval starbursts.

As an appropriate sample, we target nearby blue compact dwarf galaxies
(BCDs), some of which host active starbursts leading the formation of
super star clusters; that is, young dense star clusters with
masses as high as globular clusters \citep[e.g.,][]{turner04}.
The starbursts in BCDs are relatively isolated in that the
star formation occurs only in compact central regions and
the rest of the galaxy is quiescent. Therefore,
if we observe nearby BCDs,
we can investigate the detailed properties of starbursts by
isolating the relevant star-forming
regions without being contaminated (or bothered)
by other star-forming regions.

Another scientific merit of observing BCDs (or dwarf galaxies in general)
is that we can collect a sample with a wide metallicity rage.
Metallicity is an important factor of galaxy evolution, and
the dust abundance (or dust enrichment) is also known to be
closely related to the metallicity \citep{schmidt93,lisenfeld98}.
As explained in Section \ref{subsec:hiro_Tdust}, the
dust temperature is obtained more precisely if we
include THz data. A precise estimate of dust temperature
is a crucial step for obtaining a precise dust mass and
dust optical depth. By using these two quantities, we
will be able to discuss the dust enrichment history and
the shielding (enshrouding) properties of dust in
star forming regions at various metallicities.
Our project is complementary to the \textit{Herschel}
Dwarf Galaxy Survey (DGS) \citep{madden13}, in which the majority
of compact blue dwarf galaxy samples are not well resolved or
are unresolved in the FIR bands.

\begin{figure}
\begin{center}
\includegraphics[width=0.45\textwidth]{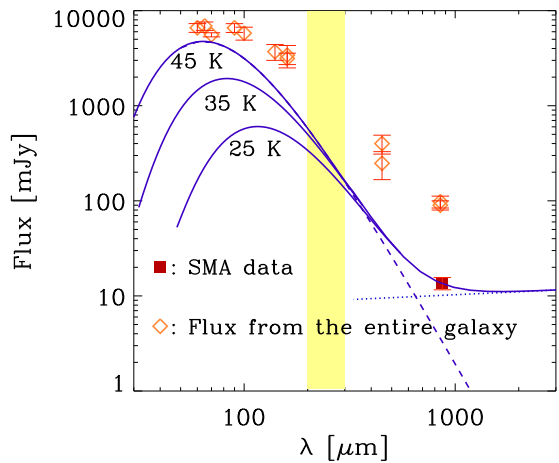}
\end{center}
\caption{FIR--submm data and model calculations for II Zw 40.
Solid, dotted, and dashed lines represent, respectively,
the total, the free--free component, and the
dust component calculated for the models for the central
starburst region in II Zw 40 by
\citet{hirashita11}. The dust mass is adjusted to
reproduce the SMA data point at 880 $\mu$m (filled
square). Three cases for the dust temperature are
examined: 45, 35, and 25 K for the upper, middle,
and lower solid lines, respectively. As upper limits for
the flux in the central region, we also show observational
data for the entire II Zw 40 system (open diamonds),
which are taken from \citet{vader93} for 60 and 100 $\mu$m
(\textit{IRAS}), \citet{hunt05} for 450 and 850 $\mu$m
(SCUBA), \citet{galliano05} for 450 and 850 $\mu$m
(SCUBA), \citet{engelbracht08} for 70 and 160 $\mu$m
(\textit{Spitzer}), and \citet{hirashita08} for 65, 90, 140,
and 160 $\mu$m (\textit{AKARI}). The shaded region shows the
THz range (200--300 $\mu$m) considered in this paper
(i.e., the wavelength range where THz atmospheric windows
are available).
\label{fig:sed_bcd}}
\end{figure}

We choose the brightest BCDs and discuss the detectability
based on our SMA observations at 880~$\micron$.
The central several-arcsec structure is well investigated by the SMA
for II Zw 40 by \citet{hirashita11}. We adopt II Zw 40 as a
standard case, although it is not visible
from the GLT site. In Figure \ref{fig:sed_bcd}, we show the
model SED constructed with the dust modified blackbody
plus free--free emission spectrum for II Zw 40
\citep[see][for the details]{hirashita11}.
The SMA data point at 850 $\mu$m for the central
star-forming region is also shown in Figure \ref{fig:sed_bcd}. In the same figure, we plot some single dish data obtained by \textit{AKARI}, \textit{Spitzer}, and the JCMT/the Submillimetre Common-User Bolometer Array
(SCUBA), noting that they should be taken as upper limits for the central region of interest.
Three cases for the dust temperature ($T_\mathrm{dust}$) is
examined: 45, 35, and 25 K. These cases correspond
to the cases where the star-forming region is concentrated
within the SMA beam ($\sim 5$ arcsec) (45 K), and slightly extended
out of the beam with the size of 10 arcsec (35 K).
Note that these temperatuers are theoretically derived by
assuming the radiative equilibrium with the ambient
stellar radiation field, so we additionally examine a case
with a lower temperture, 25 K, which is typical of the
object with moderate star formation activities such as
spiral galaxies \citep[e.g.,][]{remy14}. The SEDs are constructed so that they fit
the SMA data point at 880 $\mu$m which is decomposed into
the dust and free--free components. The level of the
free--free component is determined by the extrapolation of the
free--free spectrum that fit the data points around 10 GHz.
\citet{hirashita11}
also pointed out that the contamination of free--free
emission is significant even at 880 $\mu$m in the
central part of BCDs \citep[see also][]{hirashita13}.
This also points to the necessity of observing BCDs
at shorter FIR wavelengths, where the dust emission is
sure to be dominant.

According to the models, the central part of II Zw 40 has
a flux of 558, 486, and 347 mJy at 200 $\mu$m (1.5 THz)
for $T_\mathrm{dust}=45$, 35, and 25 K, respectively.
Since the difference is larger at 1.5 THz than at 1.0 THz,
it would be better to choose the 1.5 THz window.
(Higher THz frequencies would be better just for the purpose
of discriminating the dust temperature, but are only available
for airborne or
space facilities, which are difficult to be equipped with a
10-m class telescope.)
According to \citet{hirashita13}, 61\% of the flux is
concentrated in the SMA beam, whose size is comparable to
the GLT beam at 1.5 THz. Therefore we adopt
$347\times 0.61=211$ mJy beam$^{-1}$ for the expected GLT flux.
With the sensitivity discussed in Section \ref{subsec:capability},
it is difficult to detect such an extragalactic star-forming
region, although we target the brightest class.
However, for continuum, more sensitive bolometer-type facilities
may be available. If we
could achieve a 10-times better sensitivity
(1.9 Jy for 1-sec integration), the above source can be detected
with 5 $\sigma$ in an on-source integration time of 30 min.
This kind of sensitivity
is realistic considering the existing planning such as
ArTeMiS.
In other words, such a sensitivity as an order of 1 Jy
with 1-sec integration is crucial for extragalactic dust
studies at THz frequencies.

We list the selected sample in
Table \ref{tab:Hiro_sample}. We chose a sample from
\citet{klein91} based on bright radio continuum emission,
which is known to correlate with FIR brightness \citep{hirashita13}.
We list the BCDs brigher than $1/\sqrt{2}$ of
II Zw 40 ($21\pm 2$ mJy) at
a wavelength of 2.8 cm, expecting that such galaxies can be
detected within the twice of the expected integration time
for II Zw 40. This is for one pointing, and a multi-pixel
detector is surely required to get a panoramic view of
individual galaxies.

\begin{table*}
  \tbl{Sample of Blue Compact Dwarf Galaxies (BCDs) suitable for GLT observations.}{%
  \begin{tabular}{llcccccc}
  \hline
Object &  Other names & RA (J2000) & Dec (J2000) & Distance & $12+\log(\mathrm{O/H})$ & $S_\mathrm{2.8~cm}$ & Ref.\\
 & & [h m s] & [$\arcdeg$ $\arcmin$ $\arcsec$] & [Mpc] & & [mJy] & \\
  \hline
Haro 1  & UGC 3930, NGC 2415 & 07 36 56.7  & +35 14 31 & 52.0 & 8.40 & $21\pm 3$ & 1\\
Mrk 140 &  & 10 16 28.2 & +45 19 18 & 22.7 & 8.30 & $18\pm 6$ & 2\\
Mrk 297 & NGC 6052, UGC 10182, & 16 05 13.0 & +20 32 32 & 63.0 &    8.65 & $22\pm 5$ & 3        \\
 & Arp 209 & \\
Mrk 314 & NGC 7468, UGC 12329 & 23 02 59.2 & +16 36 19 & 31.1 & 8.10 & $25\pm 3$ & 1\\
III Zw 102 & NGC 7625, UGC 12529, & 23 20 30.1 & +17 13 32 & 25.0 & 8.49 & $17\pm 1$  & 1      \\
& Arp 212 & \\
  \hline
\end{tabular}}\label{tab:Hiro_sample}
\begin{tabnote}
References for $12+\log\mathrm{(O/H)}$: 1) \citet{cairos12};
2) \citet{izotov06}; 3) \citet{james02}.
\end{tabnote}
\end{table*}

\subsubsection{Metallicity dependence of [N\,{\small II}] 205 $\mu$m line}
\label{subsec:NII_extragal}

In Section \ref{subsec:Gal_NII}, we suggested that
[N\,\textsc{ii}] fine-structure line emission is
a useful probe of star formation, since it is emitted from
H \textsc{ii} regions developed around young massive stars.
We also proposed that the comparison with an H$\alpha$ map
or radio (free--free emission) map serves as a local calibration of
the [N\,\textsc{ii}] luminosity for a star formation indicator.
Here, we propose to extend these studies to an extragalactic sample.
In particular, [N \textsc{ii}] 205 $\mu$m emission
is poorly focused on even with \textit{Herschel},
especially because this line is located at the edge of the spectral
band \citep{cormier15}.

We especially focus on the same BCD sample as above
for the purpose of clarifying the metallicity dependence of
[N \textsc{ii}] emission as well as confirming that [N \textsc{ii}]
emission traces the star formation activities.
The correlation between the metallicity and
the hardness of stellar radiation field
\citep{hunt10} means that both these
factors enter the ``metallicity dependence'' of
the intensity of fine-structure lines. To seperate these
two factors, observations of other fine-structure lines
are necessary \citep{cormier15}.

As formulated in Section \ref{subsec:Gal_NII}, it is useful
to relate the [N \textsc{ii}] 205 $\mu$m luminosity,
$L_\mathrm{N\,II}$ to the SFR
by introducing the proportionality constant, $C_\mathrm{N\,II}$:
\begin{eqnarray}
L_\mathrm{N\,II}=C_\mathrm{N\,II}\mathrm{SFR}.
\end{eqnarray}
\citet{inoue14} derived $C_\mathrm{NII}$
as a function of metallicity using the photoionization
code \textsc{cloudy}:
$\log C_\mathrm{[N\,II]}/[\mathrm{(erg~s^{-1})/(M_\odot~yr^{-1})}]=39.82$,
39.07, 37.72, and 36.85 for
$Z/\mathrm{Z}_\odot =1$, 0.4, 0.2, and 0.02, respectively.

We take an example of II Zw 40 (the same object used in
Section \ref{subsubsec:BCD}) to estimate the detectability
of the [N \textsc{ii}] line.
For the central $\sim$4 arcsec region of II Zw 40,
\citet{beck02} derived a flux of 14 mJy at a
frequency of 15 GHz. We adopt this frequency, since
the free--free emission is probably optically thick
at lower frequencies \citep{hirashita11}. Using
equation (\ref{eq:fNII}) for $\nu =15$ GHz,
$T_e=10^4$ K, and $\Delta\nu =10^8$ Hz
(corresponding to a velocity width of 20 km s$^{-1}$) with the
reduced efficiency of [N \textsc{ii}] by 2.1 dex due to the
low metallicity (we adopt the value for 0.2 Z$_\odot$;
$C_\mathrm{H\alpha ,N\,II}=4.14\times 10^{-4}$),
we obtain
$f_\mathrm{N\, II}=58.5(S_\nu /\mathrm{mJy})$ Jy.
Since the beam size of the GLT is equal to the
radio 15 GHz beam above ($\sim 4$ arcsec), the
above 15 GHz flux ($S_\nu =14$ mJy) indicates that
$f_\mathrm{N\, II}=6.5$ Jy beam$^{-1}$ in the GLT
1.5 THz band.
Using the same capability as in Section \ref{subsec:capability},
we obtain an rms = 227 Jy beam$^{-1}$ at 1.5 THz for a velocity
resolution of 10 km s$^{-1}$ for a 1-sec integration, an
on-source integration time of 3.0 hr is required for a 3 $\sigma$
detection. Since this integration time is realistic,
we can test the metallicity dependence of [N \textsc{ii}] 205 $\mu$m
luminosity with the same BCD sample as in Section \ref{subsubsec:BCD}.

Our GLT observations of the [N\,\textsc{ii}] 205 $\mu$m line
in nearby galaxies will provide basic knowledge for the interpretation
of high-$z$ [N\,\textsc{ii}] lines by ALMA. If the metallicity is
known for the system, the [N\,\textsc{ii}]
intensity is directly used to determine the SFR of
the system. The metallicity can also be estimated
by ALMA observation for high-$z$ galaxies,
as \cite{nagao12} demonstrated that
[N\,\textsc{ii}] 205 $\mu$m/[C\,\textsc{ii}] 158 $\mu$m
flux ratio can be used to estimate the metallicity.

The CO $J=13$--12 line is also in the 1.5 THz window
(Table \ref{tab:lines}), but its intensity is much weaker.
Therefore, it is not probable that the CO line is also
detected in nearby BCDs. Such a highly excited CO line can be
emitted from PDRs with
extremely high density ($\gtrsim 10^{6.5}$ cm$^{-3}$)
and extremely high UV intensity ($\sim 10^5$ times the
interstellar radiation field in the Galaxy)
\citep{meijerink07,vanderwerf10}.
Indeed, \citet{vanderwerf10}
explained the Herschel detection of CO $J=13$--12 in 
Mrk 231 by
X-ray dominated region (XDR). If nearby AGNs have
a similar CO $J=13$--12 flux ($\sim 1$ Jy at peak
with $\Delta\nu /\nu\sim 8\times 10^{-4}$, corresponding
to a velocity width of 240 km s$^{-1}$),
an on-source integration of $\sim 5.4$ hour is necessary to detect
CO $J=13$--12 line with 3 $\sigma$ level in nearby AGNs
if we assume the same velocity resolution (240 km s$^{-1}$)
(see Table \ref{tab:capability} for the detection limit).
Probably, this is too time-consuming for the first-generation
ground-based THz telescopes.
Because such an extragalactic highly
excited region is compact and regarded as a point source,
a single-pixel detector will be sufficient for
this purpose. The requirement for angular resolution is not
severe. Therefore, airborne or space telescopes may be more
suitable for the purpose of observing extragalactic highly
excited lines.

\subsubsection{Possible targets II: the foot points of dust-enriched outflows
in starburst galaxies}\label{subsubsec:starburst}

The metal enrichment of the intergalactic medium (IGM)
occurs through the outflow of metals and dust from galaxies
\citep[e.g.,][]{adelberger03,kobayashi07}. Outflows from actively star-forming
galaxies or starburst galaxies are commonly seen
\citep[e.g.,][]{heckman90,lehnert95,lehnert96}.
They are often seen in optical emission lines or soft X-ray emission
\citep[e.g.,][]{martin98,dahlem98,strickland04,sharp10}.
Being associated with a starburst activity,
outflows are considered to be driven by
a large number of supernova explosions. This huge explosion energy blows
the surrounding ISM away from the galactic disks mostly in the perpendicular direction to the disks, and creates outflows \citep[e.g.,][]{tomisaka88,strickland00}. Such a huge energy input to the surrounding ISM affects the activities inside the ISM (i.e., star formation) in the destructive way; namely it terminates the star formation. On the other hand, the expansion of the ISM toward the galactic disk accumulates the ISM and creates new star formation or starburst \citep{matsushita05}.

\begin{figure}
\begin{center}
\includegraphics[width=0.38\textwidth]{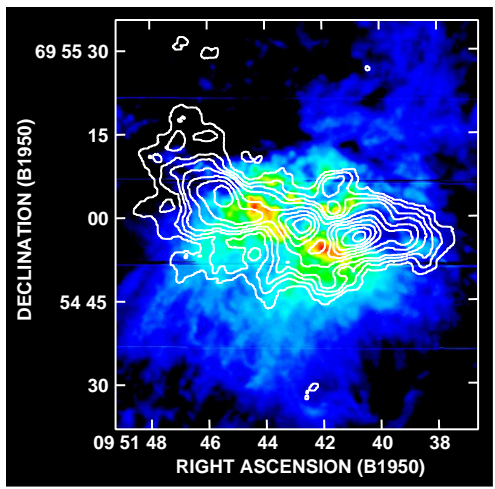}
\end{center}
\caption{The Nobeyama Millimeter Array (NMA) CO(1--0) image
(contours: \citealt{matsushita00,matsushita05}) overlaid on
the Subaru H$\alpha$ image (color map: \citealt{ohyama02})
for M82. At the distance of M82 \citep[3.9 Mpc;][]{sakai99}, $1''$ corresponds to 19 pc.
\label{fig:CO_Outflows}}
\end{figure}

The past observations of outflows have mainly focused
on plasma or ionized gas seen in the optical,
near-infrared, and X-ray. On the other hand,
observations of molecular-gas and dust outflows are
very few because of their intrinsic faintness and the limitation of the instruments.
Recent improvement of sensitive telescopes and receivers and wide-band backends
has started to make images of the molecular gas and dust outflows possible
(Figure \ref{fig:CO_Outflows}). Recent observations of molecular-gas outflows
have revealed that the outflowing mass is dominated by molecular gas, rather than the ionized gas or plasma \citep{tsai09,tsai12}. This implies that such molecular outflows
also contribute to the transport of dust to the IGM and thus to the
metal enrichment of the IGM.
Imaging of dust outflows is much less explored and is limited mostly to warm/hot dust
imaging \citep[e.g.,][]{engelbracht06,kaneda09} with cold dust imaging remaining
rare \citep*[e.g.,][]{alton99}. Moreover, multi-wavlength information, especially at the dust spectrum peak (around THz) is crucial to derive the mass and temperature of dust.

Recently, because of high sensitivity of \textit{Herschel},
dust outflows have been observed for some galaxies. In particular,
\citet{melendez15} observed a nearby edge-on starburst galaxy,
NGC 4631, by \textit{Herschel}. They found that the structures
seen in the dust outflow match extraplanar structures in other
outflow tracers such as H$\alpha$ and X-ray. However,
because of the low angular resolution of \textit{Herschel}
especially at $\geq 160~\micron$, where the dust temperature
of ``cold'' dust component can be correctly determined,
the filamentary structures of H$\alpha$ at the foot point of
the outflows were not fully traced by the dust.
For example, it is difficult to judge if the dust heating
is more associated with the shocks in dense molecular gas
(outflow) \citep{beirao15} or with the hot plasma.
On the other
hand, \textit{Herschel} was suitable for tracing the global
structure of the dust outflows. Therefore, a complementary
observation focusing on the small-scale foot point structures
of dust outflows would be desirable with a high-resolution
capability of ground-based THz telescopes.

As explained above, the largest advantage of
ground-based THz telescopes is high angular resolution.
Therefore, we aim at resolving the dust at the foot of outflow.
In particular,
we focus on relatively dense outflows, that is, molecular
outflows shown in Fig.\ \ref{fig:CO_Outflows}. 
By observing the dust emission in outflows, we will know
(1) the dust mass flowing out from galaxies due to the active star formation or
starburst activities, (2) the dust temperature and its gradient across the dust
outflows, and (3) the relation between dust, molecular gas, and ionized gas,
or the dust-to-gas ratio by comparing with the existing images of gas tracers.
Using the above information, we will obtain a comprehensive picture of the multi-component outflow.

To the above aim, we first perform THz mapping
observations of edge-on actively star-forming galaxies
that are already known to have outflows in other wavelengths. For the mapping, we target the interface between molecular outflows and
more extended outflows traced with H$\alpha$ or X-ray.
As we observe in Figure \ref{fig:CO_Outflows}, the interface has
structures well below 10 arcsec, which is only traced with the
GLT at THz frequencies, i.e., near the SED peak of dust emission.
We could also derive the variation of dust tempratures if we
observe the object in two THz windows (e.g., 1.0 and 1.5 THz).

Now we estimate the feasilibity of observing the interface
between molecular and ionized
outflows. \citet{alton99} showed the observed dust SED
in the outflow of M82 at submm wavelengths. We adopt their pessimistic 
extrapolation toward shorter wavelengths (i.e., their case with
a dust temperature of 13 K), 100 Jy at 200 $\mu$m
within a $80''\times 36''$ region. The flux density within the
$4''$ beam will therefore be about 0.5 Jy. With the sensitivity
adopted in Section \ref{subsec:capability}, an on-source integration time of
9 hr is required to detect this signal with 5$\sigma$.
However, as mentioned in Section \ref{subsubsec:BCD},
it may be possible to develop a bolometer-type instrument
with an order of magnitude better sensitivity. In this case,
we can detect the source with 6 min of on-source integration.
Moreover, we used the averaged dust brightness for the above
region; in reality, the interface associated with relatively
dense molecular outflows probably has an enhanced brightness
since the emission flux is concentrated into a small area.
Therefore, we may be able to detect the dust associated with
the interface within a reasonable time if the sensitive THz
bolometer is available.
Outflows are extended, so a multi-pixel detector is crucial.

We chose 6 nearby edge-on active star formation and starburst galaxies
listed in Table \ref{tab:Satoki_sample}.
All these galaxies are known to have H$\alpha$ and X-ray outflows
(M82, NGC 2146, and NGC 3628 are known to have molecular outflow).
The references are also listed in Table \ref{tab:Satoki_sample}.
The total 60 $\mu$m flux $f_{60}$ is also shown for each galaxy.
Note that the 60 $\mu$m flux of the galaxies other than M82
is 10--20 times smaller than that of M82; however, we target the
interface which may have similar physical conditions (i.e.,
similar brightness). Therefore, it is worth considering these
targets, although we can start observations from M82.

\begin{table*}
  \tbl{Edge-on sample for outflow studies.}{%
  \begin{tabular}{lcccccc}
  \hline
Object & RA(J2000) & Dec(J2000) & $d$ & $f_{60}\,^*$ & Ref.\\
 & [h m s] & [$\arcdeg$ $\arcmin$ $\arcsec$] & [Mpc] & [Jy] & \\
  \hline
M82  & 09 55 52.7 & +69 40 46 & 3.25 & 1314 & 1\\
NGC\,891 & 02 22 33.4 & +42 20 57 & 9.6 & 61.1 & 1      \\
NGC\,2146 & 06 18 37.7 & +78 21 25 & 17.2 & 131 & 2 \\
NGC\,3079 & 10 01 57.8 & +55 40 47 & 17.1 & 50.2 & 1       \\
NGC\,3628 & 11 20 17.0 & +13 35 23 & 10.0 & 51.6 & 1\\
NGC\,4631 & 12 42 08.0 & +32 32 29 & 7.5 & 82.9 & 1\\
  \hline
\end{tabular}}\label{tab:Satoki_sample}
\begin{tabnote}
$^*$\textit{IRAS} 60 $\mu$m flux as an indicator of the entire flux from the galaxy.\\
References --- 1) \citet{strickland04};
2) \citet{tsai09}.
\end{tabnote}
\end{table*}

\section{Sub-THz science cases}\label{sec:subTHz}

Since satisfactory weather for THz science is realized for
a fraction of the total time (Section \ref{subsec:atmosphere}),
it is worth considering ``sub-THz'' science cases, which are
executed in the non-THz weather conditions.
We will put a particular focus on the 350 $\mu$m
($\sim 850$ GHz) and
450 $\mu$m ($\sim$650 GHz) windows, for which the GLT still has an
advantage for the good atmospheric condition.

\subsection{Bolometer Surveys for High-Redshift Galaxies}\label{subsec:highz}

It has been known that the received total energy output of the entire
universe traced by the so-called extragalactic background light or  
cosmic background radiation is equally strong in the optical and 
in the FIR \citep[e.g.,][]{dole06}. This implies that half
of the galaxy and supermassive black hole growth processes are hidden
in dust, and that understanding their formation and evolution requires
the understanding of their FIR emission.  However, the FIR
part of the cosmic background has been under-explored, primarily because
of the limitation in observational techniques.

Our current understanding of the FIR universe comes from two types
of observations: ground-based bolometer-array camera imaging in
the mm and submm, and space FIR imaging.
The ground-based cameras can resolve approximately 1/3 of the
submm background into point sources \citep[e.g.,][]{coppin06}.
Such galaxies are found to be ultraluminous and primarily appear at
redshifts between $\sim$1.5 and $\sim$3.5 \citep[e.g.,][]{chapman05}.
The remaining 2/3 of the submm background is unresolved, because of the 
so-called confusion problem, that is, the low angular resolution of
ground-based submm/mm single-dish telescopes.
A small sample of them have been studied via strong gravitational 
lensing \citep[e.g.,][]{chen11}.
 {The high sensitivity of ALMA has recently resolved most of the mm background into point sources \citep{hatsukade13,fujimoto15}.
Nevertheless, most of the sources
contributing to the peak of the infrared background
at sub-THz frequencies ($\lesssim 450~\micron$) have yet to be identified. 
Moreover, as shown below, the galaxy population detected by sub-THz observations
is differently biased to the dust temperature and redshift from that traced by submm and mm
observations.}

Despite the success of ground-based bolometer-array cameras
in revealing 1/3 of the submm background population, there is
a major drawback. It is widely considered that submm observations only pick up
a special population whose dust temperature is relatively low.
This strong selection
bias is now confirmed by \emph{Herschel} \citep[e.g.,][]{magnelli12}.
\emph{Herschel} observations at 70--600 $\mu$m can
probe the peak of the infrared dust emission from redshifted galaxies,
and are much less biased by dust temperature (see Section \ref{subsubsec:confusion}).  Unfortunately,
the small telescope size of \emph{Herschel} produces an even worse angular resolution than
ground-based single-dish telescopes.  Only up to 10\%--50\% (depending
on the waveband) of the FIR background was directly resolved by 
\emph{Herschel} into individual galaxies \citep[e.g.,][]{berta10}.  The dominant
galaxy population that produces the FIR background yet remains
to be probed even after \emph{Herschel}.

\subsubsection{Beyond the Confusion Limit of \textit{Herschel}}
\label{subsubsec:confusion}

Ground-based THz telescopes can make significant
breakthrough in this field. Once equipped with
a bolometer-array camera in the sub-THz regime, 
the GLT will have an angular resolution much better than \emph{Herschel} and existing ground-based 
submm survey instruments, and thus will have much less confusion problems.
For this, the recently commissioned JCMT SCUBA-2
camera \citep{holland13}
can serve as an 
excellent pathfinder. \citet{chen13a,chen13b}
examined the 450 and 850 $\mu$m number counts obtained with
SCUBA-2 in a lensing cluster field. From the comparison with \emph{Herschel}
350 and 500 $\mu$m results \citep{oliver10}, the high angular resolution
of the JCMT helps to detect galaxies that are much less luminous (i.e., more normal) than
those detected by \emph{Herschel}. This is true even without the lensing effect.
On the other hand, this
450 $\mu$m imaging is still not confusion limited, because of the limited observing time and
the site quality.

In the 450 $\mu$m (650 GHz) window, the GLT will have a $9\farcs5$ resolution
(Section \ref{subsec:capability}), similar to that of SCUBA-2
at 450 $\mu$m. On the other 
hand, the GLT will be placed at a site that is much better than Mauna Kea for high-frequency
observations (Section \ref{subsec:atmosphere}). The SCUBA-2 450 $\mu$m observations by \citet{chen13a,chen13b}
are still noise limited and
far from confusion limited. With the GLT, the site and the substantial observing time
will allow confusion limited 350 $\mu$m or 450 $\mu$m imaging.  
With a 9$''$ resolution at 450 $\mu$m, the confusion limited source
density will be approximately 9000 deg$^{-2}$, assuming a conservative
definition of confusion limit of 30 beams per source. 
This is close to the limit probed by SCUBA-2 in lensing cluster fields for 
fully resolving the background.

\begin{figure*}
\begin{center}
\includegraphics[width=8cm]{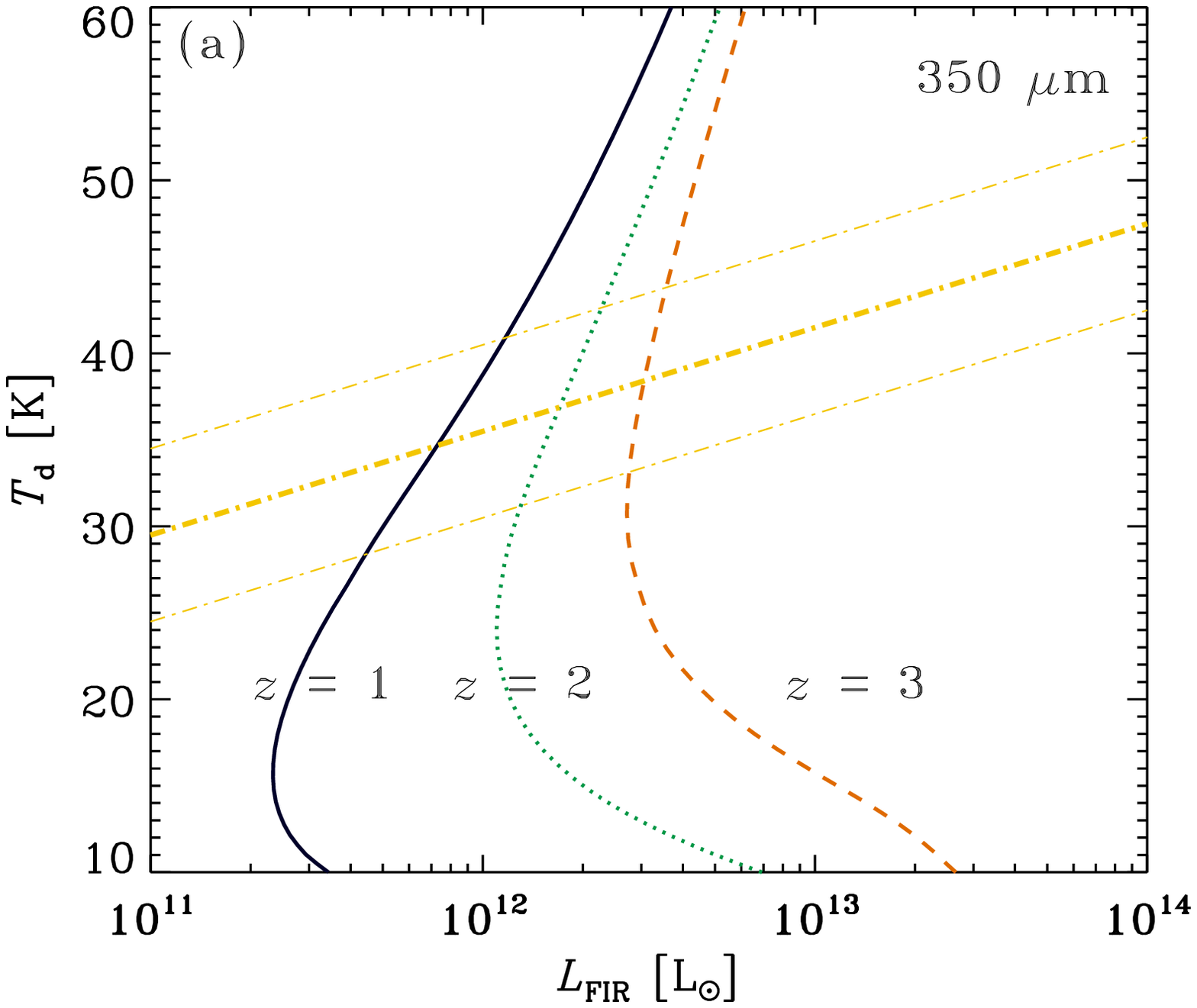}
\includegraphics[width=8cm]{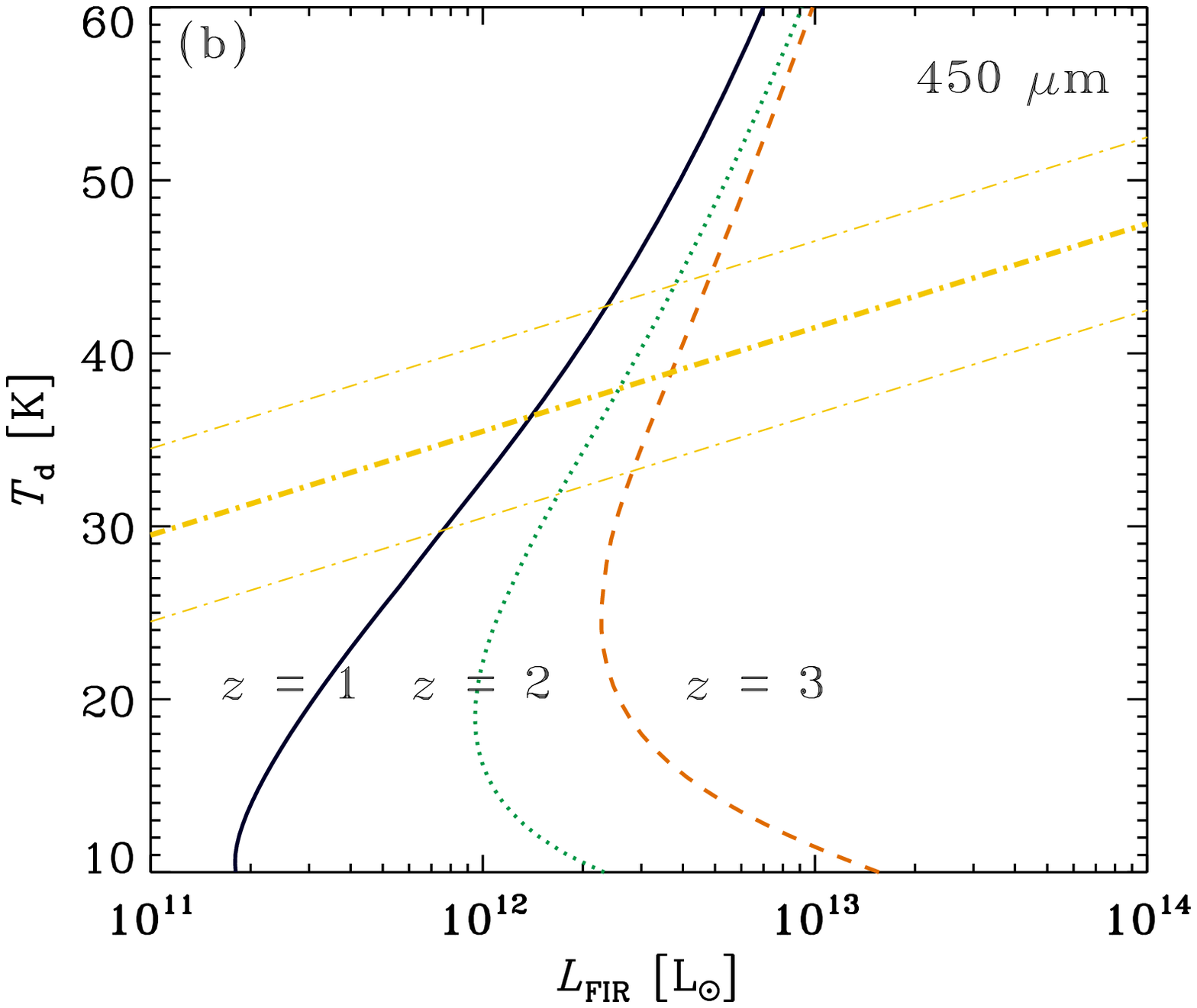}
\includegraphics[width=8cm]{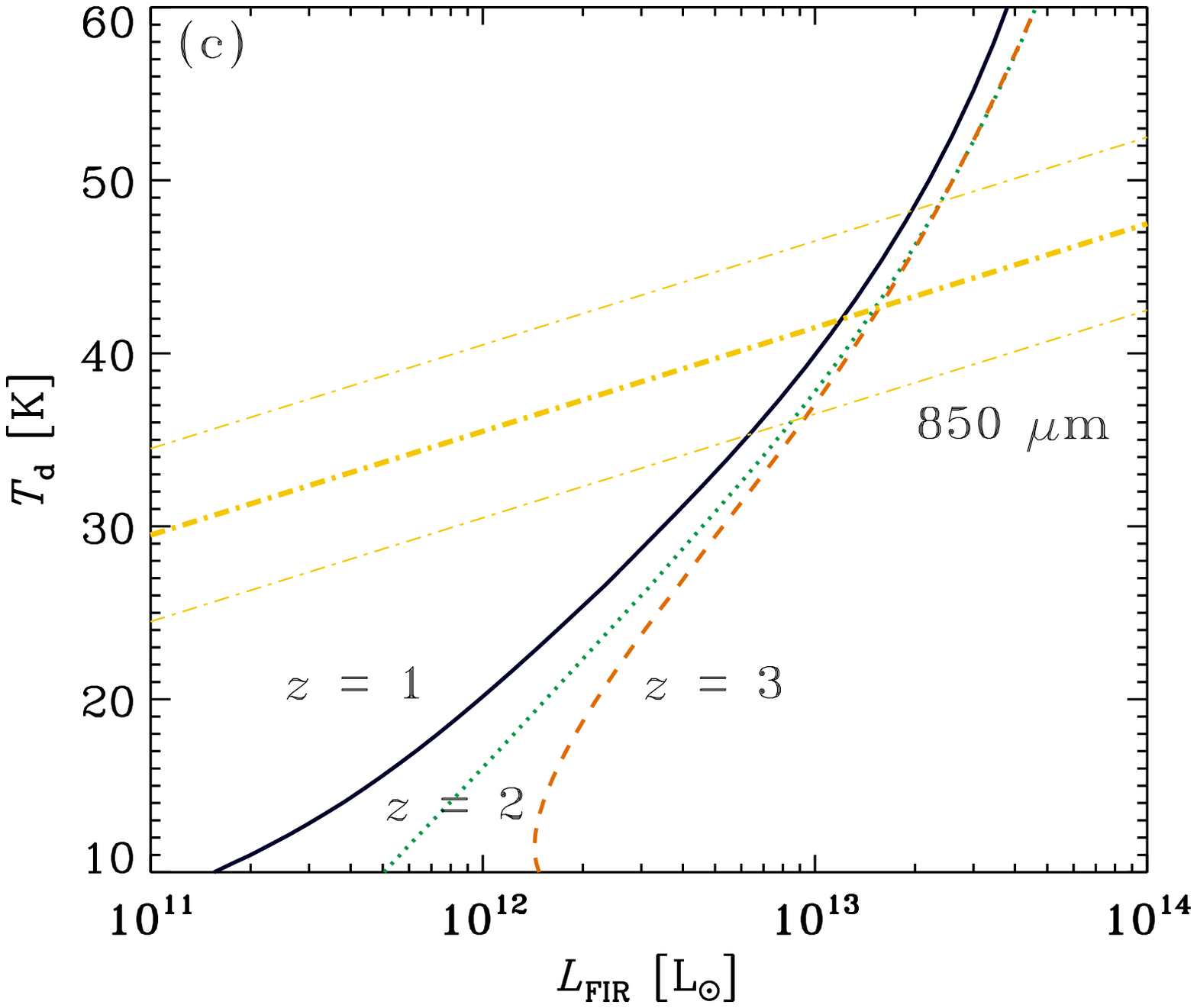}
\includegraphics[width=8cm]{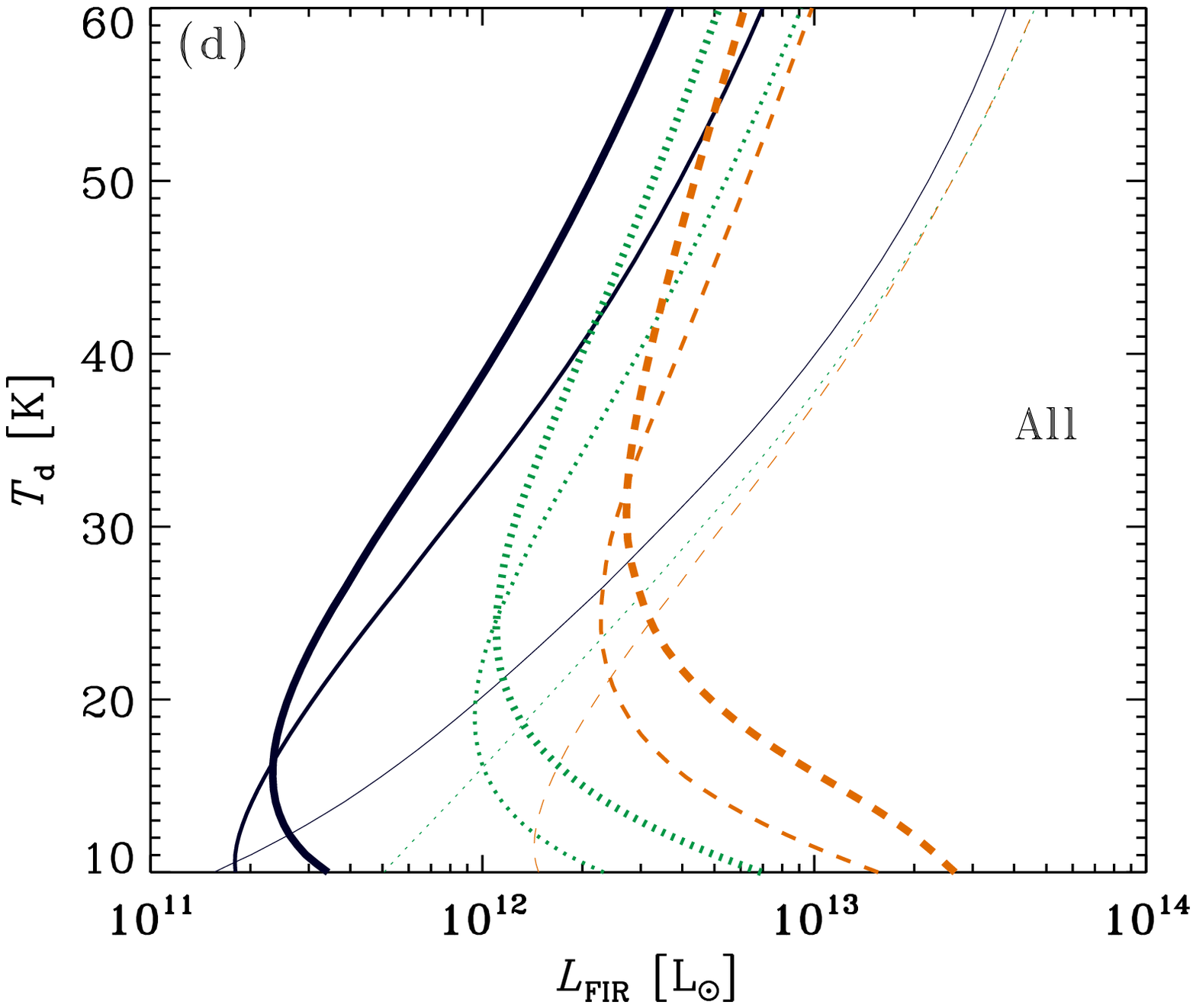}
\end{center}
\caption{Dust temperature ($T_\mathrm{d}$) vs.\
total FIR luminosity ($L_\mathrm{FIR}$) corresponding to
the detection limits assumed for each wavelength
(10 mJy, 4 mJy and 3 mJy
for 350, 450, and 850 $\mu$m, respectively).
Panels (a), (b), and (c) show the results for
350, 450, and 850 $\mu$m, respectively.
The solid, dotted, and dashed lines present the relations for
galaxies at $z=1$, 2, and 3, respectively.
Objects whose total FIR luminosity is larger than
(i.e., the right side of) each line are detected.
The thick dot-dashed lines show the $T_\mathrm{d}$--$L_\mathrm{FIR}$
relation derived for a \textit{Herschel} nearby sample \citep{symeonidis13},
for which we apply a linear fit (equation \ref{eq:Td_LIR}).
The thin dot-dashed line shows the typical range for the
dispersion of dust temperature $\pm$5 K.
All the lines
are overlaid in Panel (d) with the different thickness:
the thick, medium, and thin lines show the results for the 350,
450, and 850 $\micron$, respectively.}
\label{fig:flux_Td}
\end{figure*}

Furthermore, the 450 $\mu$m emission probes
galaxies at lower redshifts ($z\sim 1$--2) than
those producing the 850 $\mu$m background ($z \sim 2$-4) as demonstrated below.
This will allow us to study the majority of star forming galaxies at the peak of the cosmic star 
formation. As shown below, the bias to cold dust is
less at 450 $\mu$m than at 850 $\mu$m.
Thus, we can make significant 
progress in the understanding of galaxy evolution with sub-THz bolometer observations.

To clarify the population expected to be detected by
350 and 450 $\mu$m surveys, we show the relation
between the dust temperature, $T_\mathrm{d}$, and
the total IR luminoisity (total luminosity emitted
by dust), $L_\mathrm{FIR}$, in Figure \ref{fig:flux_Td}.
We adopt the detection limits of 10, 4, and 3 mJy for
350, 450, and 850 $\mu$m. For 350 and 450 $\mu$m,
we adopt the detection limits expected for the above
survey design for GLT (aiming at the confusion-limited
image; see also Section \ref{subsubsec:GLT_highz}),
while we adopt the 850 $\mu$m detection limit
based on \citet{chapman05} for
a representative 850 $\mu$m survey currently available.
We adopt the dust SED model
by \citet{totani02}, who give the dust SED
under given $T_\mathrm{d}$ and $L_\mathrm{FIR}$.
In Figure \ref{fig:flux_Td}, we plot the minimum
$L_\mathrm{FIR}$ detected for an object with a
dust temperature $T_\mathrm{d}$; that is, objects with the
FIR luminosity larger than the minimum (or
the right side of the curve) can be detected. From
Figure \ref{fig:flux_Td}, we observe that
850 $\mu$m surveys do not have
strong redshift dependence, which is due to the
so-called negative $K$-correction, and
strongly biased against high-$T_\mathrm{d}$ objects.
This kind of bias was already pointed out by
\citet{chapman05}. In contrast,
350 $\mu$m and 450 $\mu$m surveys are less
sensitive to $T_\mathrm{d}$, because they trace
the emission around the SED peak, which is primarily
determined by the total IR luminosity, rather than
the dust temperature. On the other hand, the
dependence on the redshift is strong at 350 and 450 $\mu$m,
because they are out of the wavelength range of
strong negative $K$-correction.

For a reference, we also plot in Figure \ref{fig:flux_Td}
the relation between
$T_\mathrm{d}$ and $L_\mathrm{FIR}$ derived for a
\textit{Herschel} sample at $z\lesssim 1$ by
\citet{symeonidis13}.
We applied a
linear fit to their relation and obtained
\begin{eqnarray}
T_\mathrm{d}=-36.5+6.0\log (L_\mathrm{FIR}/L_\odot ).
\label{eq:Td_LIR}
\end{eqnarray}
The typical dispersion of
the dust temprature $\sim$5 K is also shown
in the figure.
If we assume that this relation is valid for all
the redshifts of interest, the FIR luminosity at which
equation (\ref{eq:Td_LIR}) intersects with each curve
in Figure \ref{fig:flux_Td} is the
minimum detectable FIR luminosity. We observe that
the GLT survey at
350--450 $\mu$m can detect galaxies with
$L_\mathrm{FIR}\gtrsim 10^{12}$ L$_\odot$ at $z=1$--2.
Galaxies with lower dust temperatures are detectable
even if the total IR luminosity is a few $\times 10^{11}$ L$_\odot$.
We also observe that the 850 $\mu$m survey
only detects extreme galaxies with $L_\mathrm{FIR}\gtrsim 10^{13}~L_\odot$
or fainter galaxies with
much lower $T_\mathrm{d}$ than expected
from the \textit{Herschel} $T_\mathrm{d}$--$L_\mathrm{FIR}$ relation.

\begin{figure}
\begin{center}
\includegraphics[width=8cm]{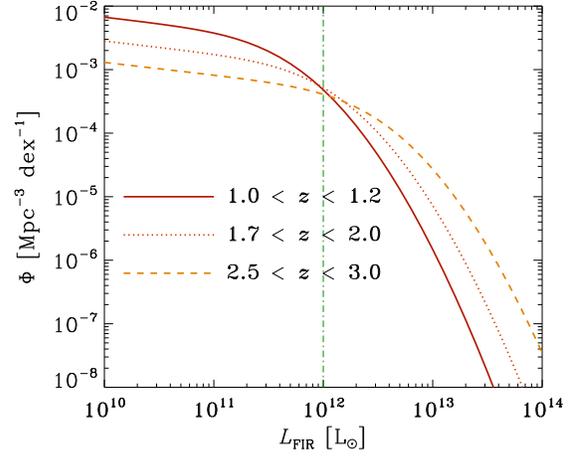}
\end{center}
\caption{Luminosity function of the total IR luminosity
for three redshift ranges, $1.0<z<1.2$, $1.7<z<2.0$, and
$2.5<z<3.0$ (solid, dotted, and dashed line, respectively),
taken from \citet{gruppioni13} (also shown in
\citealt{burgarella13}). The vertical
dot-dashed line marks a total IR luminosity of
$10^{12}$ L$_\odot$.}
\label{fig:FIR_LF}
\end{figure}

The FIR luminosities detectable at the sub-THz frequencies,
$L_\mathrm{FIR}\sim\mbox{a few}\times 10^{11}$--10$^{12}$ L$_\odot$,
are typical of galaxies at $z\sim 1$--2 in the sense that such galaxies are located at
the knees of luminosity function.
In Figure \ref{fig:FIR_LF}, we show the FIR luminosity
function derived by \citet{gruppioni13} based on
\textit{Herschel} data. We confirm that the depth of the
GLT survey at 350 and 450 $\mu$m is suitable to
detect the populations down to the knees of
luminosity function. Note that the data analysis by
\citet{gruppioni13} is based on the observational wavelengths
$\leq 160~\mu$m, so that the uncertainty in the FIR luminosity
is large. Our GLT survey will directly catch nearly the peak of
the dust SED.

\subsubsection{Capability of the GLT and a Possible Survey}
\label{subsubsec:GLT_highz}

Here we discuss the capability of the GLT by comparing it with
the JCMT/SCUBA-2. SCUBA-2 has 450 $\mu$m NEFD between
500 and 2000 mJy in a second \citep{holland13}.
The forecast of NEFD for the bolometer array to be
installed in the GLT is 110 mJy, roughly an 
order of magnitude better than SCUBA-2, making it $\sim100$ times faster than SCUBA-2 for detecting a point source.
To achieve a detection limit of 4 mJy with 3 $\sigma$ at 450 $\mu$m,
we require 1.9 hr for the integration time. We assume 3.8 hr
for the total on-source observational time per FOV.

The real limiting factor here will be the size of
the camera on the GLT. If we assume a camera of $20\times20$ pixels
and Nyquist sampling,
the FOV will be 1.65 arcmin$^2$. With 1000 hr of observational time,
we can survey 430 arcmin$^2$, which is comparable to the total
area of the famous Great Observatories Origins Deep Survey
(GOODS; \citealt{giavalisco04}). It will be
unprecedented to cover such an area at the proposed depth at 450 $\mu$m, detecting essentially
\emph{all} important dusty star forming galaxies at $z>0.5$ and nearly \emph{fully resolving} the 
background at this waveband.

The most ideal situation is to have a larger camera.
If we can increase the number of pixels by a factor of $\sim$10, then
we will be capable of covering a significant fraction of a FOV of
the Subaru Hyper-Suprime Cam (HSC) \citep{miyazaki12}, whose deep imaging will be useful to find the
optical counterpart of the GLT-detected sources.

\section{Time Variable Sources}\label{sec:monitoring}

We additionally propose science cases for time-variable sources
given that some future THz/submm telescopes such as the GLT will allow
flexible scheduling of observational time.
Long-term monitoring observations are difficult to implement in
space THz/submm facilities, which have a limited lifetime,
or existing submm interferometers, whose time allocation needs
to answer world-wide requests and proposals. In particular,
considering monitoring observations at submm (or mm) wavelengths
is useful for the purpose of making an efficient use of
the time when the
weather condition is not suitable for THz observations.
In what follows, we
explain the importance and uniqueness of this mode of
observations, focusing on AGNs.

In another paper, \citet{urata15} discuss a possibility and importance of
observing GRB afterglows at submm wavelengths. Since
such a ToO type of observational mode
is also worth considering, we also give a summary of their paper
in Section \ref{subsec:grb}.

\subsection{Monitoring of AGNs}\label{subsec:AGN}

\subsubsection{Exploring VHE $\gamma$-ray flares in Blazars}

Constraining the location and emission properties of the very high
energy (VHE; $>100$ GeV) $\gamma$-ray emission in AGNs is one of the most
important goals of the {\em Fermi} mission after 2008 because major
extragalactic sources [1017 out of 1319 in the 2nd {\em Fermi}-Large Area
Telescope (LAT) catalog; $|b|>10^{\circ}$] detected in $\gamma$-rays
are AGNs \citep{ackermann11a}. Among the VHE sources in AGNs, the so-called
blazars are identified as the dominant population (97\%), including both
flat-spectrum radio quasars (FSRQs) and BL Lacertae objects (BL Lacs);
they are recognized as the brightest AGN jets pointing very close to our
line of sight \citep{urry95}.

The $\gamma$-ray emission in blazars at GeV/TeV bands is understood
through inverse Compton (IC) scattering of ambient photons. Although
various theoretical models have been proposed, the precise location
of the $\gamma$-ray emission is poorly
constrained. {\em Or} perhaps several scenarios at different locations
in jets could be considered like in the case of M87
\citep{abramowski12}. The IC process takes place either with synchrotron photons from
inside the jet via the synchrotron
self-Compton (SSC) scattering
\citep[e.g.,][]{konigl81,marscher85}, or with external photons (external Compton
scattering or EC), where the seed photons are generated in the accretion
disk \citep[e.g.,][]{dermer92}, the circumnuclear material such
as the broad-line region (BLR) of quasars \citep[e.g.,][]{sikora94,blandford95},
or the dust torus \citep[e.g.,][]{blazejowski00,bottcher07}.


A typical broad-band shape of the SED in blazars exhibits two broad bumps 
\citep[e.g.,][]{fossati98,ghisellini98}. The peak of the lower-energy bump 
between $10^{13}$ and $10^{17}$ Hz is due to the synchrotron emission; 
moving relativistic shocks in the jet (``shock-in-jet'' model) are generally 
considered for non-thermal particle acceleration \citep{blandford79,marscher80}.
On the other hand, a high energy bump between $10^{21}$
and $10^{24}$ Hz is due to 
SSC and/or EC as mentioned above \citep[]{abdo11a},
although which of EC and SSC is the dominant mechanism of the high-energy component
is not yet conclusive \citep{ghisellini02,sikora09,zacharias12}.

In the low-energy component of SED, there is a gap in the frequency 
coverage between cm/mm and infrared wavelengths. Therefore, single dish 
observations with the GLT has a capability to fill this gap, contributing 
to completing the SED. In addition, measurements at these frequencies are important, 
since the SED peak caused by synchrotron self-absorption is typically located 
just around these frequencies. Identifying the peak leads to better estimations of 
the magnetic field strength.

\subsubsection{$\gamma$-ray flares at the ``far distant'' jet downstream}
\label{subsubsec:flare}
In fact, the VLBI observation including the GLT will play a supplemental role 
in conducting a unique science program through the GLT single dish monitoring 
observations toward $\gamma$-ray bright blazars.
The goal is to identify the production site of high energy emissions 
such as X-ray and $\gamma$-ray in blazars, which are referred to as 
the ``blazer zone'' and has been believed to be at sub-parsec distances from the nucleus. 
However, where this energetic site is located is fairly unclear, because 
a typical size of the VLBI core at mm/cm wavelengths is $\lesssim$ 0.1 mas, 
corresponding to the {\em de-projected} distances $\sim 10$ pc for FSRQs 
($\langle z \rangle \sim 1.11$) and $\sim 6$ pc for BL Lacs 
($\langle z \rangle \sim 0.37$) at average redshifts \citep[]{dermer07} 
with a viewing angle of 5$^{\circ}$ for our reference. 
Thus, if blazar zones are located inside the VLBI core ($\lesssim 10^{5}\, r_{\rm s}$ for $M_{\bullet}=10^{9} M_{\odot}$), it is normally difficult to identify the location of the shock-in-jet to energize particles for synchrotron radiations. Blazer zones, 
which may be separated from their jet nucleus (possibly far distant downstream 
$\gg 0.1$--1 pc), have been identified for limited objects in radio galaxies \citep[e.g.,][]{harris06,nagai09} and blazars 
\citep[e.g.,][]{agudo11a, agudo11b} by direct imaging in mm/cm VLBI observations.

In VLBI observations, we will aim at measuring the sizes and positions of the genuine blazar zone. 
In order to achieve this, we need to conduct observations with enough angular 
resolution to resolve 0.01 pc at the blazar distance, and {\em it can only be achieved 
with submm VLBI observations}. Combination of GLT single dish and submm VLBI observations 
will enable us to achieve that goal very effectively. Here we propose dense monitoring with 
the single dish observation mode of the GLT to detect a flare event which may indicate activity
associated with the blasar zone. If we can detect significant time
variation with single dish observations, we can trigger follow-up
submm VLBI observations in order to find a new component ejected from 
the submm VLBI core. Therefore, combined observations between
single dish monitoring and sub-mm VLBI imaging play a 
crucial role in testing the shock-in-jet model.

It was widely believed that the IC process in blazars is dominated
within a scale of BLR ($\lesssim 0.1$--1pc) by EC and/or SSC. There is,
however, a recent growing recognition that the VHE $\gamma$-ray emitting
region may be located at farther distance from the nucleus well beyond
the BLR, i.e., over parsecs or even tens of parsecs \citep[e.g.,][]{leon11}. 
The VLBA monitoring at 22 and 43 GHz of a sample of Energetic Gamma Ray Experiment Telescope (EGRET) blazars establishes a
statistical association of $\gamma$-ray flares with
ejections of superluminal components in the parsec-scale
regions of the relativistic jet nearby the VLBI radio core \citep{jorstad01}.
This remarkable finding has been also supported by the
single-dish Mets\"{a}hovi monitoring of total flux density variation at
22 and 37 GHz towards EGRET blazars \citep{valtaoja95} with
the following conclusion: the highest levels of $\gamma$-ray fluxes are
observed during the initial (or peak) stage of radio/mm-wave synchrotron
flares.

Among shock-in-jet models, two major possibilities have been
discussed; a collision of the faster shock with either the preceding
slowly moving shock (``internal shock'' model by \citealt{spada01})
or the standing shock complex \citep*{daly88,sokolov04}.
One of the important clues for positioning VHE $\gamma$-ray
flares is the emission size $r_{\gamma}$, which can be roughly constrained by the
time variability $t_{\rm var}$ (of the order of day) of flares as
\begin{equation}
r_{\gamma} \lesssim c t_{\rm var} \delta \sim 0.01 
\left(\frac{t_{\rm var}}{\rm
 day}\right)\left(\frac{\delta}{10}\right) \ {\rm pc},
\end{equation}
where $c$ is the light speed, for
a typical case in blazars with a Doppler beaming factor of $\delta \sim
\Gamma \sim 10$ ($\theta \sim 1/\Gamma$, where $\Gamma$ is the bulk
Lorentz factor $\sim 10$ and $\theta$ is the jet opening angle $\sim$ 0.1 radian).
Considering a conical geometry with $\theta\sim 6^{\circ}$, the
position of the flare can be placed as $z_{\gamma} \sim r_\gamma /\theta\sim 0.1$ pc (along
the jet axis $z$), which
is comparable to the location of BLR. Therefore, if we consider
a case where $z_{\gamma} \gg 0.1$ pc, then the internal shock model in a conical jet
may {\em not} be appropriated. Instead, non-conical (e.g.,
parabolic) MHD jet and/or standing shock model as a consequence of the
strong over-collimation (jet streamlines are focusing toward the central
axis) may be an alternative possibility for a very small region of $r_{\gamma} \lesssim
0.01$ pc at large distance $z_{\gamma} \gg 0.1$ pc.

\subsubsection{Lessons learned from the active radio galaxy M87}

\begin{figure}
\begin{center}
\includegraphics[width=0.45\textwidth,clip]{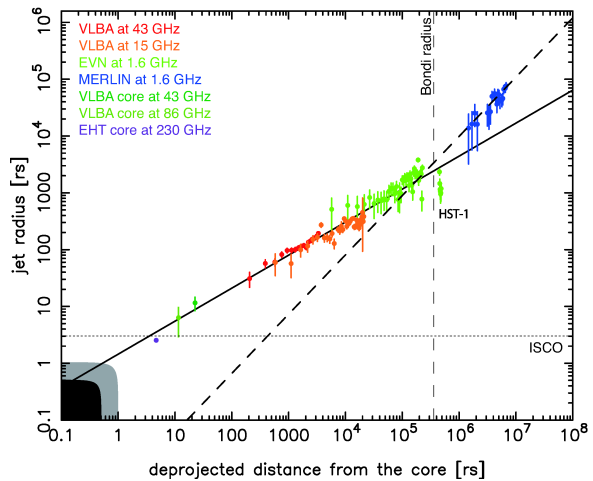}
\end{center}
\caption{
Distribution of the radius of the jet ($r$) as a function of the de-projected
distance from the core ($z$) in units of $r_{\rm s}$ \citep[]{asada12, nakamura13}.
The solid line is the linear least-square fit for the data points
except the three inner points of VLBI cores (at 43, 86, and 230 GHz), indicating the parabolic streamline $z \propto r^{a}$ 
with $a=1.73 \pm 0.05$. On the other hand, the dashed line indicates 
the conical streamline with $a$ of $0.96 \pm 0.1$. The so-called HST-1 complex, 
a bright feature with sub/super-luminal motions, is located around 
$5 \times 10^{5}\,r_{\rm s}$. The vertical (dashed) line denotes the Bondi 
accretion radius $R_{\rm B}$, located at $\simeq 3.8 \times 10^{5}\, r_{\rm s}$.
The horizontal (dotted) line indicates the inner stable circular orbit (ISCO) of the 
accretion disk for the Schwarzschild black hole.}
\label{fig:M87}
\end{figure}

{\em Why are we interested in the scenario of the far distant jet downstream 
for $\gamma$-ray flares in AGNs?} We have one solid example for VHE TeV $\gamma$-ray
event in the nearby active galaxy M87. The flare occurred at a distance 
$z_{\gamma} \gg 0.1$ pc, but it could be common for AGNs.
In what follows, let us introduce our model based on M87 and argue 
the $\gamma$-ray emission size and its corresponding time variability of the flare 
in a non-conical jet structure. 
A supermassive black hole (SMBH) mass of $M_{\bullet} \sim 6.6 \times 10^{9} M_{\odot}$ in M87 
\citep{gebhardt11}, owing to its proximity
($d=16.7$ Mpc, where 1$\arcsec$ corresponds to
$\simeq$80 pc; \citealt{blakeslee09}), gives an apparent angular size of $\sim 8\, \mu$as 
for the Schwarzschild radius $r_{\rm s} \equiv 2 GM_{\bullet}/c^{2}$,
where $G$ is the gravitational constant,
providing a unique opportunity to study the relativistic outflow with the highest 
angular resolution in units of $r_{\rm s}$ with a moderately large viewing angle 
of $\theta_{\rm v} \sim 14^{\circ}$ \citep[]{wang09}.
Recent extensive studies of the jet toward M87 have revealed
a parabolic stream at $\sim 10\, 
r_{\rm s}$--$10^{5}\, r_{\rm s}$, while it changes into a conical
stream beyond $\sim 2 \times 10^{6}\, r_{\rm s}$ \citep{asada12,nakamura13}, 
as shown in Fig.\ \ref{fig:M87}.

Correspondingly, proper motion analysis suggests that the parabolic part of 
the jet exhibits bulk acceleration while the conical part shows
deceleration. This supports the MHD jet paradigm in M87 
\citep{nakamura13, asada14}. This remarkable finding of the ``jet break''
(the jet structural transition from a parabolic to conical stream as well as 
from the acceleration to deceleration of the bulk flow) in M87 indicates 
a fundamental process of AGN jets interacting with the stratified ISM.
The jet break in M87 is presumably associated with the jet over-collimation and 
thus the formation of a narrower re-confinement nozzle is expected \citep[e.g.,][]{stawarz06}. 
Figure \ref{fig:M87} also shows that a structured complex known as ``HST-1'' 
\citep[]{biretta99} is located just downstream from the Bondi radius of 
$R_\mathrm{B}\simeq 3.8 \times 10^{5} \ r_{\rm s}$ ($\sim 250$ pc in de-projection) 
\citep[cf.][original value of $M_{\bullet}$ is replaced with 
$\sim 6.6 \times 10^{9} M_{\odot}$]{allen06, cheung07}.

HST-1 consists of bright knots, whose apparent motions are superluminal with 
a range of $4c$--$6c$, accompanied by subluminal motions 
\citep[e.g.,][]{biretta99, cheung07, giroletti12}. The furthest upstream 
component ``HST-1d'' in the complex at around $4.4\times10^{5}\,r_{\rm s}$ 
($\sim$ 290 pc in de-projection) seems to be stationary, and has been identified 
as the origin of sub/superluminal motions as well as the possible site for the TeV 
$\gamma$-ray flare event in the year 2005 \citep[]{cheung07}.
Indeed, the simultaneous rise and fall of light curves at all wavelengths (radio, optical, 
NUV, and X-ray bands) during 2005 \citep[]{aharonian06, harris06, cheung07, madrid09, abramowski12}
indicate that the flare was a local event caused by a simple compression at HST-1 
\citep[]{harris06, harris09}, which created an increase of the synchrotron energy 
at all wavelengths equally.
The component HST-1c, which had been ejected during 
2004--2005 from HST-1d, was presumably associated with the HST-1 flare occurring in 2005 
as suggested by \citet[]{cheung07}. Recently, \citet[]{liu13} proposed a model for the HST-1 flare in 2005:
the brightening/dimming fluxes on multi-wavelength light curves (from radio to X-ray) could be 
explained as an adiabatic compression/expansion under the conservation of magnetic fluxes 
when the blob passes through a hourglass or de Laval-nozzle-shaped sheath layer ({\rm i.e.}, 
the re-confinement magnetic nozzle). In their model, the maximum of the TeV flare in M87 was 
coincident with the peak of the light curves from radio to X-ray in HST-1 during 2005 
\citep[]{cheung07, abramowski12}.

As introduced above in Section \ref{subsubsec:flare}, one of
the most important  clues  for  positioning 
VHE  flares  is  the  emission  size $r_{\gamma} $, which can be  roughly constrained 
by the timescale of $\gamma$-ray flux variability ($t_{\rm  var} \sim 2$ days) in M87
around 2005 \citep[]{aharonian06}. This gives 
$r_{\gamma} \lesssim c t_{\rm var} \delta \sim 0.01\, {\rm pc} \sim  20\,r_{s}$.\footnote{
We here re-estimate $r_{\gamma}$ with the newly derived black hole mass of 
$M_{\bullet} \sim 6.6  \times 10^{9}  M_{\odot}$ \citep[]{gebhardt11} 
and the Doppler factor at HST-1 $\delta \simeq 4$ \citep[]{perlman11} 
with $\theta_{\rm v} \sim 14^{\circ}$ \citep[]{wang09}.} Based  on the  assumption 
that the  M87  jet geometry is conical, \citet[]{aharonian06} estimate 
an unrealistically small opening  angle of the jet   
$\theta \sim 1.5 \times 10^{-3} \delta$ degree at $0\farcs85$ ($\simeq$ 68 pc in projection) along the jet axis
for explaining the size of this VHE flare $r_{\gamma} \lesssim  5 \delta\,r_{s}$ with disfavoring HST-1 as a possible site of TeV $\gamma$-ray production (i.e., 
the nucleus of M87). The M87 jet is, however, not conical, but parabolic toward the upstream from HST-1
\citep[]{asada12, nakamura13}. Furthermore, a simultaneous flaring up from radio through 
optical/X-ray to $\gamma$-ray at HST-1 \citep[]{cheung07, harris09} strongly suggests 
that HST-1 could be the most promising site for the TeV $\gamma$-ray emission, 
while the nucleus exhibited quiescent behaviors at multi-wavelengths 
\citep[]{harris09, madrid09, abramowski12}. 

Narrowly focused (re-confinement) nozzle in supersonic jets could be produced 
owing to a lateral imbalance between the jet internal and external ISM pressures 
\citep[]{courant48}; the system consists of a stationary 
``oblique shocks + a Mach disk'', i.e., a standing shock complex.
Physics of the standing shock complex has been been discussed 
in the extragalactic jet community, starting from the early 1980s 
\citep[e.g.,][]{sanders83}. If we apply this stationary shock system to the HST-1 complex, 
recurrent events of ejected subluminal blobs from the furthest upstream 
of the HST-1 complex \citep[]{biretta99, cheung07, giroletti12} may 
be due to a lateral compression/expansion of the standing shock complex.

In Figure \ref{fig:M87}, the maximum radius $r_{\rm max}$ of the parabolic jet 
can be referred in the vicinity of $R_{\rm B}$. On the other hand, the radial size of 
the HST-1 complex is about 1/3 of $r_{\rm max}$ although the minimum radius $r_{\rm min}$ 
of the jet re-confinement nozzle could be much narrower at the maximal focal point 
(around the Mach disk). Note that the HST-1 complex, which consists of the ejected 
sub/superluminal blobs, is located in the downstream of the standing feature HST-1d. 
A magnetic focusing process of MHD jets in a theoretical argument makes a ratio 
$r_{\rm max}/r_{\rm min}$ by a factor of 
$\approx 10^{1-2}$ \citep[][]{achterberg83}, while it is restricted to a factor of 
$\sim  2$ in adiabatic hydrodynamic  jets \citep[]{sanders83}. 
Giving   an   upper   limit   of  $r_{\rm   max}   \simeq   3000\,r_{s}$
\citep[]{asada12} (see also Figure \ref{fig:M87}), a  range of $r_{\rm min} 
\simeq  (30-300)\,r_{s}$ would be speculated. In the  general framework of MHD jets,  
$r_{\rm min}$ cannot be smaller than the Alfv\'en radius \citep[e.g.,][]{pelletier92}. 
Correspondingly, the parabolic jet  in M87 exhibits a trans-Aflv\'enic  feature at 
around $z \sim$  a  few  of  $(100-1000)\,r_{s}$,  where the  jet  radius  is  about
$(30-300)\,r_{s}$ \citep[]{nakamura13}. Thus, we also suggest that the HST-1 complex
could be the potential site \citep[see also][]{stawarz06, cheung07}
of the $\gamma$-ray emission even 
though a short  time  variability  $t_{\rm  var}  \sim  2$  day  is  taken  into
consideration.

\subsubsection{Application of the HST-1 scenario in M87 to blazars}

Our recent efforts toward M87 may give a promising clue to the
MHD jet paradigm \citep[]{nakamura10, asada12, nakamura13, nakamura14, 
asada14}, while the role of magnetic fields in blazar jets is not well 
understood. However recent observations have shown some indications of how magnetic fields in blazars are fundamental not only
in the jet formation process 
\cite[]{zamaninasab14}, but also in their emission regions on parsec scales \cite[]{ghisellini14}.
Let us consider that AGN jets (both radio galaxies and blazars) evolve 
by the MHD process, as is generally believed in theoretical fields. 
Parabolic streamlines, as a consequence of the lateral balance between 
the jet internal (magnetic) and external (thermal) pressures
\citep[e.g.,][]{zakamska08, komissarov09, lyubarsky09}, 
are suitable for the jet bulk acceleration to the relativistic regime with
Lorentz factor $\simeq$ a few tens 
\citep[e.g.,][]{vlahakis03, beskin06, beskin09, komissarov07, tchekhovskoy08}.
Taking a SMBH mass of $M_{\bullet}=10^{9} M_{\odot}$ as our reference, the Bondi
accretion radius can be scaled as 
\begin{equation}
R_{\rm B} \simeq \ 30 \ {\rm pc} \left(\frac{kT}{{\rm keV}}\right)^{-1}
\left(\frac{M_{\bullet}}{10^{9} M_{\odot}}\right), 
\end{equation}
corresponding to $ 3.2 \times 10^{5} r_{\rm s}$,
where $kT\sim 1~{\rm keV}$ is adopted as a typical ISM value in X-ray observations.

Assuming the jet structure to be a genuine parabolic shape \citep[e.g.,][]{blandford77}, 
which originates at the innermost stable circular orbit (ISCO) of the Schwarzshild 
black hole $r_{\rm ISCO}=3 \, r_{\rm s}$, with $z \propto (r/r_{\rm ISCO})^{2}$, 
an expanding jet has the maximum radius $r_{\rm max} \simeq 0.16\, {\rm pc} \simeq 1.7 \times 10^{3}\, r_{\rm s}$ at $z_{\rm max} \simeq R_{\rm B}$. Again, in a magnetically focusing jet model 
\citep{achterberg83}, the jet radius is reduced by a factor $f_{\rm cmp}=r_{\rm min}/r_{\rm max}
\approx 0.01 - 0.1$; thus, we could infer $r_{\rm min} \approx 2 \times (10^{-3} - 10^{-2})$ pc, 
corresponding to $t_{\rm var} \approx (0.2 - 2)$ day. 
Similarly, if we consider that a lateral expansion is faster than the genuine
parabolic case (e.g., M87), say, $z \propto (r/r_{\rm ISCO})^{1.6}$ \citep[e.g.,][]{blandford82}, 
$r_{\rm min} \approx 8 \times (10^{-3} - 10^{-2})$ pc, corresponding to 
$t_{\rm var} \approx (0.8 - 8)$ day, is derived.

Based on the shock-in-jet model (a standing shock complex), which may be triggered at around 
the Bondi radius $R_{\rm B} \gtrsim 10^{5} \, r_{\rm s}$, we estimate the VHE emission size 
for AGN jets with $M_{\bullet}=10^{9} M_{\odot}$ as 
\begin{equation}
r_{\gamma} \approx (10^{-3} - 10^{-1}) \ {\rm pc},
\end{equation}
with the possible time variability range
\begin{equation}
t_{\rm var} \approx (0.1 - 10) \ {\rm day}.
\end{equation}
Therefore, it may be feasible to consider the distant jet downstream scenario toward parabolic 
jets in AGNs. It is worth to mention that recent analysis on the minimum variability 
timescales of {\em Fermi} blazars in the GeV band reveal a range $\approx$ (0.01 - 60) 
day for $M_{\bullet} \simeq (10^{7}$--$10^{10}) M_{\odot}$ \citep[]{vovk13}; none of 
the blazars exhibits variability on a timescale shorter than the black hole horizon 
light-crossing time and/or the period of rotation around the ISCO.

{\em Why do we need a flux monitoring at submm/mm wavelengths?}
FSRQ 4C +38.41 (at $z=1.807$)
has been very active in GeV bands during last five years.
Flux variability in this source from GHz frequencies to GeV energies is examined 
by \citet[]{raiteri12}; there were at least three high states of the GeV emission 
and multi-flaring peaks are associated with each event. The $\gamma$-ray and optical 
variability on a one-week timescale were tightly correlated, where the sampling is dense enough.
On the other hand, there was also some intrinsic, but weak correlation in light curves 
between the $\gamma$-ray/optical and mm fluxes (taken by the SMA 1.3 mm band). 
We suspect that the sampling rate at submm/mm wavelengths is still poor ($>$ a week). 
In the shock-in-jet scenario, it is generally expected that the flare peak in the submm/mm 
regime is followed with some time delay by the peak at longer wavelengths (i.e., mm to cm), 
as a consequence of the jet opacity, as demonstrated in radio light curves 
\citep[e.g., PKS 1510--089 (FSRQ):][]{orienti13}.
In conclusion, a dense monitoring mm/sub-mm fluxes toward $\gamma$-ray bright blazars with 
spanning less than a week is the best strategy in order to test the shock-in-jet
model and to explore the site and emission mechanism of the VHE $\gamma$-ray flares in AGNs.


\subsubsection{Sample Selection}

We started with the original sample of VLBI-detected Fermi sources
\citep{lister11}.
From this sample. we selected 17 sources in
Table \ref{can-sour} with the following criteria:

\begin{enumerate}

\item 

Submm VLBI with the GLT at 345 GHz will have an enough linear resolution to
resolve a blazar zone of 0.01 pc.
Practically, this condition limits the angular diameter distance
to 10$^{3}$ Mpc ($z<0.37$).   

\item

 Submm VLBI with GLT at 345 GHz will have an enough sensitivity to detect the source without phase-up ALMA.
Assuming a flat spectrum of blazar, it requires that the flux $>$ 0.3 Jy. 

\item
 The object is accessible from the GLT site
 with zenith angle $<60^\circ$. It gives the source declination
larger than 12$^{\circ}$.

\end{enumerate}

Measurement of the absolute flux density is expected to be limited by the calibration accuracy, 
which would be typically 10\%. In order to robustly achieve this accuracy, it is desired to achieve 
thermal noise level three-times better than the calibration error. With this condition, around 4\% degradation is expected to contribute to the final accuracy due to the thermal noise. 
As a result, we need only 30 min for the total on-source time at 345 GHz for all the sources in Table \ref{can-sour}. With taking into account the overhead, 
it will take one hour for each monitoring. Although the observable sky from 
Greenland is limited, the same object can be monitored constantly over the winter season. 
Since time variability on a $\sim$ day--week timescale is expected, we propose to conduct this monitoring 
every 3--4 days.

\begin{table*}
\tbl{Candidate AGNs for monitoring observations with the GLT.}{%
\begin{tabular}{llccc}
\hline
Name & Alias & $z$  & Optical ID & Flux at 15 GHz\\
 &  &  &  &   [Jy] \\
\hline
J0112+2244 & S2 0109+22     & 0.265 & BL Lac  & 0.48 \\
J0319+4130 & 3C 84          & 0.0176 & Galaxy  & 19.4 \\
J0721+7120 & S5 0716+71     & 0.31 & BL Lac  & 1.2 \\
J0748+2400 & PKS 0745+241   & 0.4092 & QSO  & 1.15 \\
J0854+2006 & OJ 287         & 0.306 & BL Lac & 4.67 \\
J0958+6533 & S4 0954+65     & 0.367 & BL Lac  & 1.34 \\
J1104+3812 & Mrk 421        & 0.0308 & BL Lac  & 0.33 \\
J1217+3007 & ON 325         & 0.13 & BL Lac & 0.36 \\
J1230+1223 & M87            & 0.0044 & Galaxy  & 2.51 \\
J1653+3945 & Mrk 501        & 0.0337 & BL Lac  & 0.87 \\
J1719+1745 & OT 129         & 0.137 & BL Lac  & 0.58 \\
J1806+6949 & 3C 371         & 0.051 & BL Lac  & 1.37 \\
J1927+7358 & 4C +73.18      & 0.302 & QSO  & 3.71 \\
J2022+6136 & OW 637         & 0.227 & Galaxy  & 2.26 \\
J2143+1743 & OX 169         & 0.2107 & QSO  & 1.09 \\
J2202+4216 & BL Lac         & 0.0686 & BL Lac  & 4.52 \\
J2203+3145 & 4C +31.63      & 0.2947 & QSO  & 2.6 \\
\hline
\end{tabular}}\label{can-sour}
\end{table*}

\subsection{GRBs with the GLT}\label{subsec:grb}

GRBs are among the most powerful explosions in the
Universe and are observationally characterized according to intense
short flashes mainly occurring in the high-energy band (prompt emission) and
long-lived afterglows seen from the X-ray to radio bands.
The discovery of afterglows was a watershed event in demonstrating
the following properties of GRBs:
a cosmological origin for long-soft and short-hard GRBs
\citep{metzger97,berger,fox}; collimated narrow jets with energy
$\sim10^{51}$ erg for long-soft GRBs \citep[e.g.,][]{frail};
association with star forming galaxies \citep[e.g.,][]{djorgovski};
and a massive star origin for long-soft GRBs
\citep{hjorth,stanek03}. Moreover, GRBs are now being exploited as
probes of the high redshift Universe, even at $z\gtrsim 8$
\citep{z8}. Because of their extremely high luminosity, the highest-$z$
events at the reionization epoch ($z\sim8$) have already been
observed, and their discovery at $z > 10$ is highly possible
\citep[e.g.,][]{highzrate}.

The standard fireball model \citep[e.g.,][]{fireball} predicts a
double-shock system: a forward shock (FS) 
propagating in the external medium, and a reverse shock (RS)
propagating into the ejecta.
The RS emission uniquely probes the dominant composition (i.e., whether
the composition is dominated by baryons or magnetic fields) and
properties (such as thickness and Lorentz factor) of the ejecta
\citep[e.g.,][]{RS}. These characteristics revealed by the RS can
in turn constrain the unknown nature of the central engine and
energy extraction mechanism. The brightness and spectral peak of
the RS emission depend on the
magnetization of ejecta (magnetically-dominated ejecta leads to a weaker RS),
and the shock strength (mildly relativistic shock leads to lower peak
frequency). It is also thought that the RS generates short-lived
intense emission at optical \citep[e.g.,][]{990123} and/or radio
wavelengths \citep[e.g.,][]{990123radio,130427a1}. Such emission is expected to be
2--3 orders of magnitude brighter than those generated by the FS emission.
Hence, for the purpose of probing the high-$z$ Universe with GRBs,
it is desirable to detect the emission originating from the RS
systematically, whereby we will be able to determine
their typical occurrence conditions \citep*{inoue07}.
Past searches with numerous
rapid optical follow-ups failed to detect the RS emission. A possible
reason for the missing RS component is that the typical
reverse shock synchrotron frequency is far below the optical band.
As we demonstrated with a rapid SMA ToO observation of
GRB120326A \citep{120326a}, mm/submm observations are the key
to catching the RS component and understanding the emission mechanism of GRB
afterglows.

We identify the following three scientific goals for the GRB studies
with the GLT: (i) systematic detection of bright submm emissions
originating from RS in the early afterglow phase,
(ii) characterization of FS and RS emissions by capturing
their peak flux and frequency and performing continuous monitoring,
and (iii) detections of GRBs as a result of the explosion of
first-generation stars at high redshift through
systematic rapid follow-ups.
As summarized in \citet{urata15}, the light curves and spectra
calculated by available theoretical models clearly show that the GLT
could play a crucial role in these studies.
The main workhorse wavelengths are mm and submm bands, where
observations can be managed even under the marginal weather conditions
for the THz sciences. Hence, rapid follow-ups will be flexibly
managed by the semi-automated responding system for the GRB alerts
with secure procedures at the extreme site.  We installed the prototype
system at the Kiso observatory \citep{urata03}. A dedicated GRB satellite,
the Space-based Multiband Astronomical Variable Objects Monitor
\citep[SVOM;][]{svom} will be available in the GLT era and the mission
will explore the anti-Sun direction for the GRB hunting. The
combination of the SVOM and the GLT will be able to make rapid GRB
follow-ups without any delay caused by unfavorable visibility and
to perform continuous mm/submm monitoring over days.
This combination could also address the origin of X-ray flashes
(XRFs), which are thought to represent a large portion of the entire
GRB population.  Three models have been proposed and tested for XRFs:
a high redshift origin \citep{heise03}; the off-axis jet model
\citep{offaxis,yamazaki,urata15b}, which is equivalent to the
unification scenario of AGN galaxies; and intrinsic properties (e.g.,
a subenergetic or inefficient fireball), which may also produce
on-axis orphan afterglows \citep{huang02}. The third model could be
tested by the identification of a delayed RS peak through rapid GLT
monitoring.

The flux of the RS component is typically 10--100 mJy at 230 GHz for a
GRB at $z=1$ in a day after the burst. Such a flux level is easily
detected with the 230 GHz detector planned to be used for the VLBI
observation, with an on-source integration time of a few
minutes (Section \ref{subsec:capability}). For a $z=10$ GRB,
we need to detect a flux level of a few mJy; to achieve
an rms of 1 mJy, an on-source integration time of 28 min is required
based on the median continuum NEFD at 230 GHz in
Table \ref{tab:capability}.

\section{Possible developments of instruments}\label{sec:development}

Here we summarize possible developments for the future
THz and sub-THz observations. As representative cases,
we list the efforts for the GLT based on
\citet{groppi10}, \citet{grimes14}, and
\citet{thomas15}.
We refer the interested reader to those papers
(in particular, \citealt{grimes14}).

\subsection{VLBI Receivers}

There are cartridge-type VLBI receivers to cover the 230 GHz and 350 GHz
frequency windows.
The 230 GHz receiver covers a frequency range from 224 to 238 GHz
with a circular polarizer and an IF band of 4--8 GHz. The 350 GHz receiver is a copy of the ALMA band-7 receiver with dual linear polarization and upper/lower side IF bands of 4--8 GHz. The 230 and 350 GHz receivers have been delivered and under integration and testing in the VLBI cryostat.

\subsection{350 and 650 GHz Multibeam Receivers}

Smithsonian Astrophysical Observatory (SAO) is developing a
superconductor-insulator-superconductor (SIS) multibeam array receiver
for the 325--375 GHz atmospheric window. A 48-pixel feedhorn array, populated
with six eight-pixel modules, will be built with an IF band of 4--6 GHz. The expected receiver noise temperature is 75 K, which is similar to the performance of the SMA receivers. The prototype feedhorn array (provided by Oxford Astrophysics) and the first batch of mixers (provided by Academia Sinica Institute of Astronomy and Astrophysics; ASIAA) are under evaluation. The 650 GHz multibeam array is still in the stage of conceptual studies. A receiver with $\sim$256 pixels is proposed to reach $\sim$4 times the mapping speed of Supercam, a 64-pixel receiver deployed on the 10 m Arizona Radio Observatory Submillimeter Telescope.

\subsection{THz Multi-beam Receivers}\label{subsec:THzreceiver}

Small multibeam THz receivers are planned to be built
based on the hot-electron-bolometer (HEB) technique. There are
two plans for the development: one is a 4-pixel 1.4 THz receiver fed by smooth-wall feedhorns, and the other is a 4-pixel receiver, fed by a Si-lens, at frequency 1.45--1.55 THz. The latter will be extended to 9 pixels in the future. The power limitation of the local oscillator (LO) is the biggest challenge for THz multibeam receivers.
SAO is investigating the use of a polarization multiplexing technique with a balanced mixer operation to reduce the power requirement of the HEB array. ASIAA is using a cooled multiplier to increase the total LO power.
The expected receiver noise temperature is around 1000--2000 K with an IF band of 0.5--4 GHz.

\subsection{Bolometer Arrays}

As emphasized in Section \ref{sec:subTHz}, it is worth
considering bolometer arrays at sub-THz frequencies for
large surveys. The NEFD of 40, 110, and 400 mJy beam$^{-1}$
with 1-sec integration at 350, 650, and 850 $\micron$,
respectively, is feasible
with the current
technology. 

Another effort worth mentioning is the CAMbridge Emission Line Surveyor
(CAMELS), which is being developed by 
SAO and the Detector Physics Group at Cavendish Laboratory.
CAMELS is an on-chip spectrometer at frequencies from 103 to 114.7 GHz,
providing 512 channels with a spectral resolution of $R=3000$. The CAMELS integrates a 4-probe orthogonal-mode-transducer (OMT), integrated filter band spectrometer (IFBS) and microwave kinetic inductance detectors (MKIDs) on one chip using superconductor technology. The targeted NEPs (noise equivalent power) at the low and the top end of the band are $1\times 10^{-18}~\mathrm{W/\sqrt{Hz}}$ and $4\times 10^{-18}~\mathrm{W/\sqrt{Hz}}$, respectively. The expected line sensitivities based on our atmospheric conditions are from $5\times 10^3~\mathrm{mJy~km~s^{-1}}$ to $3\times 10^4~\mathrm{mJy~km~s^{-1}}$.

\section{Summary}\label{sec:summary}

Our scientific cases described in this paper are broadly
categorized into\\
(A) chemistry and evolution in the diffuse to dense
interstellar medium (ISM);\\
(B)  collective
or integrated effects of star formation seen in
extragalactic objects; and\\
(C) the time-variable submm Universe.\\
We review these three categories below. The requirements
are briefly summarized in Table \ref{tab:requirement}.

\begin{table*}
  \tbl{Specific requirements for each scientific
  objective.}{%
  \begin{tabular}{@{}lcccccccc@{}}
  \hline
 Target & Frequency & $\theta$\,$^\mathrm{a}$ & $\Delta v\,^\mathrm{b}$ & Sensitivity$^\mathrm{c}$ & $t\,^\mathrm{d}$ & Instrument$\,^\mathrm{e}$ &
 Section$^\mathrm{f}$ & Cat.$^\mathrm{g}$\\
  \hline
  ISM CO(13--12) & 1.5 THz & a few$''$ & 0.2\,km\,s$^{-1}$ & 83\,Jy\,beam$^{-1}$ & 30\,min & Spectrometer & \ref{subsubsec:THz_SF} & A\\
  ISM [N \textsc{ii}] & 1.5 THz & a few$''$ &
  2\,km\,s$^{-1}$ & 18 Jy beam$^{-1}$& 13 min & Spectrometer & \ref{subsec:Gal_NII} & A\\
  ISM continuum & 1--1.5 THz & a few$''$ & --- & 0.4\,Jy\,beam$^{-1}$ & 34\,min & THz bolometer & \ref{subsubsec:cont_SF} & A\\
  Polarization & 1--1.5 THz & a few$''$ & --- & ---$^\mathrm{h}$ & 10\,hr & Polarimetry & \ref{subsubsec:pol} & A\\
  BCDs (dust) & 1.5 THz & a few $''$ & --- & 0.2\,Jy\,beam$^{-1}$ & 30\,min & THz bolometer & \ref{subsubsec:BCD} & A\\
  BCDs ([N\,\textsc{ii}]) & 1.5 THz & a few$''$ & 10\,km\,s$^{-1}$ & 6.5\,Jy\,beam$^{-1}$ & 3\,hr & Spectrometer & \ref{subsec:NII_extragal} & A \\
  Starburst interface & 1--1.5 THz & a few$''$ & --- & 0.5\,Jy\,beam$^{-1}$ & 6\,min & THz bolometer & \ref{subsubsec:starburst} & A\\
  High-$z$ galaxies & 850/650 GHz & $\sim$10$''$ & --- & 4\,mJy\,beam$^{-1}$ & 1000\,hr & Bolometer array & \ref{subsec:highz} & B\\
  AGNs  & submm & --- & --- & $\sim$1\,Jy\,beam$^{-1}$ & 30\,min & VLBI detector & \ref{subsec:AGN} & C\\
  GRBs & submm & --- & --- & $\sim$a few mJy & 3--48 hr & VLBI detector & \ref{subsec:grb} & C\\
  \hline
\end{tabular}}\label{tab:requirement}
\begin{tabnote}
$^\mathrm{a}$Angular resolution. $^\mathrm{b}$Velocity (or
frequency) resolution, or continuum. $^\mathrm{c}$1 $\sigma$ noise level
requested for detection. $^\mathrm{d}$Integration time per pointing.
For polarization, the time per source. For the high-$z$ galaxy survey,
the total survey time. For AGN monitoring, the total on-source
time for all the objects in Table \ref{can-sour}. For GRBs, the total
duration of monitoring of a source.
$^\mathrm{e}$Spectrometer or bolometer array. $^\mathrm{f}$Section discussing
the case.
$^\mathrm{g}$Category in Section \ref{sec:summary}.
$^\mathrm{h}$See Section \ref{subsubsec:pol} for the
sensitivity estimate.
\end{tabnote}
\end{table*}

\subsection{Category (A)}

For category (A), it is recommended that
we explore the new window, that is, terahertz (THz), because of
its uniqueness. The scientific importance is
summarized in what follows.

The CO $J=13$--12 line at 1.4969 THz is a unique tracer of
warm (300--500 K) regions in the vicinity of newly formed stars
(Section \ref{subsubsec:THz_SF}). An example of
detection by APEX \citep{wiedner06} indicates that
this observation is feasible with currently available
technology. In the diffuse ISM, there are other lines
emitted from the species shown in Figure \ref{fig:PDR_Oscar}.
Among them, [N\,\textsc{ii}] 205 $\mu$m emission is strong
enough for the first-generation THz facilities to detect
(Section \ref{subsec:Gal_NII}).
Observations of [N\,\textsc{ii}] 205 $\mu$m give a useful
``calibration'' of fine structure lines as they are often used
for star-formation indicators in high-redshift galaxies by ALMA.
Using a nearby extragalactic sample with a variety of metallicities,
THz spectroscopic observations can test the metallicity
dependence of [N\,\textsc{ii}] 205 $\mu$m emission
(Section \ref{subsec:NII_extragal}).

Ground-based THz continuum observations have importance in
high angular resolutions. For example, the resolution achieved
by the GLT is suitable for resolving the individual star-forming
site (Section \ref{subsubsec:cont_SF}). THz continuum polarization
is also an important scientific target (Section \ref{subsubsec:pol}).
Since THz is near to the dust spectral energy distribution (SED) peak,
we can trace the magnetic field structure
of the major dust component responsible for
the far-infrared emission. Although it is reasonable to start observations
with relatively bright Galactic objects, some possibilities of observing
extreme extragalactic star-forming
activities are also discussed
(Section \ref{subsec:extragal}).
For those objects, the high spatial resolution
of the GLT is crucial to extract/resolve the regions of interest.
For extragalactic continuum studies, development of
sensitive bolometer-type THz detector is desirable.

\subsection{Category (B)}

For this category, survey-type observations are crucial
to obtain a general picture, because only a large number of
extragalactic objects can provide a general picture of
the star formation activities occuring in the history of
the Universe. To utilize the good atmospheric
conditions in the GLT site, it is preferable to concentrate
on relatively short wavelengths (high frequencies) in
the submillimeter (submm) regime. Thus, we
target the 350 and 450 $\mu$m ($\sim$850 and $\sim$650 GHz, respectively)
windows. The largest advantage of the ground-based THz telescopes
is again their high angular resolutions, which are useful to
overcome the confusion limit of \textit{Herschel}
(Section \ref{subsec:highz}).
An area larger than 400 arcmin$^2$ can be surveyed with a reasonable
observational time ($\sim$1000 hr). Installing the most
advanced bolometer arrays is cruicial to maximize the survey
efficiency.

\subsection{Category (C)}

Some ground-based submm--millimeter (mm) telescopes such as the GLT
have possibilities of flexible scheduling. This category is unique
in the sense that it opens up time-domain submm--mm astronomy.
In addition to the target-of-opportunity mode for $\gamma$-ray
bursts already discussed in another paper by \citet{urata15},
we discussed a possibility of monitoring blasars, which are
strongly time-variable (Section \ref{subsec:AGN}). The
collaboration with submm very long baseline interferometry (VLBI)
is interesting: if an interesting event is detected in the single-dish
observation, a submm VLBI observation mode is triggered.
This is a unique science case
that connects the VLBI and single dish modes.
We should also emphasize that
the submm (or THz) emission traces
synchrotron emission, which gives
constraints on magnetic field strengths. As a consequence,
we can address the dynamical properties in which magnetic fields play a central
role (e.g., collimation).

\subsection{Requirements for instruments}

For category (A), because of the diffuse nature of the ISM, a multi-pixel
detector is strongly favored. The current status of
development of THz detectors was described in Section \ref{subsec:THzreceiver}.
Targetting the 1.5 THz window rather than the 1.0 THz window
is better because interesting
lines exist (Section \ref{subsec:line}), and the separation from
already explored submm wavelengths is larger (Section \ref{subsec:cont}).

For category (B), a large submm detector array is required.
Multi-band detectors are preferred,
considering dust temperature determination.
The uniqueness of the GLT site is the good atmospheric condition so that
relatively high frequencies referred to as sub-THz
($\sim$850 GHz and $\sim$650 GHz) are
preferred to make the maximum use of this uniqueness.
These frequencies also have an advantage that
they are also covered by \textit{Herschel}; i.e.,
the \textit{Herschel} catalog or survey data can be utilized as input.

For category (C), no particular detector capabilities are required.
Thus, for the GLT, we can simply use
the mm/submm detectors developed for the VLBI observations.
Flexibility of time allocation should be kept if we want
to pursue this category.

\section*{Acknowledgments}

We are grateful to the anonymous referee for useful comments
and R. Blundell and S. Paine for continuous support.
We thank the Ministry of Science and Technology
(MoST) in Taiwan for support through grants 102-2119-M-001-006-MY3 (HH),
103-2112-M-001-032-MY3 (SM) and
102-2119-M-001-007-MY3 (WHW).
PMK acknowledges support from grant MoST 103-2119-M-001-009
and an Academia Sinica Career Development Award.



\appendix

\section{Simulation of Dust Temperature Estimate}

We examine how useful adding THz data is to determine
the dust temperature. 
We assume that the dust emission is optically thin, which
is valid for the objects of interest in this paper.
We adopt a power-law wavelength
dependence of the mass absorption coefficient ($\kappa_\nu$) of
dust at THz frequencies.
Then, the flux at frequency $\nu$ is written as
\begin{eqnarray}
f_\nu =C\nu^\beta B_\nu (T),\label{eq:grey}
\end{eqnarray}
where $C$ is the normalizing constant, which is proportional
to the dust mass multiplied by $\kappa_\nu$
[see the footnote for equation (\ref{eq:F_comp})], and
$B_\nu (T)$ is the Planck function.
In principle, it is possible to determine both
$\beta$ and $T$ if we have data at three or more
wavelengths (since $C$ is also unknown).
However, as shown by \citet{schnee07},
errors as small as $\sim 2$\% is required to determine
both. Thus, we fix
$\beta =1.5$ \citep*{hirashita07}. Since we focus
on the goodness of temperature estimate, the value of
$\beta$ chosen does not affect our conclusion. We
survey a larger range of $T$ than \citet{schnee07},
since we target objects with
a wide dust temperature range. Our focus is also different
from theirs in that we emphasize the importance of
THz frequencies; in particular, we include frequencies
$>$850 GHz or wavelengths $<$350 $\mu$m into our analysis.

\begin{figure*}
\begin{center}
\includegraphics[width=0.45\textwidth]{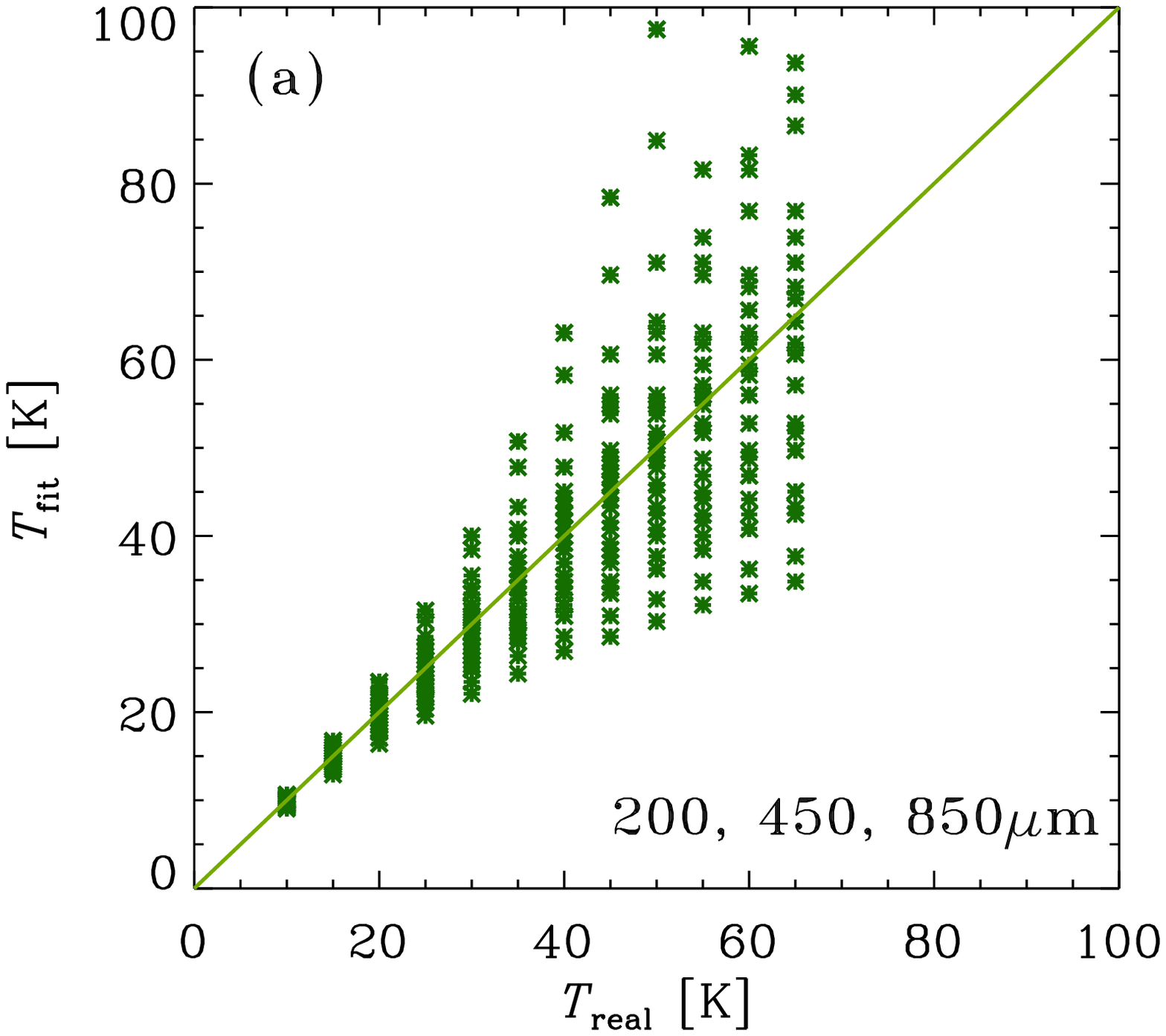}
\includegraphics[width=0.45\textwidth]{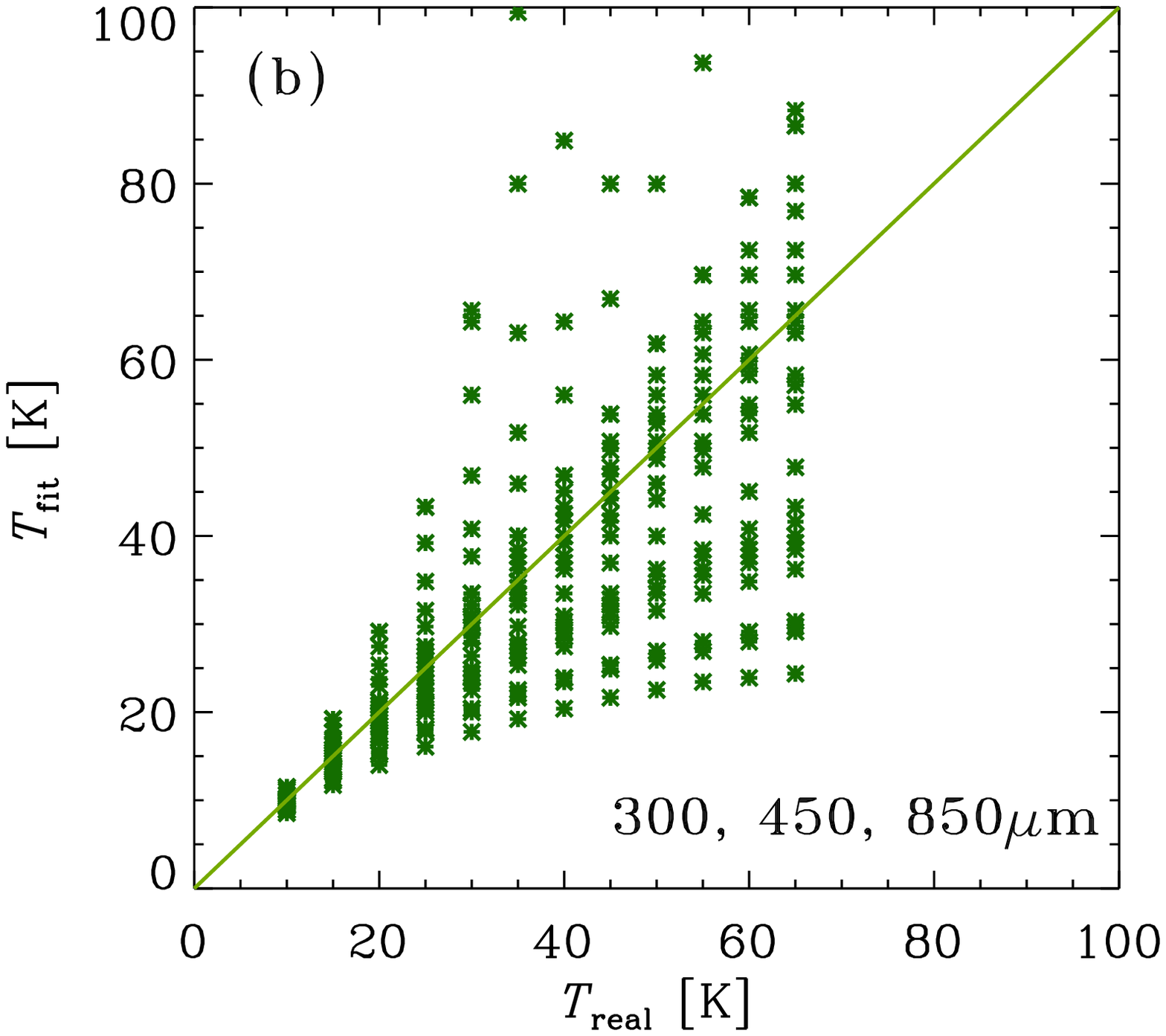}
\includegraphics[width=0.45\textwidth]{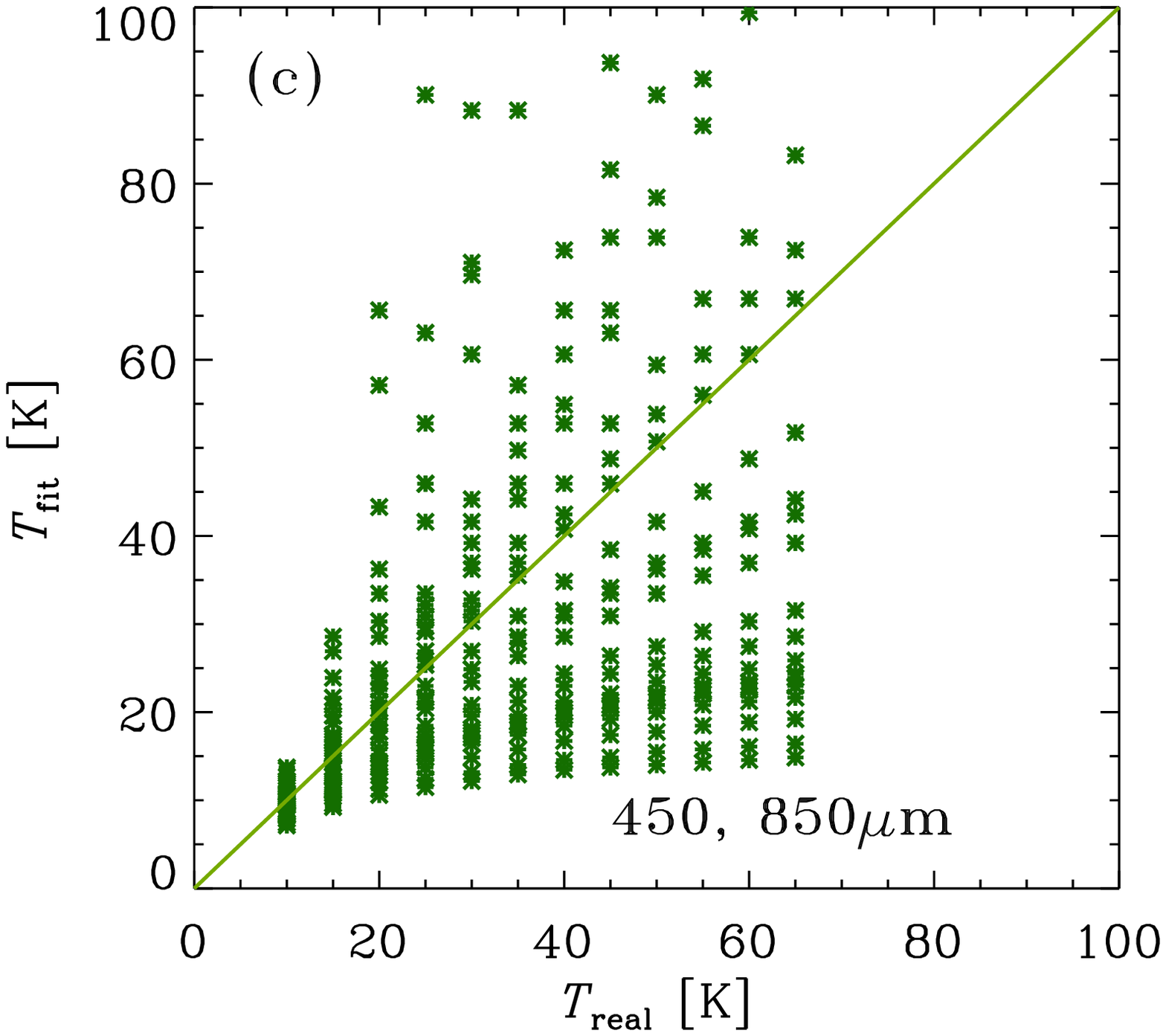}
\includegraphics[width=0.45\textwidth]{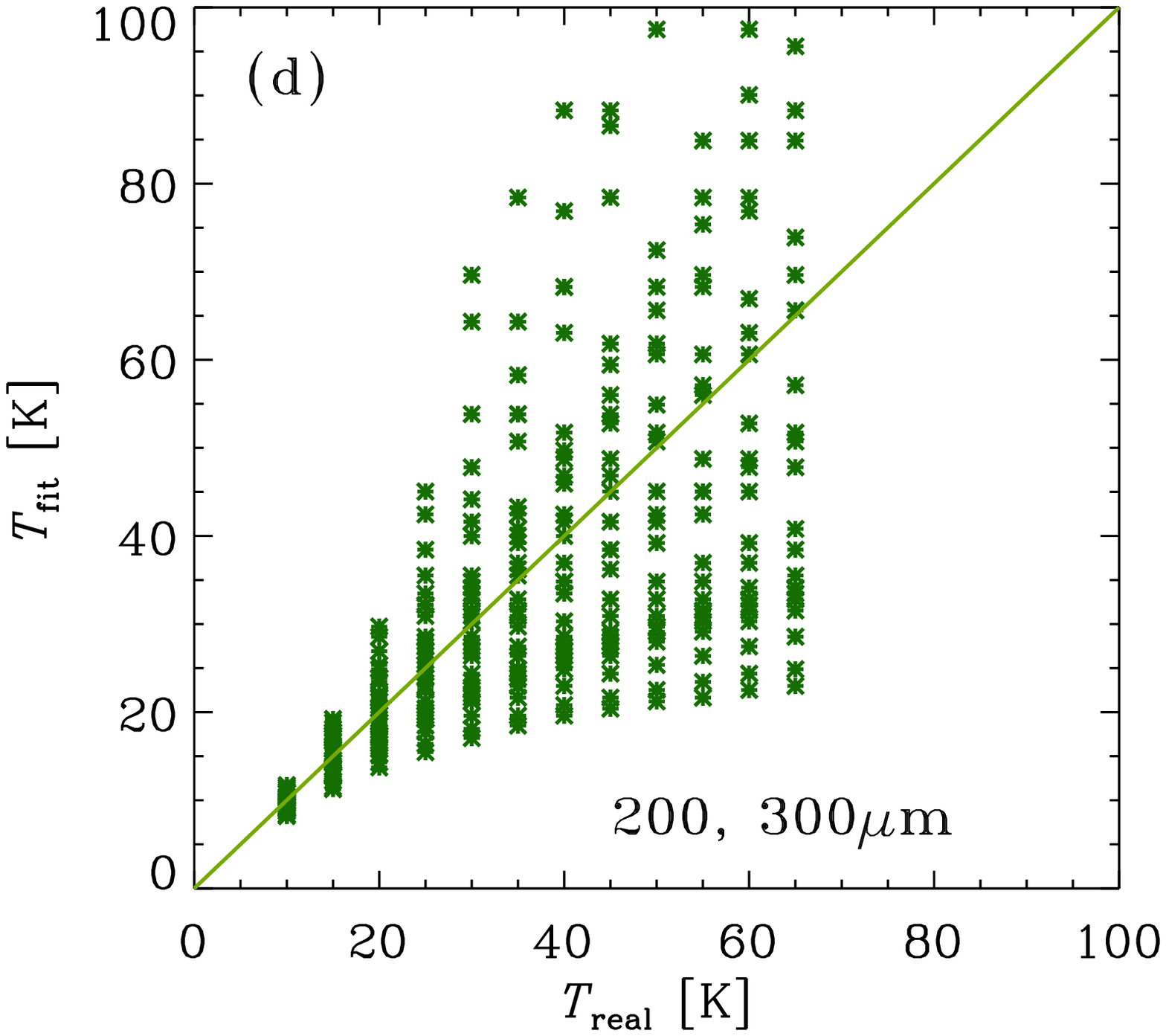}
\end{center}
\caption{Simulations of dust temperature estimate for maximum
error $\Delta =30$\%.
$T_\mathrm{real}$ is the input dust temperature while
$T_\mathrm{fit}$ is the dust temperature derived by the
SED fitting. We use the data at
(a) 200, 450, and 850 $\mu$m;
(b) 300, 450, and 850 $\mu$m; (c) 450 and 850 $\mu$m,
and (d) 200 and 300 $\mu$m.
\label{fig:Tdust_err}}
\end{figure*}

First, we add random noise to the flux calculated by
equation (\ref{eq:grey}) and produce ``observed fluxes''
at a few given wavelengths. Next, based on
these artificially made observed fluxes, we
estimate the dust temperature. This
estimated dust temperature is compared with
the input dust temperature to
examine how precisely we can obtain the dust
temperature. To realize this procedure, we run the
following Monte Carlo simulations:
\begin{enumerate}
\item With a certain value of $T$, we produce $f_\nu$
at given frequencies. Note that the value of $C$ does not
affect the temperature estimate so we fix $C$.
The given dust temperature
is denoted as $T_\mathrm{real}$. For the same value of
$T_\mathrm{real}$, we produce thirty realizations according
to the following processes.
\item We add a random independent error to $f_\nu$
at each frequency. The percentage of
the error $\delta_f$ is chosen randomly between $-\Delta$ and
$+\Delta$ with $\Delta =30$\% chosen unless otherwise stated.
The data with the error is denoted as
$\tilde{f}_\nu =f_\nu (1+\delta_f)$.
\item We fit $\tilde{f}_\nu$ at all the wavelengths
with equation (\ref{eq:grey}) with two free parameters
($C$ and $T$) by minimizing the sum of the squares of logarithmic
difference. The obtained dust temperature is denoted
as $T_\mathrm{fit}$, which is compared with the input dust temperature,
$T_\mathrm{real}$.
\end{enumerate}

We examine the following four sets of wavelengths in units of
$\mu$m: (200, 450, 850), (300, 450, 850), (450, 850)
and (200, 300).
The first and second cases are compared to
judge which of the THz atmospheric windows, 200 $\mu$m
(1.5 THz) or
300 $\mu$m (1.0 THz), in addition to often used submm
wavelengths (450 and 850 $\mu$m), is useful to obtain
precise dust temperatures. The third case represents
the case where we only have submm observations.
The last is for the case with only THz observations.

In Figure \ref{fig:Tdust_err}, we show the comparison between
the input dust temperature $T_\mathrm{real}$ and the obtained
dust temperature $T_\mathrm{fit}$.
Comparing the wavelength sets of
(200, 450, 850) and (450, 850) shown in
Figures \ref{fig:Tdust_err}a and c, respectively,
we find that the addition of 200 $\mu$m
improves the dust temperature estimate significantly,
especially for the range of dust temperature
typical of normal spiral galaxies (15--30 K).
This confirms the conclusion drawn by \citet{schnee07}
that adding a data point
at a short wavelength (in their case, 350 $\mu$m)
improves the dust temperature estimate.
Comparing Figures \ref{fig:Tdust_err}a and b,
we find that 200 $\mu$m rather than 300 $\mu$m
is preferable to suppress the error at
20--40 K; this is because adding a wavelength farther
from the submm wavelengths is more effective.
Therefore, the addition of a THz frequency
to submm frequencies actually improves the
dust temperature estimate and the 1.5 THz (200 $\mu$m)
band rather than
the 1.0 THz band (300 $\mu$m) is preferable.

The temperature estiamtes with only two THz bands
(Figure \ref{fig:Tdust_err}d) is not as good as
the combination of a single THz band and submm bands
(Figure \ref{fig:Tdust_err}a). This is because
the wavelengths of the two THz bands are too close.
Therefore, the best scenario is that, in addition to
a THz data point,
we have multiple submm data with currently available
submm interferometers whose angular resolution
at submm is comparable to or better than that in the
THz bands.

In the above, we assumed conservatively large errors.
We expect that, if we increase the number of data
points, the temperature estimate becomes more
precise. In reality, because of the limited atmospheric
window, we cannot increase the number of data points
arbitrarily, but rather we can increase the integration
time or the band width within technical constraints. This
leads to an improvement in the signal-to-noise ratio.
Therefore, it is worth examining the effects of
signal-to-noise ratio (or error) on the dust
temperature estimate.

\begin{figure*}
\begin{center}
\includegraphics[width=0.45\textwidth]{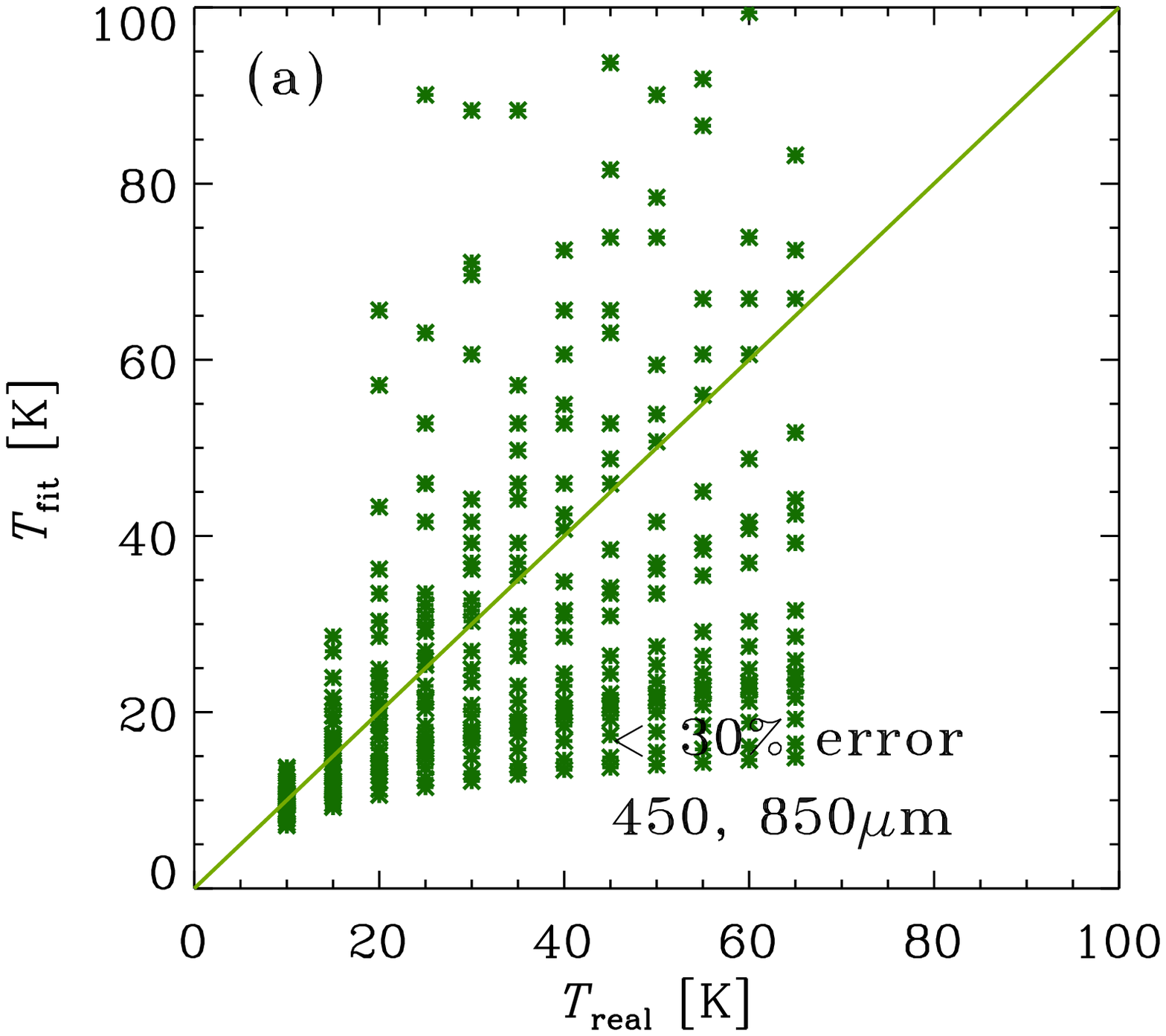}
\includegraphics[width=0.45\textwidth]{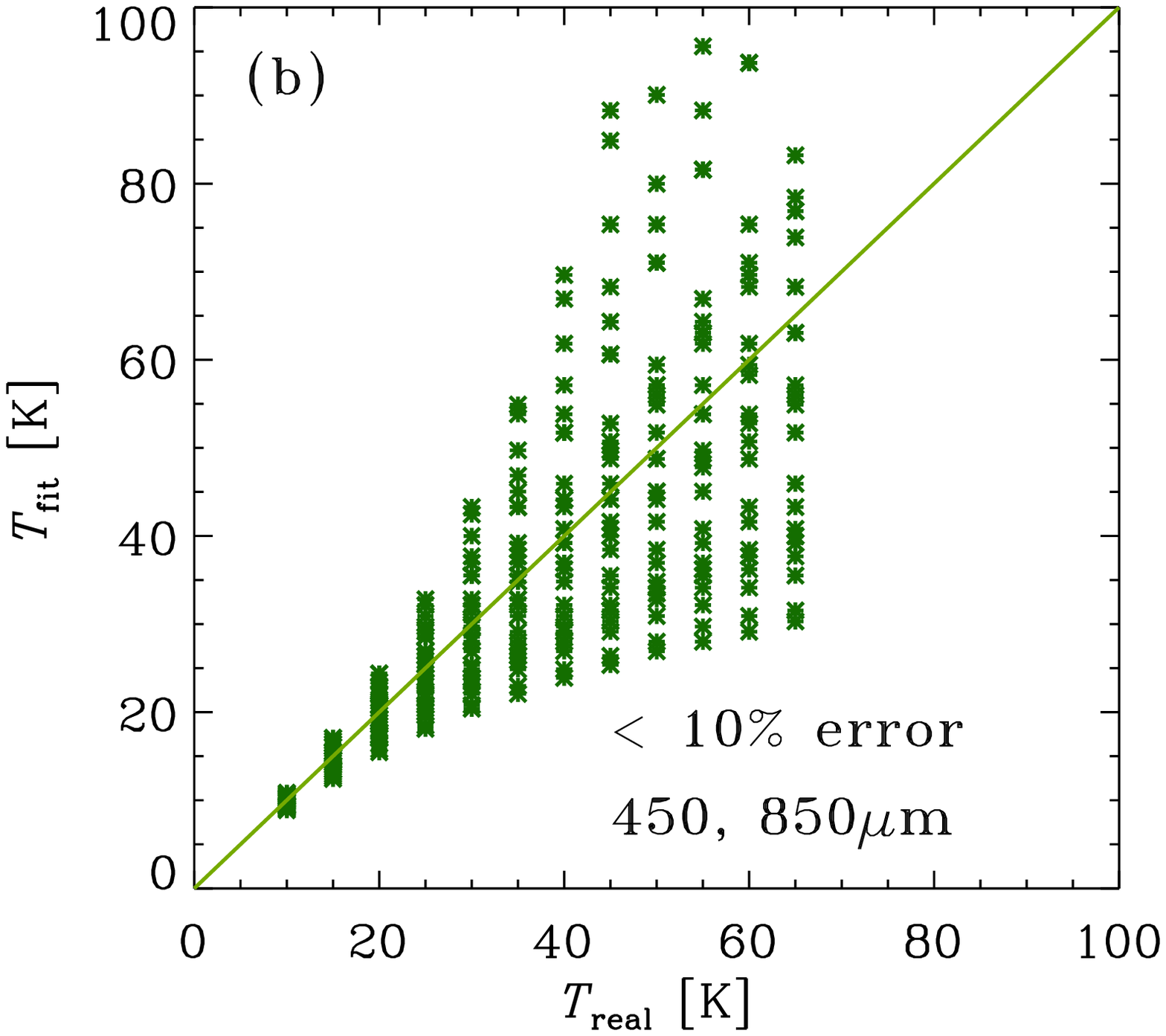}
\includegraphics[width=0.45\textwidth]{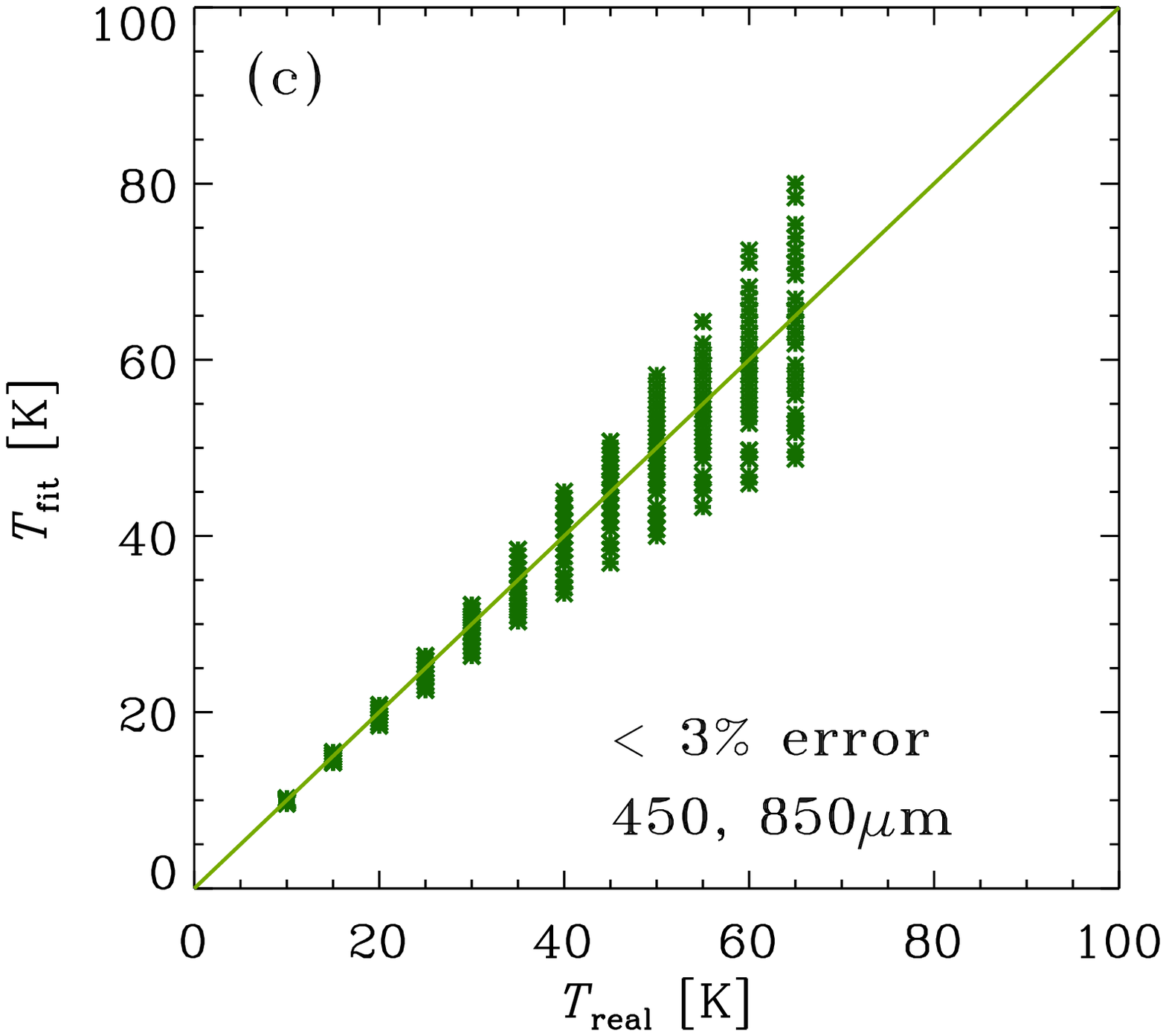}
\includegraphics[width=0.45\textwidth]{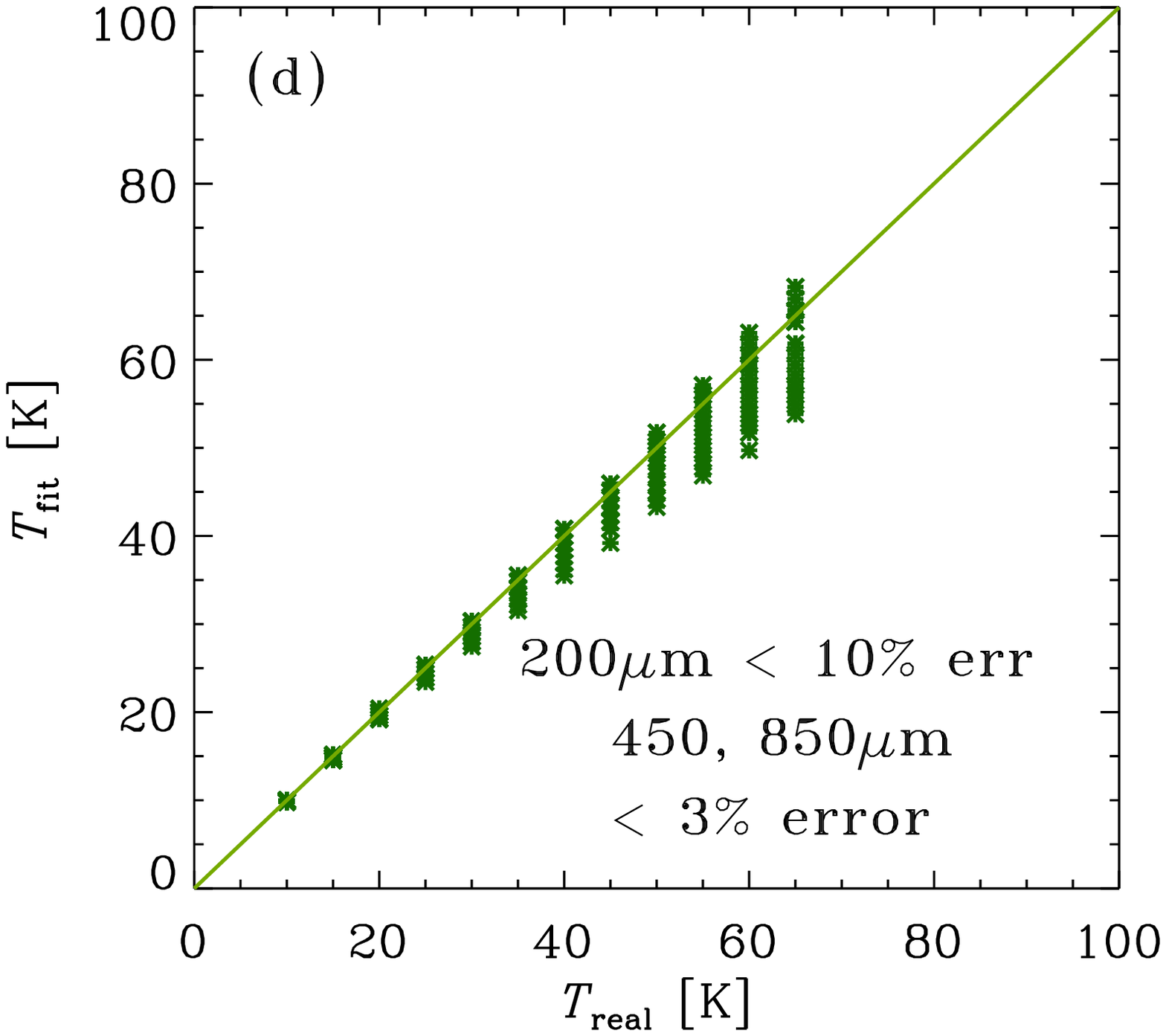}
\end{center}
\caption{Same as Figure \ref{fig:Tdust_err}, but for various
errors $\Delta$. Panels (a), (b), and (c) represent
the temperature determinations with 450 $\mu$m and
850 $\mu$m data for $\Delta =30$\%, 10\%, and 3\%,
respectively. Panel (d) presents the result with
an addition of 200 $\mu$m data to Panel (c). 
\label{fig:Tdust_450_850}}
\end{figure*}

In Figure \ref{fig:Tdust_450_850}, we show three cases
for various errors $\Delta$: 30\%
(same as Figure \ref{fig:Tdust_err}c), 10\% and 3\%.
We observe that, if the measurement error is very small,
we can get a precise dust temperatures for
$T_\mathrm{dust}\lesssim 40$ K. Yet, if we add a THz
data point with an error $<$10\%, the temperature estimate at
$T_\mathrm{dust}\gtrsim 40$ K is much improved.

\begin{figure*}
\begin{center}
\includegraphics[width=0.45\textwidth]{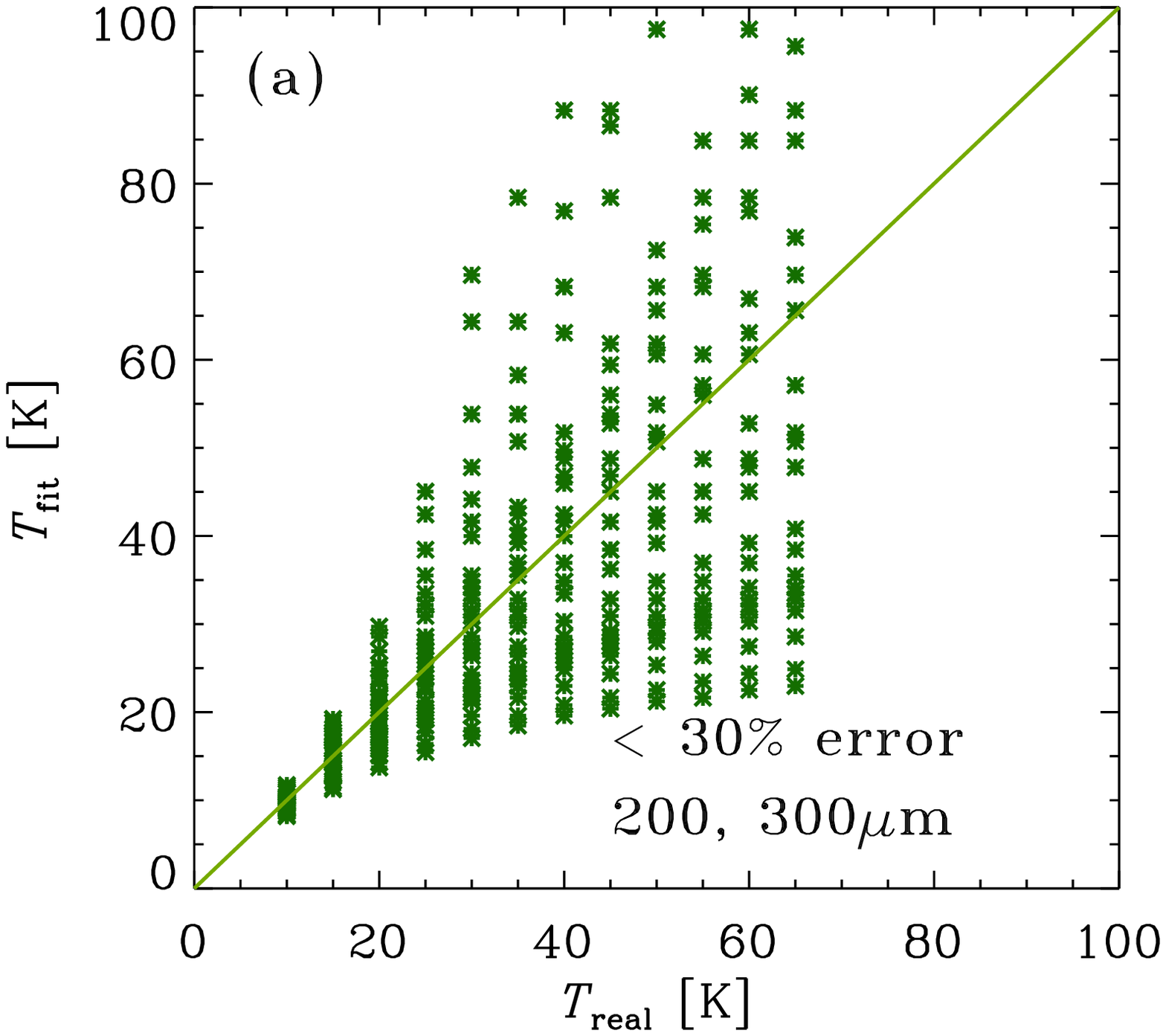}
\includegraphics[width=0.45\textwidth]{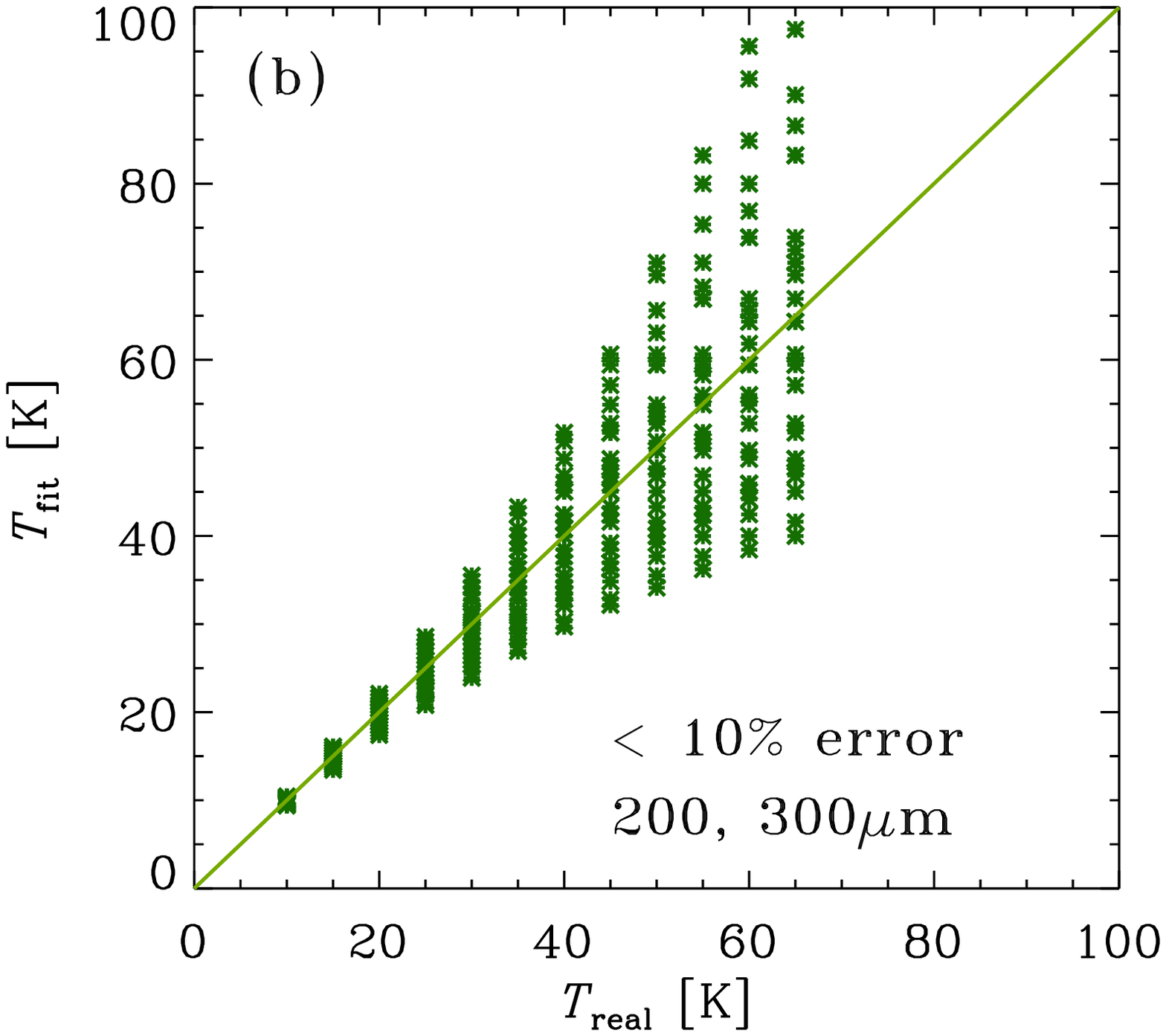}
\includegraphics[width=0.45\textwidth]{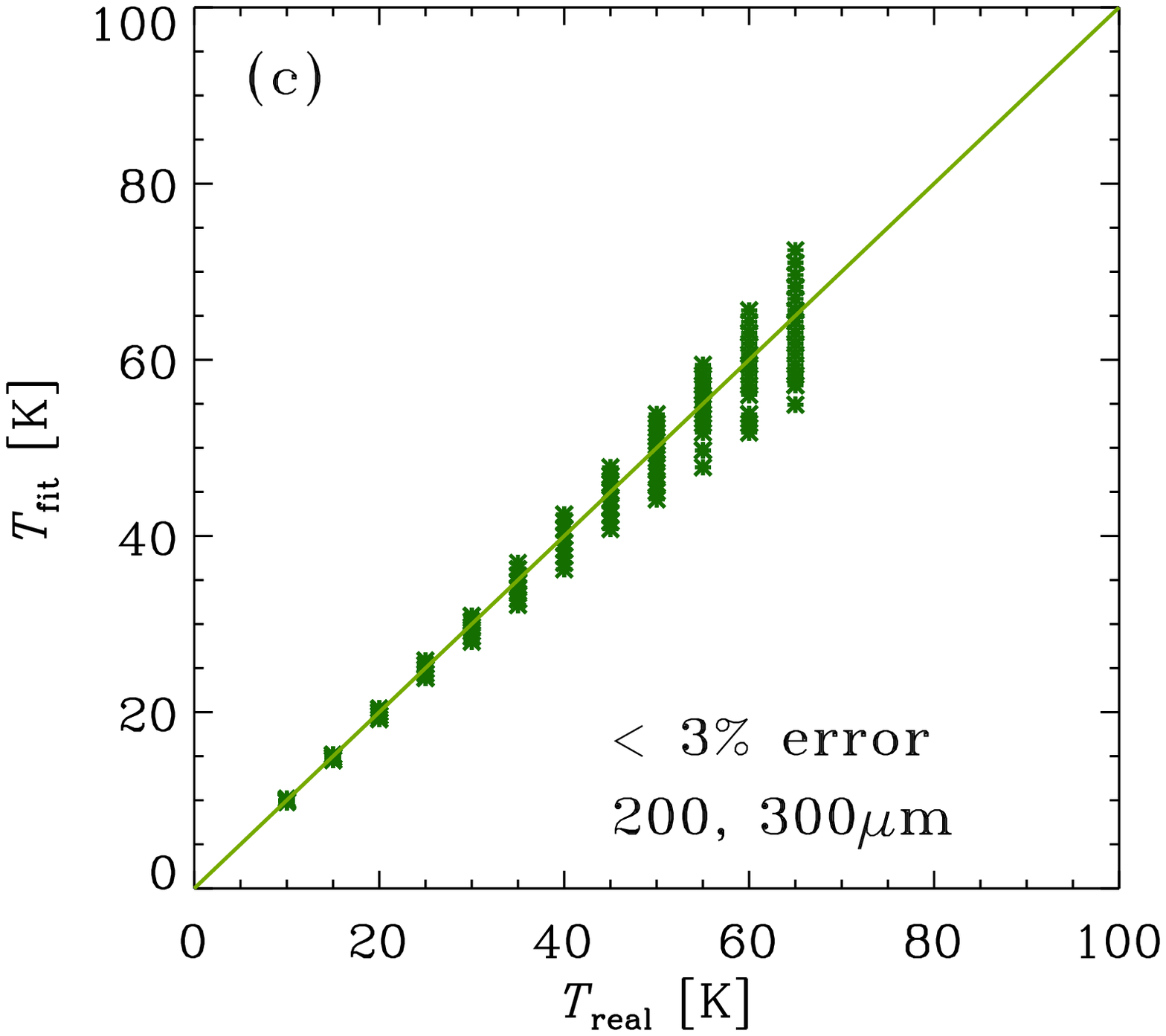}
\end{center}
\caption{Same as Figure \ref{fig:Tdust_450_850}, but for
the set of wavelengths of (200~$\mu$m, 300 $\mu$m).
Panels (a), (b), and (c) represent
the temperature determinations with $\Delta =30$\%
10\%, and 3\%,
respectively.
\label{fig:Tdust_200_300}}
\end{figure*}

Finally, we also examine the effect of errors on the temperature determination
only by the two THz bands in Figure \ref{fig:Tdust_200_300}. As already shown in
Figure \ref{fig:Tdust_err}, if the errors are as large
as 30\%, we cannot obtain precise values.
To obtain a precise enough dust temperature, we need
very precise measurements
whose typical errors are $\lesssim 3$\%. Therefore,
we recommend to combine THz observations with
submm data obtained by existing submm interferometers.

\end{document}